\documentclass[
aps,
nofootinbib,
prd,
superscriptaddress,
tightenlines,
notitlepage,
twocolumn,
showpacs,
floatfix
]{revtex4-1}
\usepackage{graphicx}
\usepackage{dcolumn}
\usepackage{bm}

\usepackage{orcidlink}
\usepackage{amsmath}
\usepackage{latexsym}
\usepackage{amsfonts}
\usepackage{graphicx}
\usepackage{mathrsfs}
\usepackage{epstopdf}
\usepackage{subfigure}
\usepackage[utf8]{inputenc}
\usepackage{CJK}
\usepackage{float}
\usepackage{xcolor}
\usepackage{color}
\usepackage{hyperref}
\usepackage{diagbox}
\usepackage{bm}
\usepackage{lipsum}
\usepackage[normalem]{ulem}
\hypersetup{
colorlinks=true,
linkcolor=red,
citecolor=blue,
}
\newcommand{\barlambda}{%
  \mathbin{%
    \ooalign{%
      $\lambda$\cr
      \hidewidth\raisebox{0.7ex}{\rule[0.5pt]{0.9ex}{0.4pt}}\hidewidth%
    }%
  }%
}
\newcommand{\reply}[1]{#1}

\begin{document}

\title{Scalarized neutron stars with a highly relativistic core in scalar-tensor
gravity}

\author{Peixiang Ji}
\affiliation{Department of Astronomy, School of Physics, Peking University,
Beijing 100871, China}
\affiliation{Kavli Institute for Astronomy and Astrophysics, Peking University,
Beijing 100871, China}

\author{Lijing Shao}\email[Corresponding author: ]{lshao@pku.edu.cn}
\affiliation{Kavli Institute for Astronomy and Astrophysics, Peking University,
Beijing 100871, China}
\affiliation{National Astronomical Observatories, Chinese Academy of Sciences,
Beijing 100012, China}


\begin{abstract}
Compact stars in scalar-tensor (ST) gravity have been extensively investigated, but
relatively few studies have focused on highly relativistic neutron stars (NSs)
with an extremely dense core region where the trace of the energy-momentum
tensor reverses its sign. In this regime, we identify the origin of the phenomenon where {\it multiple} scalarized solutions exist for a {\it fixed} central density, arising from the oscillatory profile of the scalar field inside the star. This origin further indicates that the multi-branch structure emerges for both negative and positive
$\beta$, the quadratic-term coefficient in the effective coupling function between the scalar field and conventional matter in the Einstein frame. By comparing the
Damour--Esposito-Far\`ese and Mendes-Ortiz models of the ST gravity,
we demonstrate that their distinct scalarization behaviors stem from whether the effective
coupling function is bounded. We also compute for scalarized NSs with a highly relativistic dense core in ST theories the moment of inertia and
tidal deformability that are relevant to pulsar-timing and
gravitational-wave experiments.
\end{abstract}

\maketitle

\section{Introduction}\label{sec:1}

Theoretical motivations for modifying general relativity (GR)‌ have long been
pursued to address persistent challenges: ‌fundamental conflicts‌ with quantum
field theory~\cite{tHooft:1974toh} and unresolved issues including spacetime
singularities~\cite{Hawking:1970zqf} and the information loss
paradox~\cite{Hawking:1976ra}, as well as observational anomalies‌ such as
galaxies' rotation curves~\cite{Zwicky:1933gu, Zwicky:1937zza} and accelerated
cosmic expansion~\cite{SupernovaSearchTeam:1998fmf,
SupernovaCosmologyProject:1998vns}, commonly attributed to dark
matter~\cite{Kolb:1990vq, Bergstrom:2012fi, Bertone:2016nfn} and dark
energy~\cite{Weinberg:1988cp, Peebles:2002gy}, respectively.  ‌Scalar-tensor
(ST) theories‌, as prominent alternatives to GR, incorporate one or more
non-minimally coupled scalar fields alongside the spin-2 metric
tensor~\cite{Damour:1992we, Fujii:2003pa, Faraoni:2004pi}.  These theories
emerge naturally in unified candidate theories~\cite{Kaluza:1921tu} and provide
mechanisms for cosmic inflation~\cite{La:1989za, Steinhardt:1990zx}. Crucially,
ST theories also offer theoretical pathways to explain the cosmic acceleration
without invoking dark energy~\cite{Tsujikawa:2010zza, Ji:2024gdc}.

The general ST gravity action with non-minimal coupling can be expressed
as~\cite{Bergmann:1968ve, Nordtvedt:1970uv, Wagoner:1970vr}
\begin{align}\label{eq:actionJ}
    S=&\ \frac{1}{16\pi G}\int d^4x\sqrt{-\tilde g} \Big[F(\Phi)\tilde
    R-Z(\Phi)\tilde g^{\mu\nu}\partial_\mu\Phi\partial_\nu\Phi\nonumber\\
    &-2U(\Phi) \Big]+S_\mathrm{m}[\Psi_\mathrm{m};\tilde g_{\mu\nu}],
\end{align}
where $G$ denotes the bare gravitational coupling constant, $\tilde R$ is the
Ricci scalar curvature of the metric $\tilde g_{\mu\nu}$, and $\tilde g$ is its
determinant; $F$, $Z$, and $U$ are three arbitrary functions of $\Phi$.  Since
in the action~\eqref{eq:actionJ} matter is universally coupled to $\tilde
g_{\mu\nu}$, this ``Jordan frame (JF) metric", where the tildes denote the
metric and the metric-related quantities in, defines the length and time
actually measured by laboratory rulers and clocks that are made of ordinary
matter.  Thus, all experimental data maintain standard interpretations in this
frame. We 

However, it is usually much clearer to analyze in the so-called Einstein frame
(EF), defined by diagonalizing the kinetic terms of the metric and scalar
fields.  This is achieved thanks to a redefinition of the scalar field,
\begin{eqnarray}
    \Phi\to\varphi,
\end{eqnarray}
and a conformal transformation of the metric field,
\begin{eqnarray}
    \tilde g_{\mu\nu}\to g_{\mu\nu}=F(\Phi)\tilde g_{\mu\nu}.
\end{eqnarray}
Here, $\varphi$ and $g_{\mu\nu}$ are the new field variables, which are related
to the original ones by the following equations,
\begin{eqnarray}
   \frac{d\varphi}{d\Phi}&=&\pm\sqrt{\frac{3}{4}\left(\frac{d\ln
   F(\Phi)}{d\Phi}\right)^2+\frac{Z(\Phi)}{2F(\Phi)}},\\
   A(\varphi)&=&F^{-1/2}(\Phi).
\end{eqnarray}
With the transformation of the potential,
\begin{eqnarray}
    V(\varphi)&=&\frac12F^{-2}(\Phi)U(\Phi),
\end{eqnarray}
the action~\eqref{eq:actionJ} takes the form
\begin{align}\label{eq:actionE}
    S=&\ \frac{1}{16\pi G}\int d^4 x\sqrt{-g} \Big[R-2g^{\mu\nu} \partial_\mu
    \varphi \partial_\nu \varphi-4V(\varphi) \Big]\nonumber\\
    &+S_\mathrm{m}[\Psi_\mathrm{m};A^2(\varphi)g_{\mu\nu}],
\end{align}
where $g$ is the determinant of the EF metric $g_{\mu\nu}$, and $R$ is the
corresponding Ricci curvature scalar.  Within EF, the non-minimal coupling
manifests as an effective coupling between the gravitational field and matter,
as characterized by the coupling function,
\begin{eqnarray}\label{eq:alpha}
    \alpha(\varphi)=\frac{d\ln A(\varphi)}{d\varphi}.
\end{eqnarray}
Therefore, a ST theory is entirely determined by the two functions,
$\alpha(\varphi)$ and $V(\varphi)$.\footnote{Although the behaviors of the
scalar field $\Phi$ depends on a prior on three functions in
action~\eqref{eq:actionJ}, one can set $Z(\Phi)$, or one of any other two
functions, to unity by redefinition of $\Phi$.
The well-known $f(R)$ theory is equivalent to a ST theory with the specific choice of functions $F(\Phi)$ and $U(\Phi)$~\cite{Sotiriou:2008rp}, and the study of NS structures within this framework can be found in Refs.~\cite{Astashenok:2021peo, Capozziello:2015yza, Astashenok:2014nua}.}

Cosmological evolution naturally establishes a constant background value
$\varphi_0$ for the scalar field at the present epoch.  We therefore expand the
coupling function $\alpha(\varphi)$ about this background scalar field value,
\begin{eqnarray}\label{eq:alpexp}
    \alpha(\varphi) = \alpha_0+\beta_0 (\varphi-\varphi_0)+
    \mathcal{O}(\varphi-\varphi_0)^2.
\end{eqnarray}
We set $\varphi_0=0$ throughout this paper in accordance with: (i) observational
constraints from the current cosmological epoch in the massless ST
theories~\cite{Damour:1992kf}, and (ii) theoretical requirements for vacuum
solutions in the massive ST gravity theories.  Notably, when retaining only the
leading term in this expansion, the theory reduces to the original
Jordan-Fierz-Brans-Dicke formulation~\cite{jordan1955schwerkraft,Brans:1961sx}.
Current Solar System experiments impose stringent constraints on such theories,
requiring the coupling parameter to satisfy
$|\alpha|\lesssim3.4\times10^{-3}$~\cite{Bertotti:2003rm}.  The linear
coefficient $\beta_0$ is, however, nearly not constrained by Solar System
experiments because that its contribution to the post-Newtonian expansion
appears with a multiplier of $\alpha_0^2$~\cite{Damour:1995kt}, making it
naturally suppressed by the tightly constrained $\alpha_0$ parameter.

Rigorous constraints on ST gravity theories arise from both Solar System
experiments~\cite{Will:2014kxa} and cosmological observations~\cite{Ji:2024gdc},
substantially restricting the viable parameter space for these models and deeply
reducing the diversity of permissible neutron star (NS) and black hole (BH)
solutions within these gravity theories~\cite{Quiros:2019ktw}.  Even so, there
are several mechanisms to hide the scalar field in the weak field regime but
still allow the theory to deviate significantly from GR in cosmology or in the
strong-gravity regime, such as the Vainshtein model~\cite{Vainshtein:1972sx},
chameleon screening~\cite{Khoury:2013yya}, as well as the nonperturbative
mechanism introduced below.

\begin{figure}
\includegraphics[width=\linewidth]{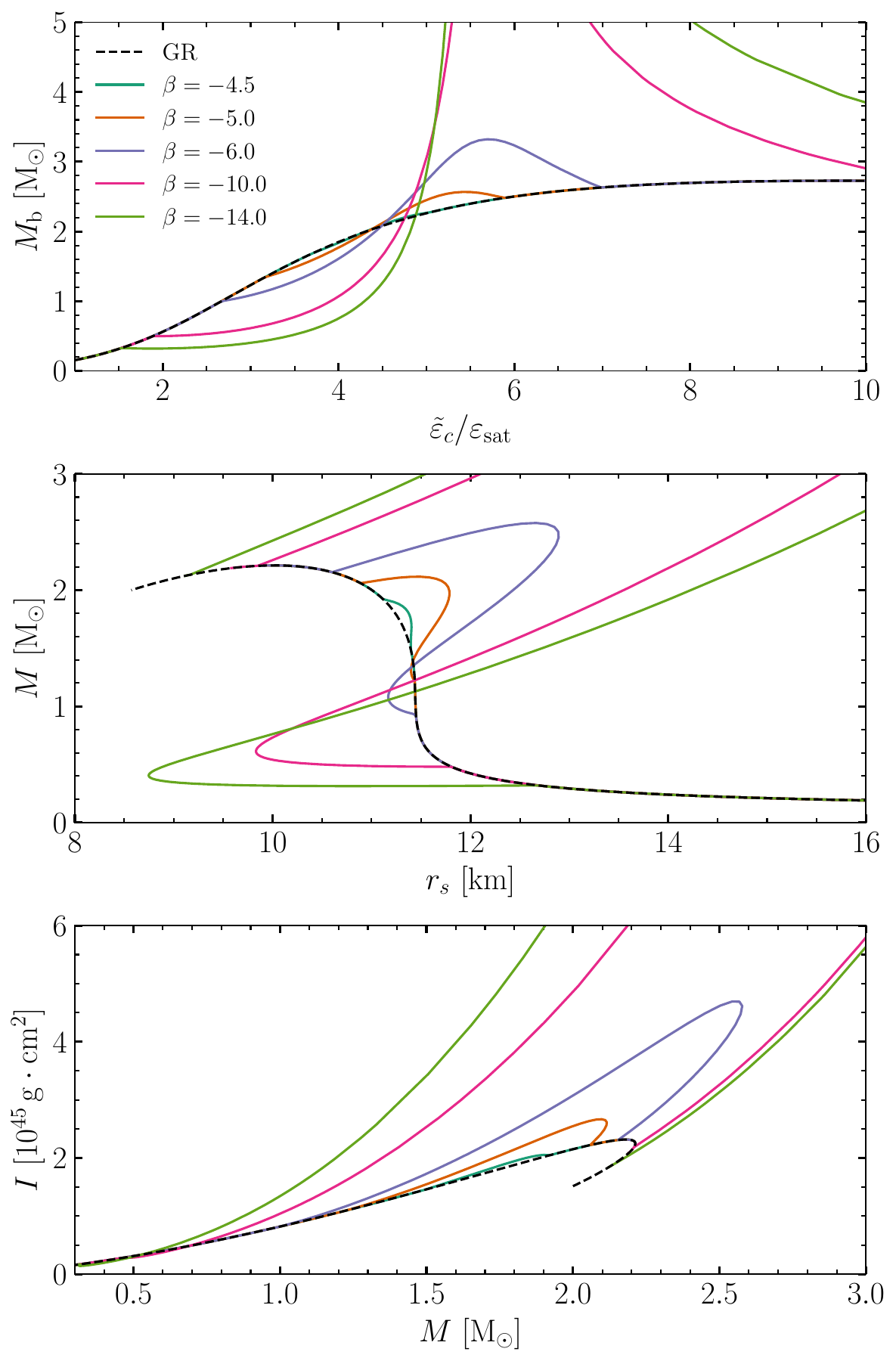}
\caption{\label{fig:mlessscalarization} Illustration of scalarization in the
massless DEF theory for various negative $\beta$ values.  The top, middle, and
bottom panels display the baryonic mass $M_{\rm b}$ as a function of (rescaled)
central density $\tilde{\varepsilon}_c$, the mass-radius relation, and the
moment of inertia $I$ of the NS, respectively.}
\end{figure}

It has been shown by Damour and Esposito-Far\`{e}se~\cite{Damour:1993hw, 
Damour:1996ke, Doneva:2022ewd} that there exist ST
theories where in principle both $\varphi$ being a constant and non-trivial
$\varphi$ solutions exist for spherically symmetric star.  Which of the two
configurations will be favored after gravitational collapse depends roughly on
the compactness of the star~\cite{Doneva:2022ewd}, which is a dimensionless
number defined by the stellar mass $M$ over the stellar radius $r_s$,
\begin{eqnarray}
    \mathscr{C}=\frac{GM}{r_s}.
\end{eqnarray}
For ordinary stars, like the Sun, the constant scalar solution is realized and
the metric describing their exterior is then the Schwarzschild
solution~\cite{Schwarzschild:1916uq}, which makes the ST theories
indistinguishable from GR in the Solar System.  For compact stars instead, such
as NSs, the non-trivial scalar configuration becomes energetically
favorable~\cite{Damour:1993hw} and the metric significantly deviates from the
one GR would yield.  The importance of this result lies on the fact that it was
the first demonstration that one can construct a  gravity theory which agrees
with GR in the weak-field limit but still gives distinct and testable
predictions in the strong-field regime~\cite{Damour:1996ke}.

There is a very sharp transition from the constant-$\varphi$ to the
nontrivial-$\varphi$ configurations as one increases the compactness of the
star, thereby the mechanism that causes this transition has been dubbed {\it
spontaneous scalarization}~\cite{Damour:1993hw}.  This phenomenon could be
understood as a typical Landau {\it phase transition}, with baryonic mass being
the order parameter~\cite{Damour:1996ke, Sennett:2017lcx}.  In the prototype
Damour--Esposito-Far\`{e}se (DEF) theory as an example, the coupling function is
\begin{eqnarray}\label{eq:alpDEF}
    \alpha_\mathrm{DEF}(\varphi)=\beta\varphi,
\end{eqnarray}
with $\beta$ being a constant number.  Basically, a negative parameter $\beta
\lesssim -4 $ would lead to {\it tachyonic instability} and trigger the
scalarization~\cite{Harada:1997mr}.  An illustration of the scalarization
mechanism is shown in Fig.~\ref{fig:mlessscalarization}.

The positive $\beta$ scenario for NSs was first investigated by Mendes and
Ortiz~\cite{Mendes:2014ufa,Mendes:2016fby}, and they discovered that multiple
branches of scalarized NSs emerge in the $\beta>0$ regime for the first time.
The presence of multiple branches of scalarized solutions indicates that, for a
given central density $\tilde{\varepsilon}_c$, there can be more than one
scalarized configuration.  The ST theory that they considered was derived from
an analytical approximation within a specific theoretical framework that is
well-motivated by the inflationary cosmology~\cite{Salopek:1988qh}.  The
coupling function of the Mendes-Ortiz (MO) theory is
\begin{eqnarray}\label{eq:alpMO}
    \alpha_\mathrm{MO} (\varphi) =
    \frac{1}{\sqrt{3}}\tanh\big(\sqrt{3}\beta\varphi\big),
\end{eqnarray}
where $\beta$ is also a constant number and we use the same symbol as in the DEF
theory for simplicity. The coupling functions of both DEF and MO theories have
the same expansion (\ref{eq:alpexp}) to the linear order of $\varphi$, i.e.
$\beta_0^\text{DEF}=\beta_0^\text{MO}=\beta$, which indicates that behaviors of
two theories should be identical in the limit of $\varphi\to0$.

The scalarized NSs in massless ST theories carry scalar charge, triggering a
gravitational dipole radiation in a binary system~\cite{Damour:1996ke}. The
dipole radiation will make the orbital energy loss faster than that in GR, which
in turn can be constrained by pulsar-timing observations~\cite{Damour:1996ke,
Shao:2017gwu, Anderson:2019eay, Kramer:2021jcw, Zhao:2022vig, Shao:2022izp}.  In
addition, gravitational waves (GWs) from binary NS mergers can also be used to
test the dipole radiation~\cite{Barausse:2012da,Freire:2012mg,
LIGOScientific:2018dkp, Zhang:2017sym, Zhao:2021bjw} and put constraints on ST
theories~\cite{Shao:2017gwu, Anderson:2019eay, Zhao:2019suc, Guo:2021leu}. 
Currently the best limit comes from a combination of binary pulsar systems that
demands the scalar charge of NSs to be smaller than $\sim 10^{-2}$ no matter of
the yet-uncertain equation of state (EOS) for NSs~\cite{Shao:2017gwu,
Zhao:2022vig}. Moreover, massless scalar fields could lead to complete
cosmological scalarization, a scenario definitively excluded by combined Solar
System and cosmological observations~\cite{Anderson:2016aoi,Anderson:2017phb},
unless GR serves as a cosmological attractor in the massless ST theories. This
generally requires either a positive $\beta$~\cite{Damour:1992kf, Damour:1993id}
or an additional topological~\cite{Antoniou:2020nax} or quasi-topological
term~\cite{Erices:2022bws}.

\begin{figure}
\includegraphics[width=\linewidth]{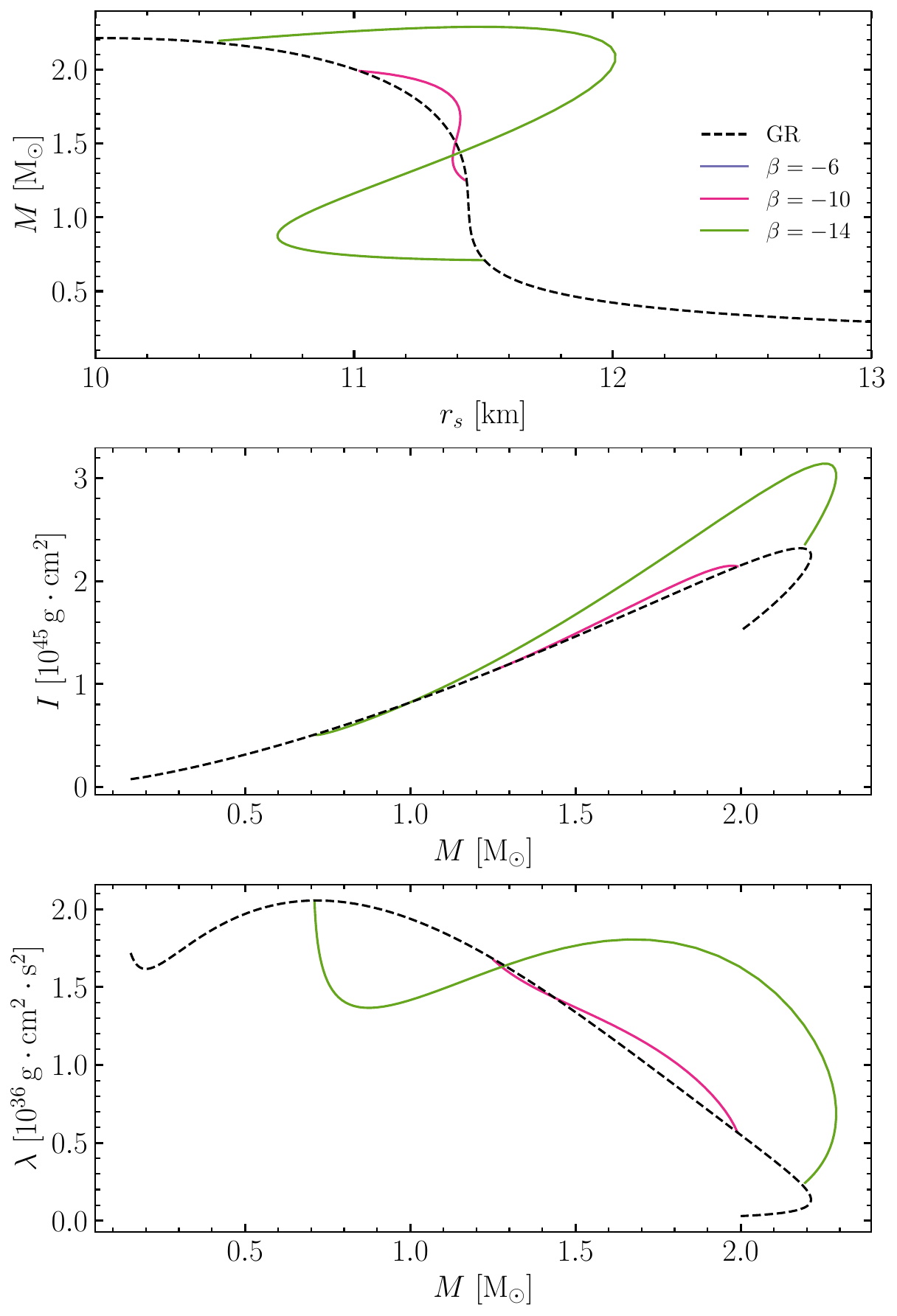}
\caption{\label{fig:msivescalarization} Illustration of scalarization in massive
DEF theory for various negative $\beta$ values.  The scalar mass corresponds to
a Compton wavelength of $2\pi\cdot10\,\mathrm{km}$.  The top, middle, and bottom
panels display the mass-radius relation, the moment of inertia, and the tidal
deformability of  NSs, respectively.  The $\beta=-6$ case is not sufficient to
trigger scalarization in this case.}
\end{figure}

After considering the observational constraints, the significantly reduced
theory parameter space for massless ST theories makes massive ST theories
particularly appealing~\cite{Ramazanoglu:2016kul, Yazadjiev:2016pcb, Xu:2020vbs,
Hu:2021tyw}, where the potential is generally modeled as the quadratic form  in
the EF,\footnote{Conformal transformation connecting the two frames (JF and EF)
is functionally dependent on the scalar field.  Hence, a scalar field mass
constant in a specific conformal frame would no longer be constant when
transitioning to the other frame~\cite{Faraoni:2009km}. The potential modeled as
quadratic form in the JF is discussed in Ref.~\cite{Xu:2020vbs}.}
\begin{eqnarray}\label{eq:V}
    V(\varphi)=\frac12m_\varphi^2\varphi^2,
\end{eqnarray}
where $m_\varphi$ is the mass of the scalar field, yielding a characteristic
Compton wavelength
\begin{eqnarray}\label{eq:lambda}
    \lambda_\varphi=\frac{2\pi}{m_\varphi}.
\end{eqnarray}

The massive ST theory induces a Yukawa-type suppression of the scalar field
outside stars, thereby generally weakening  the scalarization, as illustrated in
Fig.~\ref{fig:msivescalarization}.  The mass term also automatically suppresses
scalar contributions to GWs. Consequently, current high-precision pulsar-timing
observations have not yet placed stringent constraints on the parameter space of
massive ST theories~\cite{Alsing:2011er, Liu:2020moh}.    Massive ST theories
also work well with cosmology. They not only naturally prevent the universal
scalarization mentioned above, but also emerge as promising dark matter
candidates, since their massive scalars interact with baryonic matter solely
through gravity~\cite{Chen:2015zmx,Morisaki:2017nit}.  Consequently, in this
work we investigate the scenario where the scalar field possesses a
non-vanishing mass, with the moment of inertia and the tidal deformability
calculated as supplementary components that are relevent to pulsar-timing and GW
observations.

The paper is organized as follows.  In Sec.~\ref{sec:2}, we present the setup of
the problem, including the mechanism of scalarization, the choice of the EOS,
and the transformation of physical quantities between JF and EF.  The numerical
equations for solving the structure of scalarized NSs are given in
Sec.~\ref{sec:3}, where we also investigate the origin of the multiple
scalarized NS solutions that share the same central pressure and density.  We
find that this phenomenon stems from oscillations of the scalar field inside the
star, which also occurs for $\beta>0$, as explained in Appendix~\ref{sec:app}.
In Sec.~\ref{sec:4}, we introduce static perturbations to the background
solutions and compute the moment of inertia and tidal deformability of
scalarized NSs, followed by a presentation of the full numerical results and the
stellar properties.  Finally, we summarize our findings in Sec.~\ref{sec:5}.

We adopt the $(-,+,+,+)$ metric signature convention and set $c=\hbar=1$
throughout this work, with $G=1$ in subsequent sections.  The cgs unit system is
restored when discussing various properties of NSs.‌

\section{SETUP OF THE PROBLEM}\label{sec:2}

\subsection{Spontaneous scalarization}

We consider the ST theory in EF given by the action~\eqref{eq:actionE}, and the
variation of it gives the equations of motion~\cite{Damour:1996ke,
Doneva:2022ewd},
\begin{align}
	\label{eq:eomG}R_{\mu\nu}& = 2\partial_\mu\varphi \partial_\nu
	\varphi+2g_{\mu\nu}V+8\pi\left(T_{\mu\nu}-\frac12g_{\mu\nu}T\right),\\
	\label{eq:eomP}\square\varphi&=\frac{dV}{d\varphi}-4\pi\alpha(\varphi)T.
\end{align}
Here, $\alpha(\varphi)$ is the coupling function defined in Eq.~\eqref{eq:alpha}, $\square=g^{\mu\nu}\nabla_\mu\nabla_\nu$ is the wave operator in EF, and
\begin{align}\label{eq:TmnE}
	T_{\mu\nu}\equiv-\frac{2}{\sqrt{-g}} \frac{\delta \mathcal
	L_\mathrm{m}}{\delta g^{\mu\nu}},
\end{align}
is the energy-momentum tensor of conventional matter in EF with $T\equiv
g^{\mu\nu}T_{\mu\nu}$ its trace.

When incorporating the potential~\eqref{eq:V} within DEF theory,\footnote{The
exact same approaches in the following paragraphs can be applied to the MO
theory or other theories that have a different conformal function.} the scalar
field's equation of motion takes the form,
\begin{eqnarray}\label{eq:DEFsfeq}
    \square\varphi=(-4\pi\beta e^{2\beta\varphi^2}\tilde T+m_\varphi^2)\varphi,
\end{eqnarray}
where we have applied Eq.~\eqref{eq:tracetransf} that relates the trace of
energy-momentum tensor between two conformal frames (see below).  Considering a
small perturbation $\delta\varphi$ around the GR metric solution with
$\varphi=0$, we expand to the linear order in $\delta\varphi$ as
\begin{eqnarray}\label{eq:pertdphi}
    \square\delta\varphi=(-4\pi\beta\tilde T+m_\varphi^2)\delta\varphi.
\end{eqnarray}
For non-relativistic matter that can be modeled as a perfect fluid,  one has
$\tilde T=-\tilde\varepsilon+3\tilde p\approx-\tilde\varepsilon$, where
$\tilde\varepsilon$ and $\tilde p$ are the physical rest-frame mass density and
pressure of the fluid, respectively.  When the first term on the right-hand side
of Eq.~\eqref{eq:pertdphi} is effectively negative, i.e. $\beta\tilde T>0$
occurs in certain regions within the extended body in the gravitational field,
the theory develops a tachyonic instability at the linear level, characterized
by $\lambda_\mathrm{eff}<\lambda_\varphi$, where the effective wavelength is
given by~\cite{Ramazanoglu:2016kul, Doneva:2016xmf, Doneva:2022ewd}
\begin{eqnarray}
    \lambda_\mathrm{eff}=\sqrt{\frac{\pi}{|\beta|\tilde\varepsilon}}.
\end{eqnarray}
Consequently, all Fourier modes with wavelengths 
\begin{eqnarray}
    \lambda>\frac{\lambda_\mathrm{eff}}{\sqrt{1 -
    (\lambda_\mathrm{eff}/\lambda_\varphi)^2}}
\end{eqnarray}
that fit within the region where $\lambda_\mathrm{eff}<\lambda_\varphi$ will
initially undergo an exponential growth~\cite{Ramazanoglu:2016kul}.  While this
behavior would be catastrophic for the theory in isolation \reply{if the theory were considered without further including nonlinear effects}, the nonlinear term
$e^{2\beta\varphi^2}$ in Eq.~\eqref{eq:DEFsfeq} eventually becomes significant,
suppressing the growth and saturating $\varphi$ at a value of order
$1/\sqrt{\beta}$~\cite{Doneva:2022ewd}.

For a quantitative analysis, one expands the scalar perturbation under the
spherically symmetric assumption through standard spherical harmonics $Y_{\ell
m}$ as
\begin{eqnarray}\label{eq:shexp}
    \delta\varphi = \sum_{\ell=0}^\infty \sum_{m=-\ell}^\ell \frac{\psi_{\ell
    m}(\rho)}{\rho}Y_{\ell m}(\theta,\phi)e^{-i\omega t},
\end{eqnarray}
where $\psi_{\ell m}$ depends solely on the radial coordinate $\rho$ in the EF,
and the temporal dependence is given by $e^{-i\omega t}$ with $\omega$ being a
complex frequency parameter.  A Schr\"{o}dinger-like
equation~\cite{Maggiore:2018sht,Ferrari:2020nzo}
\begin{eqnarray}
    \frac{d^2\psi_{\ell m}}{d r^2_*}+\big[\omega^2-V(r_*)\big]\psi_{\ell m}=0,
\end{eqnarray}
is derived by substituting the spherical harmonic expansion in
Eq.~\eqref{eq:shexp} into the perturbation equation \eqref{eq:pertdphi}.  Here,
the generalized tortoise coordinate transformation is defined by
$dr_*=e^{(\mu-\nu)/2}d\rho$, and the effective potential is
\begin{eqnarray}
    V(r_*) = \frac{\nu'-\mu'}{2\rho}e^{\nu-\mu} +
    \left[\frac{\ell(\ell+1)}{\rho^2}+m_\mathrm{eff}^2\right]e^{\nu},
\end{eqnarray}
where the effective mass squared is
\begin{eqnarray}
    m_\mathrm{eff}^2=m_\varphi^2-4\pi\beta e^{2\beta\varphi^2}\tilde T.
\end{eqnarray}
Here, the prime denotes the derivative with respect to $\rho$, and $\nu=\ln
g_{tt}$, $\mu=\ln g_{rr}$.  The stability of perturbations is determined by the
sign of the imaginary part of the frequency $\omega$, where a positive imaginary
part indicates instability that leads to an exponential growth of perturbations,
while a negative one corresponds to stability with an exponential decay.  The
specific parameter space for stable stars has been given in
Refs.~\cite{Harada:1997mr,Mendes:2016fby}.

\subsection{EOS and the positive $\beta$ case}

It appears that only when $\beta < 0$ it enables the first term on the
right-hand side of Eq.~\eqref{eq:pertdphi} to become negative for
non-relativistic fluids whose $\tilde{T} < 0$.  However, given the extremely
high density and pressure within the NS, particularly in the core, the trace of
the energy-momentum tensor may reverse sign under stiff EOSs, in which
circumstance a positive $\beta$ is required.

\begin{figure}
\includegraphics[width=\linewidth]{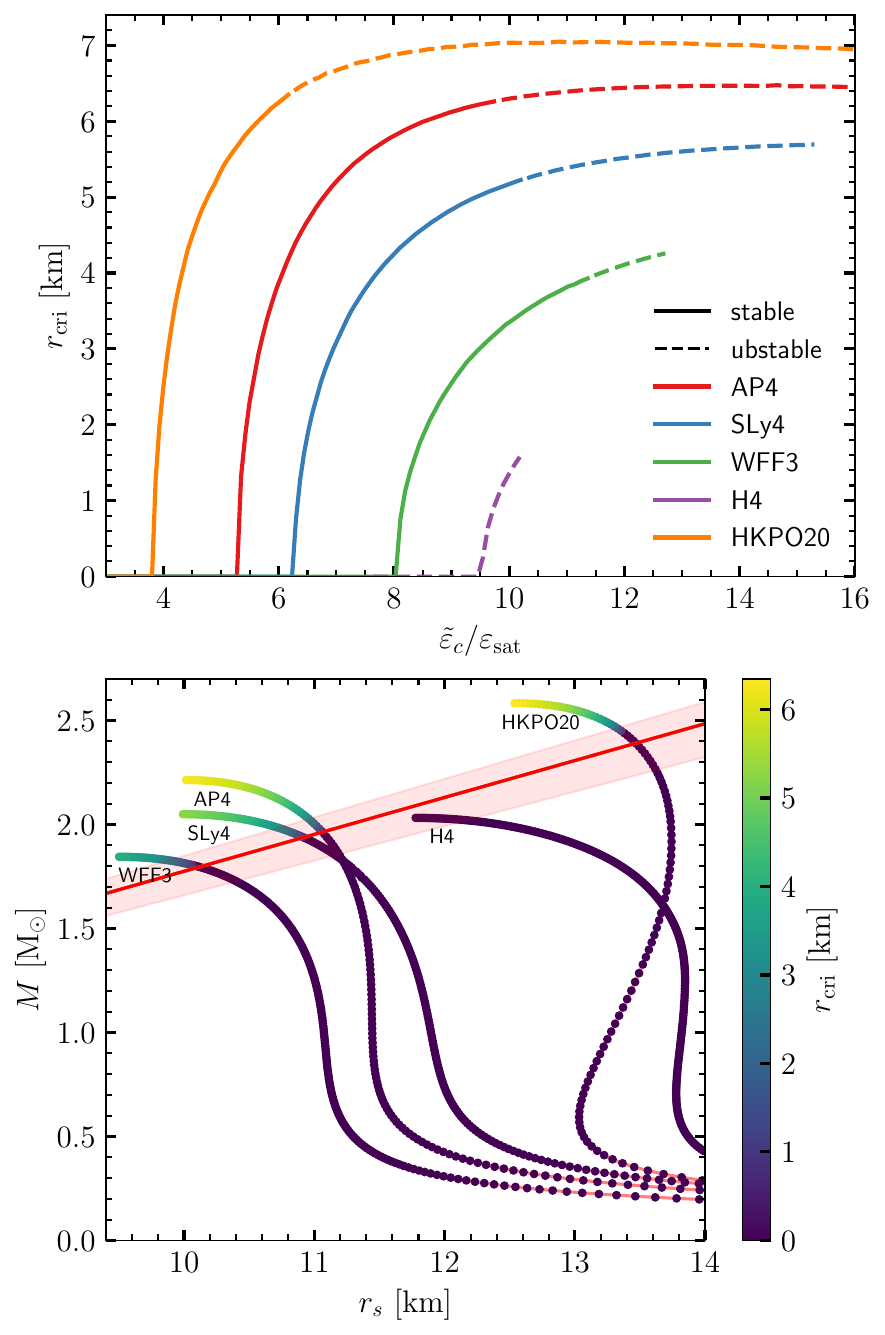}
\caption{\label{fig:eos} The figure illustrates properties for different EOSs of
NSs.  In the top panel, the critical radius is plotted as a function of the
central mass density normalized by the nuclear saturation density, with unstable
configurations indicated by dashed lines.  The bottom panel displays a
conventional mass-radius diagram for NSs, where the critical radius is
represented by the colorbar.  The red line and shaded region correspond to the
constraint value and uncertainty, $\mathscr{C}=0.262^{+0.011}_{-0.017}$, from
Ref.~\cite{Podkowka:2018gib}, indicating the minimum compactness that NSs should
have in order to possess a region where $\tilde T>0$.}
\end{figure}

Several typical EOSs from nuclear physics simulations~\cite{Lattimer:2000nx} are
displayed in Fig.~\ref{fig:eos}.  The top panel shows the critical radius
$r_\mathrm{cri}$, inside which sphere the trace $\tilde{T}$ is reversed in sign
relative to the non-relativistic convention, as a function of the central
density normalized by the nuclear saturation density $\varepsilon_{\mathrm{sat}}
= 2.8\times10^{14}\,\mathrm{g\cdot cm^{-3}}$.  We observe that at low central
densities, no trace-sign-reversed core exists within the NS.  As density
increases, however, this core emerges and expands until reaching its maximum
extent, i.e. the onset of stellar instability.  Relatively soft EOSs like H4
produce no stable NS containing a region where $\tilde T>0$.  The mass-radius
diagram in the bottom panel encodes critical radii $r_\mathrm{cri}$ via point
colors, while the red line and region demarcate the fitting compactness limit
$\mathscr{C}=0.262^{+0.011}_{-0.017}$~\cite{Podkowka:2018gib} required for a
trace-sign-reversed core.

Figure~\ref{fig:eos} demonstrates that, among the five EOSs, the AP4 EOS
exhibits the broadest central density range supporting stable NSs with a $\tilde
T>0$ region.  Hence, we utilize this EOS for later numerical computations.  To
avoid inherent limitations in the tabulated EOS, specifically the constrained
upper and lower bounds of density and pressure that might lead to inaccuracy in
later calculations, we implement the analytical polytropic approximation
model~\cite{Haensel:2007yy}, 
\begin{eqnarray}
	\tilde p(\tilde n)=Km_\mathrm{b}n_0\left(\frac{\tilde n}{n_0}\right)^\Gamma,
\end{eqnarray}
where $m_\mathrm{b}=1.66\times10^{-24}\,\mathrm{g}$ is the average baryon mass
and $n_0=0.1\,\mathrm{fm}^{-3}$ is a reference value for the baryonic number
density, while $K$ and $\Gamma$ are dimensionless constants.  The mass density
(or energy density by multiplying a factor of $c^2$) is then fixed by the first
law of thermodynamics,
\begin{eqnarray}
    \tilde\varepsilon(\tilde n) = m_\mathrm{b}\tilde n+\frac{\tilde p(\tilde
    n)}{\Gamma-1}.
\end{eqnarray}
The polytropic parameter $\Gamma$ is constrained from both above and below by
the existence of the trace-sign-reversed region as well as the forbidden
superluminal propagation of sound.  The two conditions, namely $3\tilde
p-\tilde\varepsilon>0$ and $v_s^2=d\tilde p/d\tilde\varepsilon<1$, lead to
\begin{eqnarray}
    \left[\frac{\Gamma-1}{K(3\Gamma-4)} \right]^{\frac{1}{\Gamma-1}}
    <\frac{\tilde n}{n_0}<\left[\frac{\Gamma-1}{K\Gamma(\Gamma-2)}
    \right]^{\frac{1}{\Gamma-1}},
\end{eqnarray}
where $\Gamma>2$ is assumed.  For the trace-sign-reversed region to be
subluminal, i.e., for the upper limit of $\tilde n$ to be larger than the lower
limit, one must set $\Gamma<4$.

\begin{table}
\caption{\label{tab:eospara} Parameters for the three polytropic segments of the
AP4 EOS based on Ref.~\cite{Read:2008iy}.}
\renewcommand\arraystretch{1.3}
\begin{ruledtabular}
\begin{tabular}{llll}
\textrm{Region}&
\textrm{$m_\mathrm{b}\tilde n$ $(\mathrm{g\cdot cm^{-3}})$}&
\textrm{$K_i$}&
\textrm{$\Gamma_i$}\\
\colrule
I $(i=1)$ & $<10^{14.7}$  & $0.00546$ & $2.830$ \\
II $(i=2)$ & $\big[ 10^{14.7}, 10^{15} \big]$ & $0.00277$ & $3.445$ \\
III $(i=3)$ & $>10^{15}$ & $0.00329$ & $3.348$\\
\end{tabular}
\end{ruledtabular}
\end{table}

Apparently, the simple polytropic EOS cannot be considered entirely realistic,
as it does not match the relatively well-understood microphysical models below
the nuclear density~\cite{Douchin:2001sv}.  The problem can be effectively
addressed by implementing slight modifications to the EOS at both low and high
density regimes, specifically through the application of a piecewise polytropic
EOS approach.  Building upon the foundational data from \citet{Read:2008iy}, we
have calculated and tabulated the parameters for the three polytropic segments
of the AP4 EOS, as presented in Table~\ref{tab:eospara}.  Given our primary
focus of this work on the trace-sign-reversed region, where the sophisticated
EOS describing the crust region induces only minor modifications to NS
structural properties, and considering that the lower density limit for the
$\tilde T>0$ region is at approximately $m_\mathrm{b}\tilde
n\sim10^{14.91}\,\mathrm{g\cdot cm^{-3}}$, we adopt the parameters 
\begin{eqnarray}
    K=0.003,\quad\Gamma=3.4,
\end{eqnarray}
and designate this configuration as the polytropic AP4 EOS when implemented.

\subsection{Relations between quantities in JF and EF}

Due to the coupling between the scalar field and matter in the EF, the local
conservation of the energy-momentum tensor no longer holds, and it changes 
to~\cite{Damour:2007uf}
\begin{eqnarray}
    \nabla^\nu T_{\mu\nu}=\alpha(\varphi) T\nabla_\mu\varphi.
\end{eqnarray}
As a result, the test particle no longer moves along the geodesics of
$g_{\mu\nu}$, as if it suffers from a kind of new force.  On the contrary, using
the equations of motion in the JF, one has
\begin{eqnarray}
    \tilde\nabla^\nu\tilde T_{\mu\nu}=0,
\end{eqnarray}
with the physical energy-momentum tensor defined by
\begin{eqnarray}\label{eq:TmnJ}
    \tilde T_{\mu\nu}\equiv -\frac{2}{\sqrt{-\tilde g}}
    \frac{\delta\tilde{\mathcal L}_\mathrm{m}}{\delta\tilde g^{\mu\nu}},
\end{eqnarray}
indicating the validity of the weak equivalence principle in the JF.  Here,
$\tilde\nabla$ and $\nabla$ are covariant derivatives compatible with the JF and
EF metrics, respectively.  Given the theoretical framework established, a
self-consistent treatment necessitates transforming results in the EF into
corresponding JF quantities that are physically measured.

The matter action explicitly represented by JF and EF quantities are 
\begin{align}
	S_\mathrm{m}^{\rm (JF)}[\Psi_\mathrm{m};\tilde{\mathbf{g}}] &=\int
	d^4x\sqrt{-\tilde g}\tilde{\mathcal L}_\mathrm{m}(\Psi_\mathrm{m},
	\tilde\nabla_\lambda\Psi_\mathrm{m},\tilde{\mathbf{g}}),\\
    S_\mathrm{m}^{\rm (EF)} [\Psi_\mathrm{m};\mathbf{g}] &=\int d^4x
    \sqrt{-g}{\mathcal L}_\mathrm{m}(\Psi_\mathrm{m},
    \nabla_\lambda\Psi_\mathrm{m},{\mathbf{g}}).
\end{align}
The variations with respect to the two metrics are 
\begin{align}
	\text{JF}:\ \quad &\int d^4x\sqrt{-\tilde
	g}\left(\frac{\partial\tilde{\mathcal L}_\mathrm{m}}{\partial\tilde
	g^{\mu\nu}}-\frac12\tilde{\mathcal L}_\mathrm{m}\tilde
	g_{\mu\nu}\right)\delta \tilde g^{\mu\nu},\\
	\text{EF}:\ \quad &\int d^4x\sqrt{-g}\left(\frac{\partial{\mathcal
	L}_\mathrm{m}}{\partial g^{\mu\nu}}-\frac12{\mathcal L}_\mathrm{m}
	g_{\mu\nu}\right)\delta g^{\mu\nu}.
\end{align}
According to the energy-momentum tensor definitions in Eqs.~\eqref{eq:TmnE}
and~\eqref{eq:TmnJ}, one has,
\begin{eqnarray}
    T_{\mu\nu}=A^2\tilde T_{\mu\nu},\quad T^{\mu\nu}=A^6\tilde T^{\mu\nu},
\end{eqnarray}
leading to
\begin{eqnarray}\label{eq:tracetransf}
    T=A^4 \tilde T.
\end{eqnarray}

For a perfect fluid in the EF, its energy-momentum tensor can be written as
\begin{align}
	T^{\mu\nu}& =(A^4\tilde\varepsilon+A^4\tilde p) (A\tilde u^\mu)(A\tilde
	u^\nu)+(A^4\tilde p) (A^2\tilde g^{\mu\nu})\nonumber\\
	&=(\varepsilon+p) u^\mu u^\nu+pg^{\mu\nu},
\end{align}
where we can connect the mass density, pressure, and 4-velocity of fluid in JF
and EF by the following relations~\cite{Damour:2007uf},
\begin{eqnarray}\label{eq:epu}
    \varepsilon\equiv A^4\tilde\varepsilon,\quad p\equiv A^4\tilde p\quad
    u^\mu\equiv A\tilde u^\mu.
\end{eqnarray}
We take a slowly-rotating star as example, then the 4-velocities in JF and EF are
\begin{align}\label{eq:4v}
	\tilde u^\mu=\frac{dt}{d\tilde\tau}(1,0,0,\Omega),\quad
	u^\mu=\frac{dt}{d\tau}(1,0,0,\Omega),
\end{align}
which satisfy the relation~\eqref{eq:epu} by noticing $d\tilde\tau^2=A^2d\tau^2$
that is given by the conformal transformation.

To acquire the transformation of quantities of stars, we write down the static
and spherically symmetric metric ansatz in JF and EF as
\begin{align}
	\label{eq:ansatzJ}d\tilde s^2&=\tilde g_{tt}(r)dt^2+\tilde
	g_{rr}(r)dr^2+r^2(d\theta^2+\sin^2\theta d\phi^2),\\
	\label{eq:ansatzE}ds^2&=g_{tt}(\rho) dt^2
	+g_{\rho\rho}(\rho)d\rho^2+\rho^2(d\theta^2+\sin^2\theta d\phi^2),
\end{align}
where $r$ and $\rho$ constitute the areal radii of 2-spheres in their respective
metric representations.  It is straightforward that the metric functions are
related via the conformal transformation,
\begin{align}
	\label{eq:r2r}r&=A\rho,\\
	\tilde g_{tt}(r)&=A^2g_{tt}(\rho),\\
	\tilde g_{rr}(r) &=\left(1+\alpha \frac{d\varphi}{d\rho}
	\rho\right)^{-2}g_{\rho\rho}(\rho).
\end{align}
By setting the conformal function $A(\varphi)$ to unity at spatial infinity,
which corresponds to $\alpha_0=0$ in Eq.~\eqref{eq:alpexp}, we ensure $\tilde
g_{tt}=g_{tt}$ at spatial infinity.

For the spatial component of the metric, we further expand them at spatial
infinity,
\begin{eqnarray}
	\tilde g_{rr} =\sum_{n=0}^\infty\frac{\tilde g_{rr}^{(n)}}{r^n}, \quad
	g_{\rho\rho}=\sum_{n=0}^\infty\frac{g_{\rho\rho}^{(n)}}{\rho^n}.
\end{eqnarray}
In the massless ST theory, the scalar field $\varphi$ admits an asymptotic
expansion at spatial infinity,
\begin{eqnarray}
    \varphi=\sum_{n=0}^\infty \frac{\varphi^{(n)}}{\rho^n},
\end{eqnarray}
where $\varphi^{(1)}$, representing an additional degree of freedom in static
spherically symmetric NS spacetimes, is identified as the scalar charge $q$ in
the EF.\footnote{The physically meaningful scalar charge should be the JF one
whose definition corresponds to the coefficient of the $1/r$ term in the
asymptotic expansion of the scalar field $\Phi$.} The following relations are
then derived,
\begin{align}
	\label{eq:gttrelataion}\tilde g_{rr}^{(0)}&=g_{\rho\rho}^{(0)},\\
	\label{eq:grrrelataion}\tilde g^{(1)}_{rr}& =\Big[g^{(1)}_{\rho\rho}
	+2\alpha_0qg_{\rho\rho}^{(0)}\Big]A_0,\\
	\tilde g_{rr}^{(2)}&=\Big[g^{(2)}_{\rho\rho} +\alpha_0
	q\big(3g^{(1)}_{\rho\rho}+5\alpha_0q g^{(0)}_{\rho\rho}\big)\nonumber\\
    &\qquad +2\beta_0q^2g^{(0)}_{\rho\rho}+4g^{(0)}_{\rho\rho}
    \varphi^{(2)}\Big]A_0^2,
\end{align}
where $\alpha_0$ and $A_0$ denote the asymptotic values of $\alpha(\varphi)$ and
$A(\varphi)$ respectively as $\rho\to\infty$.

Equations~\eqref{eq:gttrelataion} and~\eqref{eq:grrrelataion} show that in the
JF spacetime is asymptotically flat if and only if in the EF spacetime is
asymptotically flat when $\alpha_0=0$.  According to the Arnowitt-Deser-Misner
(ADM) mass~\cite{Arnowitt:1959ah}, $M=g_{rr}^{(1)}/2$, we can further acquire
\begin{eqnarray}
    \tilde M=A_0(M+\alpha_0q)=(1+\eta\alpha_0)A_0M,
\end{eqnarray}
where $\eta=q/M$ is usually defined as the charge-to-mass ratio in the EF.

In the massive ST theory, however, the scalar field exhibits an exponentially
decaying  behavior with radial distance in the asymptotically spatial infinity,
\begin{eqnarray}
    \varphi(\rho)\to\frac{e^{-\lambda_\varphi/\rho}}{\rho} .
\end{eqnarray}
As a consequence, the scalar charge is not well-defined, and the asymptotic
field values remain identical in both frames.  In all subsequent text and
graphical representations, the tilde notation will be systematically omitted
when referring to the physical ADM mass.

\section{NUMERICAL SOLUTIONS OF SCALARIZED NSs IN ST THEORIES}
\label{sec:3}

In this section, we give the differential equations governing NS spacetimes in
general massive ST theories~\cite{Damour:1996ke, Damour:2007uf}.  The
corresponding equations for the massless case are readily obtained by setting $V
= 0$.  A detailed analysis of the multi-branch scalarized solutions is provided
thereafter.

\subsection{Theoretical framework to calculate NS structures}

We begin with the static spherically symmetric line element in the EF,
\begin{eqnarray}\label{eq:SSm}
    \reply{ds^2_{\text{sph}} = -e^{\nu(\rho)}dt^2 + \frac{d\rho^2}{1-\frac{2m(\rho)}{\rho}} +
    \rho^2d\Omega_{(2)}^2},
\end{eqnarray}
where $\nu(\rho)$ and $m(\rho)$ characterize the gravitational potential and
gravitational mass respectively, \reply{and 
\begin{eqnarray}
    d\Omega_{(2)}^2=d\theta^2 + \sin^2\theta\, d\phi^2
\end{eqnarray}
is the line element of a 2-sphere.}
Substituting this metric ansatz into
Eqs.~\eqref{eq:eomG} and \eqref{eq:eomP} yields the modified
Tolman-Oppenheimer-Volkoff (TOV) equations~\cite{Damour:2007uf, Xu:2020vbs},
\begin{align}
	\label{eq:tovn}\nu' &=\frac{2m}{\rho(\rho-2m)} +\frac{8\pi\rho^2\tilde
	pA^4}{\rho-2m}+\rho\psi^2-\frac{2\rho^2V}{\rho-2m},\\
    \label{eq:tovm}m' &=4\pi\rho^2\tilde\varepsilon
    A^4+\frac12\rho(\rho-2m)\psi^2+\rho^2V,\\
	\varphi'&=\psi,\\
	\label{eq:tovs}\psi' &=\frac{4\pi\rho
	A^4}{\rho-2m}\big[\rho(\tilde\varepsilon-\tilde p)\psi
	+\alpha(\tilde\varepsilon-3\tilde p)\big]\nonumber\\
    &\quad-\frac{2(\rho-m)}{\rho(\rho-2m)}\psi
    +\frac{\rho}{\rho-2m}\left(2\rho\psi V+\frac{dV}{d\varphi} \right),\\
	\label{eq:tovp}p'&=-(\tilde\varepsilon +\tilde
	p)\left(\frac{4\pi\rho^2\tilde
	pA^4}{\rho-2m}+\frac{m}{\rho(\rho-2m)}\right)\nonumber\\
    &\quad-(\tilde\varepsilon+\tilde p) \left(\alpha\psi+
    \frac12\rho\psi^2-\frac{\rho^2V}{\rho-2m}\right),
\end{align}
where the prime denotes the derivative with respect to radial coordinate $\rho$.
In addition, one acquires the ordinary differential equations for the baryonic
mass,
\begin{eqnarray}\label{eq:baym}
    \bar m'=\frac{4\pi\rho^2m_\mathrm{b}\tilde nA^3}{\sqrt{1-2m/\rho}}.
\end{eqnarray}

The solution of the modified TOV equations can be expanded near the center
as~\cite{Damour:1996ke}
\begin{align}
	\nu&=\nu_c+\left[ \frac{4\pi}{3}(3\tilde p_c+\tilde\varepsilon_c)
	A^4_c-\frac23V_c\right] \rho^2+ \mathcal O(\rho^3),\\
    \label{eq:mexp}m&=\left( \frac{4\pi}{3}\tilde \varepsilon_cA_c^4
    +\frac13V_c\right)\rho^3+\mathcal O(\rho^4),\\
	\label{eq:phiexp}\varphi& =\varphi_c-\frac16 \big(4\pi\tilde
	T_cA_c^4\alpha_c-V_{\varphi c}\big)\rho^2 +\mathcal O(\rho^3),\\
	\label{eq:pexp}\tilde p& =\tilde p_c-\frac16(\tilde p_c +\tilde
	\varepsilon_c) \Big[4\pi A_c^4 (3\tilde p_c+\tilde\varepsilon_c-\tilde
	T_c\alpha_c^2)\nonumber\\
    &\qquad-2V_c +\alpha_c V_{\varphi c} \Big]\rho^2+\mathcal O(\rho^3),
\end{align}
where the subscript `$c$' represents the values at the stellar center where
$\rho=0$, and $A_c=A(\varphi_c)$, $\alpha_c=\alpha(\varphi_c)$,
$V_c=V(\varphi_c)$, $V_{\varphi c}=dV(\varphi_c)/d\varphi$, and $\tilde
T_c=3\tilde p_c-\tilde\varepsilon_c$.  The GR corresponding NS solutions are
acquired once setting $A_c=1$ and $\alpha_c=V_c=V_{\varphi c}=0$.

We start the integration at the stellar center (actually a very small radius
$\rho_c$ to avoid singular behaviors), and all the way down to the spatial
infinity.  We set two adjustable parameters, $\varphi_c$ and $p_c$, as the
degree of freedom of the system, and leave the others fixed to
zero~\cite{Damour:1996ke},
\begin{eqnarray}
    m_c=0,\quad\psi_c=0,\quad\nu_c=0.
\end{eqnarray}
The first two variables, $m_c$ and $\psi_c$, have to vanish because of the
expansions~\eqref{eq:mexp} and~\eqref{eq:phiexp}, while $\nu_c$ actually can be
any value as the modified TOV equations are independent of $\nu$, and $\nu_c=0$
is a common choice.\footnote{\reply{For the same reason, the choice $\nu_c=0$ is also employed in the perturbation calculations presented in Sec.~\ref{sec:4}.}}

Integration temporarily suspends when $\tilde p=0$, representing the case
reaching the stellar surface. Using Eq.~\eqref{eq:r2r} and integrating
Eq.~\eqref{eq:baym}, the baryonic mass and the radius of NS are then obtained by
\begin{eqnarray}
    M_\mathrm{b}=\bar m_s,\quad r_s=A_s\rho_s,
\end{eqnarray}
where the subscript `$s$' represents the values at the stellar surface
$\rho=\rho_s$, and $A_s=A(\varphi_s)$.  To complete the integration, we continue
integrating the vacuum version of Eqs.~\eqref{eq:tovn}--\eqref{eq:tovs} outwards
to the spatial infinity.

For massless ST theories, Damour and Esposito-Far\`{e}se have derived analytical
solutions~\cite{Damour:1992kf,Damour:1992we}, enabling direct analytic relations
between NS properties and surface values (those with the subscript `$s$').
However, spatial infinity remains computationally inaccessible without
coordinate transformations that map infinity to finite values (see e.g.,
Ref.~\cite{Ji:2024aeg}), if no analytical solution in vacuum region has been
found.  We therefore integrate to a sufficiently large radius $\rho_i$, treated
as effectively spatial infinity due to the Yukawa suppression of the scalar
field that renders spacetime asymptotically indistinguishable from the
Schwarzschild solution.  The ADM mass and asymptotic scalar field value (also
known as the cosmological background value) are consequently extracted as
\begin{eqnarray}
M = m_i, \quad \varphi_0 = \varphi_i,
\end{eqnarray}
where the subscript `$i$' represents the values at numerical infinity
$\rho=\rho_i$.  It is noteworthy that while the scalar field must theoretically
vanish at infinity ($\varphi\sim e^{-\rho/\lambda_\varphi}/\rho$) to avoid
divergence ($\varphi\sim e^{\rho/\lambda_\varphi}/\rho$), numerical
implementations inevitably exhibit {\it numerical} divergence because that
achieving exactly zero coefficients is impossible in the exponentially growing
branch.  To identify the critical initial value $\varphi_c$ that yields a scalar
field vanishing at spatial infinity, we employ a shooting method to determine
the central value $\varphi_c$ maximizing $\rho_i$, which is defined as the
radial coordinate of the minimal scalar field value exterior to the
NS.\footnote{A straightforward and simpler choice for numerical infinity is the
radial position where the scalar field value minimizes, denoted as $\rho_{i'}$.
It is evident that $\rho_{i'} \leq \rho_i$ while $\varphi(\rho_{i'}) \leq
\varphi(\rho_i)$. From our numerical tests, $\varphi(\rho_i) \lesssim 10^{-5}$
proves small enough, since a difference negligible enough to justify equating JF
and EF quantities due to $A(\varphi(\rho_i)) - 1 \lesssim 10^{-7}$ for $\beta
\sim \mathcal{O}(10^3)$ in both DEF and MO theories. The choice of a larger
$\rho_i$ is more physically significant for numerical accuracy, which motivates
our adoption of the $\rho_i$ definition presented in the main text.}

\subsection{Multi-branch scalarized solutions}

The scalarized NS solution describes configurations exhibiting a non-trivial
scalar field profile, fundamentally distinct from those in GR.  Since the scalar
field must asymptotically approach the cosmological background value, assumed
zero at the present cosmic epoch, this boundary condition restricts the
arbitrariness in the central scalar field value $\varphi_c$.



\begin{figure}
\includegraphics[width=\linewidth]{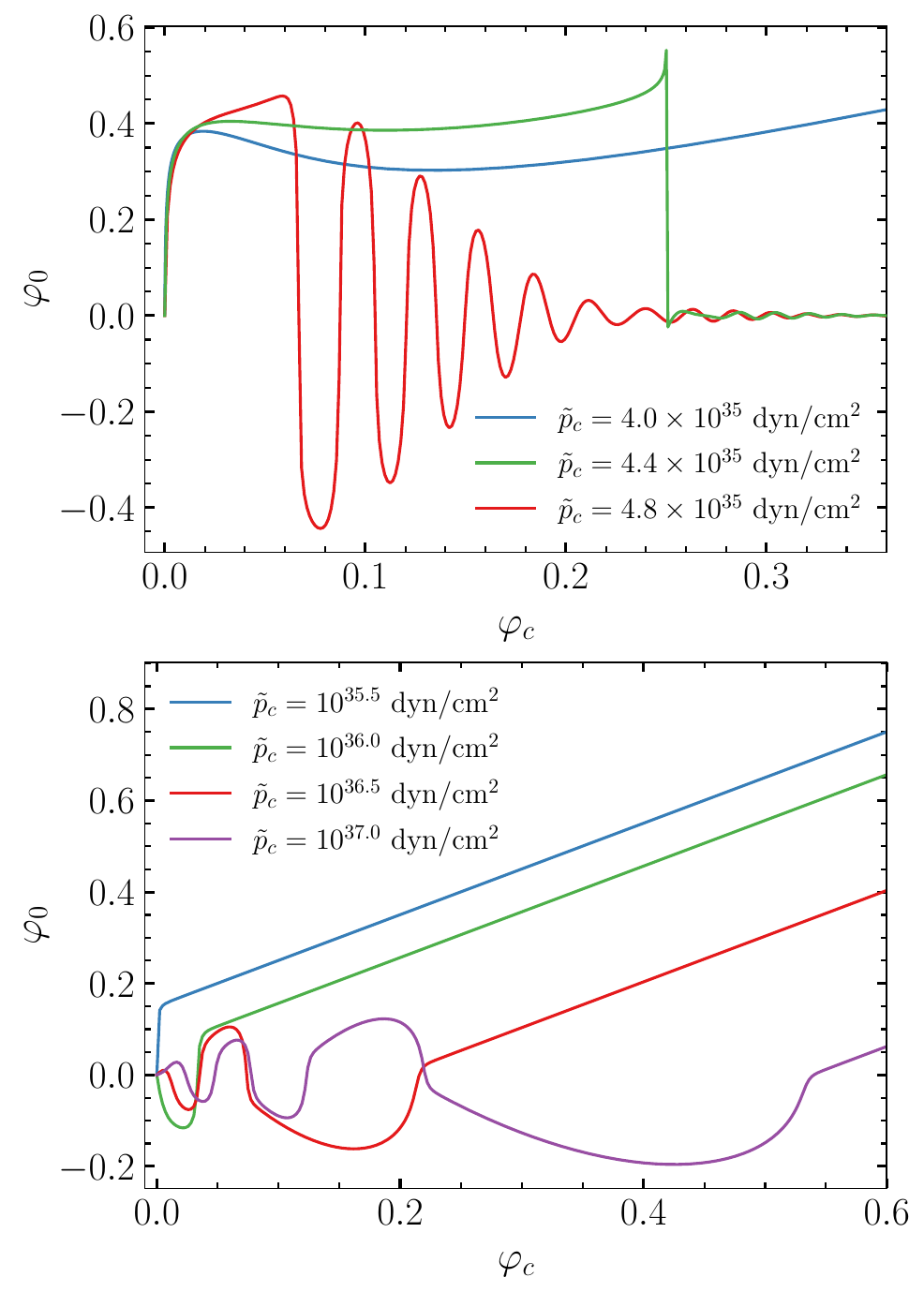}
\caption{\label{fig:pp} The asymptotic scalar field value $\varphi_0$ versus the
central scalar field value $\varphi_c$ for various central pressure of NSs in
the massless DEF theory (upper panel) and massless MO theory (bottom panel) with
$\beta=100$.  The EOS is taken to be the polytropic AP4 model.}
\end{figure}

Figure~\ref{fig:pp} illustrates the functional dependence of the scalar field's
asymptotic value $\varphi_0$ on its central value $\varphi_c$ for the massless
DEF theory (top panel) and the massless MO theory (bottom panel), with the
coupling parameter $\beta=100$.\footnote{In massive ST theories, the asymptotic
scalar field value $\varphi_0$ always diverges, and similar structures emerge
when the direction of divergence is plotted as a function of the central field
value $\varphi_c$.  Since the mass term contributes only marginally, we focus
primarily on the massless case in the following analysis.} All curves for
different central pressures converge at the origin, corresponding to the trivial
GR solution where the scalar field vanishes across the entire spacetime.  It
demonstrates that the scalarized solutions emerge when the central pressure of a
NS exceeds a critical threshold in both DEF and MO theories.  Nevertheless, the
differences observed between the upper and lower panels are entirely
attributable to the distinct coupling functions specified in
Eqs.~\eqref{eq:alpDEF} and~\eqref{eq:alpMO}.

We confirm that the threshold pressure for the DEF theory precisely satisfies
$\tilde T=3\tilde p-\tilde\varepsilon=0$, indicating that the scalarization
occurs exclusively when a trace-sign-reversed region exists within the stellar
interior.  Even if the region is very small, the unbound coupling function,
$\alpha_\mathrm{DEF}$, ensures the occurrence of scalarization as long as
$\varphi_c$ is sufficiently large.  In contrast, scalarization in the MO theory
requires a trace-sign-reversed core with a finite size, since
$\alpha_\mathrm{MO}$ is bound above by $1/\sqrt{3}$.  As $\varphi_c$ increases,
all curves tend to behave similarly in the bottom panel, precisely because that
the scalar field inside the star approaches this upper bound.

We now turn our attention to the pressure profile, rewriting the scalar field
contribution to $d\tilde{p}/d\rho$, while neglecting the contribution from the
mass term, as
\begin{eqnarray}\label{eq:p'}
    \left.\frac{d\tilde p}{d\rho} \right|_\mathrm{scalar\;field}
    \approx-(\tilde\varepsilon+\tilde p)
    \left(\alpha\psi+\frac{1}{2}\rho\psi^2\right).
\end{eqnarray}
Without loss of generality, we assume a positive $\varphi_c$ and hence a
negative $\psi$ near the stellar center.  The term $\alpha\psi$ is negative in
both the DEF and MO theories with a positive $\beta$, which slows down the decay
of pressure, or may even lead to an increase in pressure and thereby potentially
rendering the NS unstable~\cite{Mendes:2016fby}.  According to Eq.~\eqref{eq:pexp}, the
the sign of $3\tilde p_c+\tilde\varepsilon_c-\tilde T_c\alpha_c^2$ approximately
determines the radial pressure gradient at $\rho=0$.  In the MO coupling, the
coefficient remains positive, leading to a monotonic decrease in pressure inside
the NS.  However, in the DEF theory, this coefficient changes sign from positive
to negative, a transition that directly triggers the appearance of scalarized
solutions.\footnote{Including the mass term makes it more difficult to achieve
$p' > 0$ at the center in the DEF theory, whereas in the MO theory, this becomes
possible if and $\varphi_c > 1/\sqrt{3}$ and $m_\varphi \gg 0$.} This is
evidenced by the precise match between the vanishing point of $3\tilde{p}_c +
\tilde{\varepsilon}_c - \beta^2 \tilde{T}_c \varphi_c^2$ and the discontinuity
(cliff) in the red curve in the top panel of Fig.~\ref{fig:pp}, further
indicating that all scalarized NSs have non-monotonic pressure profiles in DEF
theories with a negative $\beta$.

\begin{figure}
\includegraphics[width=\linewidth]{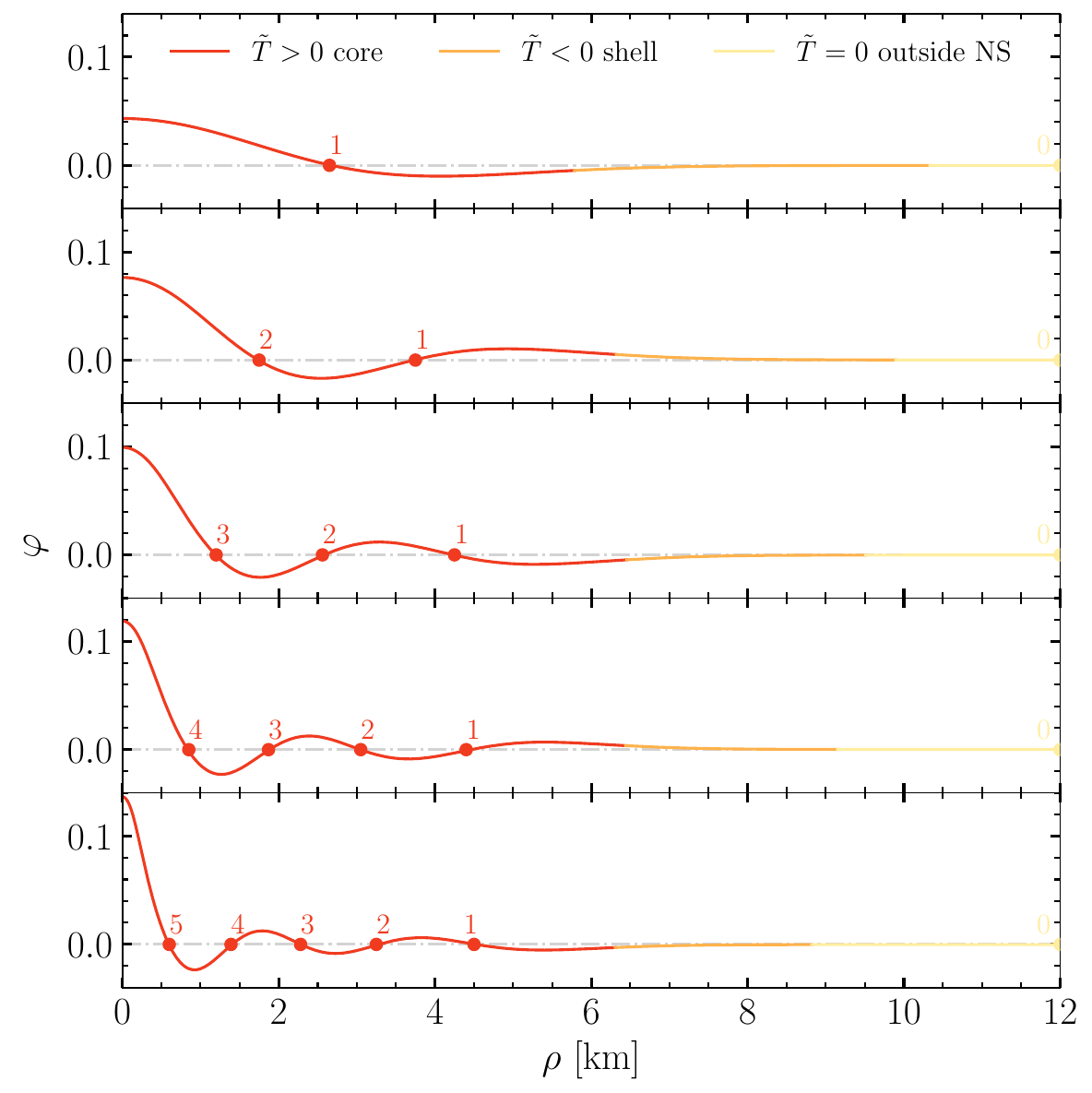}
\caption{\label{fig:msDEF} The scalar field profiles of five scalarized NS
solutions are shown, corresponding to the five smallest central values of the
scalar field (from top to bottom, $\varphi_c$ increases successively) in the
massless DEF theory with $\beta = 100$, $\tilde\varepsilon_c = 2 \times
10^{15}\,\mathrm{g\cdot cm^{-3}}$, and the AP4 EOS.  Red dots denote the zero
points of the scalar field profiles.  The yellow dots labeled with `0' mark the
designated boundary condition $\varphi(\rho\to\infty)=0$, while each positive
integer indicates the number of nodes in the corresponding solution.  Each
subsequent scalarized solution possesses one more node than the previous one.}
\end{figure}
\begin{figure}
\includegraphics[width=\linewidth]{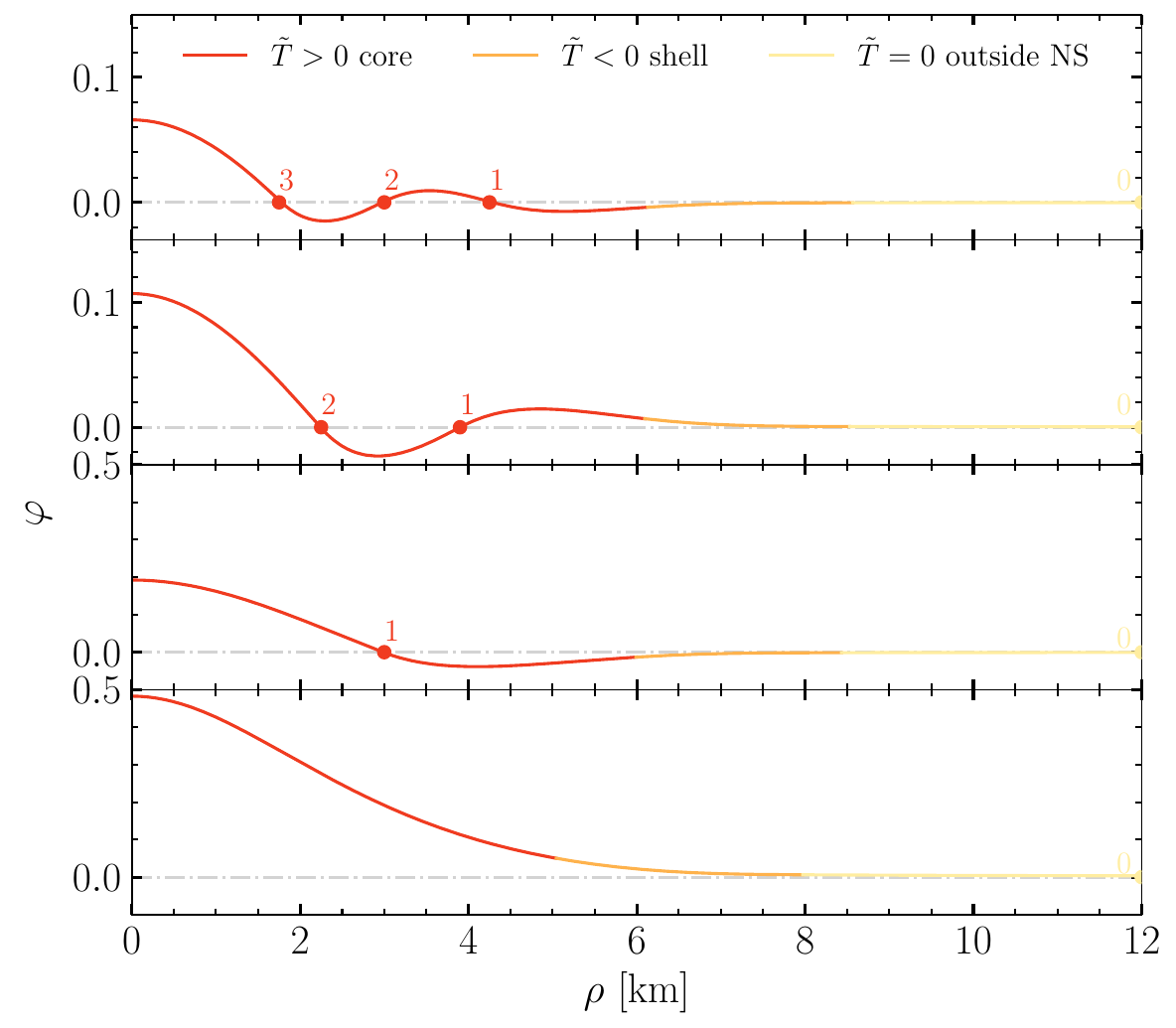}
\caption{\label{fig:msMO} The scalar field profiles of four scalarized NS solutions in massless MO theory with $\beta=100$ and
$\tilde\varepsilon_c=8\times10^{15}\,\mathrm{g\cdot cm^{-3}}$  (from top to
bottom, $\varphi_c$ increases successively).  We adopt the AP4 EOS, and the
meanings of the dots and numerals are the same as those in Fig.~\ref{fig:msDEF}.
Each subsequent scalarized solution possesses one less node than the previous
one.  It should be noted that the vertical axis range in the top two panels
differs from that in the bottom two panels.}
\end{figure}

To explain the multiple zero-crossings in the $\varphi_0$-$\varphi_c$ plots,
which underlie the existence of multiple branches of scalarized solutions, we
show in Fig.~\ref{fig:msDEF} and Fig.~\ref{fig:msMO} the scalar field profiles
for solutions with the same central density, in both DEF and MO theories.
Within the $\tilde T>0$ core region, the scalar field fluctuates, while
throughout the $\tilde T<0$ shell and external domains, the absolute value of
the scalar field decays monotonically.  Red dots mark the nodes of the scalar
field profile, with adjacent integers indicating the node number.  All nodes lie
within the $\tilde{T} > 0$ core, except for the `0' node, which corresponds to
the boundary condition of a vanishing scalar field at spatial infinity.

The shape of the scalar profile can be qualitatively elucidated through the
subsequent simplified model for the DEF theory first used
by~\citet{Damour:1993hw}. Let us neglect the curvature of metric $g_{\mu\nu}$
and treat the trace $T$ as a constant denoted by $\bar T$, which can be either
positive or negative.  Under this approximation, Eq.~\eqref{eq:eomP} for the DEF
theory reduces to
\begin{eqnarray}\label{eq:DEFapp}
    \Delta\varphi=(-4\pi\beta\bar T+m_\varphi^2)\varphi,
\end{eqnarray}
where $\Delta=\partial_x^2+\partial_y^2+\partial_z^2$ represents the Laplacian
operator in the three-dimensional Euclidean space, and the
potential~\eqref{eq:V} has been applied.  The behavior of the solution to
Eq.~\eqref{eq:DEFapp} is determined by the value of the parameter combination
$\beta\bar T$   (see also Ref.~\cite{Doneva:2022ewd}).  For $\beta\bar
T<m_\varphi^2/4\pi$, the solution exhibits an exponential behavior, while for
$\beta\bar T>m_\varphi^2/4\pi$, the solution shows an oscillatory behavior with
a characteristic half-wavelength,
\begin{eqnarray}\label{eq:lamo2}
    \bar\lambda_{1/2}=\frac{\pi}{\sqrt{4\pi\beta\bar T-m_\varphi^2}}.
\end{eqnarray}
From the above analysis, we conclude that the multi-branch structure is not
unique to the positive $\beta$ case, as the condition $4\pi\beta\bar{T} >
m_\varphi^2$ can also be readily satisfied for some negative values of $\beta$.
See Appendix~\ref{sec:app} for an illustration with $\beta = -200$ in the
massless DEF theory.

Let us now concentrate on the core region where $\tilde T>0$, and consider how
the scalar profile evolves with increasing $\varphi_c$.  In Fig.~\ref{fig:msDEF}
the red part of curves resembles a serpentine path of a ``snake" starting from
its cave ($\rho=0$) and advancing along the positive $\rho$ direction in the DEF
theory.  Whenever the ``snake's head'' reverses its direction, indicated by a
sign change in the slope at its leading position, an additional scalarized
solution emerges due to the strictly monotonous field residual.  Concurrently,
as the ``snake's elongated tail" persistently remains above zero, a new red node
invariably appears simultaneously.  Theoretically, new nodes invariably emerge
as the central scalar field increases.  However, the curves in the upper panel
of Fig.~\ref{fig:pp} usually terminate at a large $\varphi_c$ because that the
increasing pressure reaches its maximum value---either the  maximum in the
tabulated  EOS or a theoretical upper limit for NSs---thus preventing the
existence of infinitely many scalarized solutions at a fixed central density.

\reply{In contrast to the DEF theory, the number of nodes decreases with increasing $\varphi_c$, as illustrated in Fig.~\ref{fig:msMO}.}
As a result, at a
fixed $\tilde\varepsilon_c$, the total number of scalarized NS solutions
corresponds to the total node count of the scalar profile (including the one at
spatial infinity) with the minimal $\varphi_c$ (the scalar profile in the top
panel).  Even so, in theory the number of nodes could becomes infinite if the
central density diverges and the $\tilde T > 0$ core becomes infinitely large.
In conclusion, from a purely mathematical perspective, both the DEF and MO
theories admit an infinite number of scalarized branches unless an upper limit
on the density or pressure is imposed in the EOS modeling, which will also be
evident from Fig.~\ref{fig:scalarization} (see below).

\section{STATIC PERTURBATIONS AND NUMERICAL RESULTS}\label{sec:4}

In this section, we introduce static perturbations on the background scalarized
solutions in both massless and massive theories, and obtain the moment of
inertia and tidal deformability for them, which are relevant to pulsar-timing
and gravitational-wave observations.

\subsection{Moment of inertia and tidal deformability}

The moment of inertia of a NS is a fundamental parameter that probes its
internal structure and  provides insights into the behavior of ultra-dense
matter~\cite{Raithel:2016vtt, Breu:2016ufb, Gao:2021uus}.  It is sensitive to
the EOS, also a powerful tool for testing general relativity and alternative
theories of gravity~\cite{Yagi:2013awa, Shao:2022koz}.  Moreover, the moment of
inertia influences the dynamics of  binary systems, including GW emission and
orbital evolution, offering a critical astrophysical
probe~\cite{Landry:2018jyg}.  On the other hand, in binary star systems, the
point-mass approximation remains valid at large orbital separations.  As stellar
components approach, inhomogeneous gravitational fields induce tidal deformation
by distorting both stellar bodies.  This tidal effect introduces measurable
signatures in GW waveforms~\cite{Flanagan:2007ix, Damour:2009vw,
Hinderer:2009ca}. It also enables empirical tests of gravitational theories,
including both GR~\cite{Silva:2020acr} and ST theories~\cite{Pani:2014jra,
Brown:2022kbw}, through binary NS coalescence
observations~\cite{LIGOScientific:2017vwq, LIGOScientific:2018hze}.  Combining
(dimensionless) moment of inertia, tidal deformability, as well as spin-induced
quadrupole moment of slowly rotating NSs, Yagi and Yunes~\cite{Yagi:2013awa,
Yagi:2013bca} established nearly universal relations independent of the NS EOS. 
The existence of such universal relations in ST theories remains an open
question under active investigation~\cite{Doneva:2013rha, Pani:2014jra,
Doneva:2014faa, Hu:2021tyw}.  These relations potentially enable distinguishing
GR from ST theories and testing ST theory validity without requiring detailed
EOS knowledge of NSs.

To calculate the moment of inertia of a NS,\footnote{\reply{Calculations of the moment of inertia, together with pulsar timing observables, were carried out for the massless DEF model in Ref.~\cite{Damour:1996ke} and for the massless MO model in Ref.~\cite{Mendes:2019zpw}.}} we write the slowly-rotating,
stationary, axisymmetric metric first considered by \citet{Hartle:1967he},
\begin{eqnarray}
    \reply{g_{\mu\nu}^\mathrm{axi}dx^\mu dx^\nu =ds_{\text{sph}}^2+2\rho^2\sin^2 \theta
    [\omega(\rho,\theta)-\Omega]dtd\phi },
\end{eqnarray}
which can be regarded as a perturbation on ansatz~\eqref{eq:SSm}, together with
the 4-velocity of the perfect fluid $u^\mu=(u^t,0,0,u^\phi)$ already shown in
Eq.~\eqref{eq:4v}.  Here, $\omega(\rho,\theta)$ is a function of the order of
angular velocity, $\Omega \equiv u^\phi/u^t$.  Owing to the rotational symmetry,
odd-parity terms in $\Omega$ vanish identically within all diagonal components
of both the gravitational and scalar field equations, thereby
Eqs.~\eqref{eq:tovn}--\eqref{eq:tovp} remain unaltered.  The supplementary
differential equation governing $\omega(\rho,\theta)$ arises exclusively from
the non-trivial off-diagonal $t\phi$-component of the field equation
\begin{eqnarray}\label{eq:rot}
    R_{t\phi} = 8\pi T_{t\phi},
\end{eqnarray}
at the linear order in $\Omega$.  Remarkably, axisymmetry ($\partial_\phi=0$)
renders Eq.~\eqref{eq:rot} independent of scalar field contributions.  This
leads to dual consequences in equations: (i) indistinguishability between
massless and massive ST theories, and (ii) structural similarity with
the formula in GR, differing solely through a conformal transformation of the
matter energy-momentum tensor.

When expanding the $\omega(\rho,\theta)$ function in a Legendre polynomial
basis,
\begin{eqnarray}
    \omega(\rho,\theta) = \sum_{\ell=1}^\infty
    \omega_\ell(\rho)\left(-\frac{1}{\sin\theta}\frac{dP_\ell}{d\theta}\right),
\end{eqnarray}
the asymptotic behavior in the large-$\rho$ regime reveals that only the dipole
component $\ell=1$ contributes, which implies an angular independence of
$\omega$.  Consequently, the partial differential equation of
$\omega(\rho,\theta)$ reduces to an ordinary radial differential
equation~\cite{Damour:1996ke,Hu:2021tyw},
\begin{eqnarray}\label{eq:moi}
	\omega''= \frac{4\pi\rho A^4}{\rho-2m} (\tilde\varepsilon+ \tilde
	p)(\rho\omega'+4\omega)+\left(\rho\psi^2- \frac4\rho\right)\omega'.
\end{eqnarray}
By the virtue of its homogeneity, Eq.~\eqref{eq:moi} permits arbitrary choice of
$\omega_c$ subject to $\omega'_c=0$, as mandated by the small-$\rho$ expansion,\footnote{\reply{Throughout the perturbation analysis, the same subscript ‘$c$’, denoting the value at $\rho=0$, is employed both here and in Eqs.~\eqref{eq:td1} and~\eqref{eq:td2}.}}
\begin{eqnarray}
    \omega= \omega_c+\frac{8\pi}{5} (\tilde p_c+\tilde\varepsilon_c)
    A_c^4\omega_c\rho^2+\mathcal O(\rho^3).
\end{eqnarray}

To calculate how the external field distorts the NS, we perturb the metric field
and the scalar field as
\begin{align}
	\label{eq:RWg}\delta g_{\mu \nu}^{(2 m)}&=\left[\begin{array}{cccc}
    -e^\nu H_0 & H_1 & 0 & 0 \\
    H_1 & H_2 /\left(1-\frac{2 m}{\rho}\right) & 0 & 0 \\
    0 & 0 & \rho^2 K & 0 \\
    0 & 0 & 0 & \rho^2 \sin ^2 \theta K
    \end{array}\right]\nonumber\\
    &\qquad\times Y_{2 m}(\theta, \phi),\\
    \delta\varphi^{(2m)}&=\mathcal{J}Y_{2m}(\theta,\phi),\label{eq:RWs}
\end{align}
where $H_0$, $H_1$, $H_2$, $K$ and $\mathcal{J}$ are functions depending only on
$\rho$, and $Y_{2m}$ is the spherical harmonic function.  Note that
Eq.~\eqref{eq:RWg} is just the $\ell=2$ polar perturbation in the Regge-Wheeler
gauge~\cite{Regge:1957td}.

Substituting the perturbation equations~\eqref{eq:RWg} and \eqref{eq:RWs} into
the field equations~\eqref{eq:eomG} and \eqref{eq:eomP}, the off-diagonal terms
of the perturbed gravitational field equation give
\begin{align}
	&H_0=\mathcal{H}(\rho), \\
	&H_1=0,\\
    &H_2=-\mathcal{H}(\rho),\\
	&K=-\int(H'+H\nu'+4\mathcal{J}\varphi')d\rho,
\end{align}
where $\mathcal{H}(\rho)$ is a function of $\rho$.  Moreover, the diagonal terms
of the perturbed metric field equation, together with the perturbed scalar field
equation gives the following coupled second-order differential equations,
\begin{align}
	\label{eq:H}\mathcal{H}''+c_1\mathcal{H}'+c_0\mathcal{H}&=c_s\mathcal{J},\\
	\label{eq:J}\mathcal{J}''+d_1\mathcal{J}'+d_0\mathcal{J}&=d_s\mathcal{H},
\end{align}
where
\begin{widetext}
    \begin{align}
	c_0 &=-\frac{6-6x+x^2}{(1-x)^2\rho^2} -\frac{4}{(1-x)^2} \Big[\pi A^4\tilde
	p(16\pi\rho^2A^4\tilde p-9+13x-8\rho^2V)+ (1-2x+\rho^2V)V\Big]\nonumber\\
	&\quad+\frac{2}{1-x}\left[2\pi A^4 \left(5\tilde\varepsilon+ (\tilde
	p+\tilde\varepsilon) \frac{d\tilde\varepsilon}{d\tilde p} \right)-
	(x+8\pi\rho^2A^4\tilde p-2\rho^2V) \varphi'^2\right]-\rho^2\varphi'^4,\\
	d_0& =-\frac{6}{(1-x)\rho^2} +\frac{4\pi A^2}{1-x} \left\{A(3\tilde p-\tilde
	\varepsilon)\frac{d^2A}{d\varphi^2}+ \left[\left( \frac{d\tilde
	\varepsilon}{d\tilde p} +6\right) \tilde p +\left(\frac{d\tilde
	\varepsilon}{d\tilde p}-6\right) \tilde\varepsilon \right]\left(
	\frac{dA}{d\varphi} \right)^2\right\}-\frac{1}{1-x}
	\frac{d^2V}{d\varphi^2}-4\varphi'^2,\\
	c_1=d_1 &=\frac{2-x-2\rho^2\big[ V-2\pi A^4(\tilde p-\tilde\varepsilon)
	\big]}{(1-x)\rho},\\
	c_s =4d_s &=\frac{4x\varphi'}{(1-x)\rho} -\frac{8\pi A^3}{1-x}
	\left\{\left[\left(\frac{d\tilde \varepsilon}{d\tilde p} -9\right) \tilde
	p+\left(\frac{d\tilde\varepsilon}{d\tilde p} -1\right) \tilde\varepsilon
	\right]\frac{dA}{d\varphi}-4\rho A\tilde p\varphi'\right\}-\frac{4}{1-x}
	\left(\frac{dV}{d\varphi}+2\rho V\varphi'\right)+4\rho\varphi'^3,
\end{align}
\end{widetext}
and the variable $x$ is defined as $2m/\rho$.  To go back to GR, one sets
$A(\varphi)=1$, $V(\varphi)=0$, and $\varphi(\rho)=0$, indicating $c_s=d_s=0$,
which decouples the equations of $\mathcal{H}$ and $\mathcal{J}$, giving the tidal
equation~\cite{Hinderer:2007mb},
\begin{align}\label{eq:grtide}
    &\left[-\frac{6 e^\mu}{r^2}+4 \pi e^\mu\left(5\tilde\varepsilon+9\tilde
    p+\frac{\tilde\varepsilon+\tilde p}{d\tilde p /
    d\tilde\varepsilon}\right)-\nu^{\prime 2}\right]\mathcal{H}\nonumber\\
    &\quad+\left[\frac{2}{r}+e^\mu\left(\frac{2m}{r^2}+4 \pi r(\tilde p- \tilde
    \varepsilon)\right)\right]\mathcal{H}^{\prime}+\mathcal{H}^{\prime \prime}=0,
\end{align}
and an unrelated equation of $\mathcal{J}$,
\begin{align}
	\mathcal{J}''+\left[\frac{2}{r} +e^\mu\left(\frac{2 m}{r^2}+4 \pi r(\tilde
	p-\tilde\varepsilon)\right) \right]\mathcal{J}'
	-\frac{6\mathcal{J}}{r(r-2m)}=0,
\end{align}
where $\mu$ is defined by
\begin{eqnarray}
    e^{-\mu(r)}=1-\frac{2m(r)}{r}.
\end{eqnarray}
We replace the variable $\rho$ with the physical radial coordinate $r$ here
since they are indistinguishable now.  Note that Eqs.~\eqref{eq:H} and
\eqref{eq:J} are linear equations of $\mathcal{H}$ and $\mathcal{J}$.  To solve
them, one can integrate the system twice with the initial values
\begin{eqnarray}\label{eq:td1}
    \mathcal{H}_c=a_0^2, \quad\mathcal{H}_0' =2a_0,\quad
    \mathcal{J}_c=0,\quad\mathcal{J}_c'=0,
\end{eqnarray}
and
\begin{eqnarray}\label{eq:td2}
    \mathcal{H}_c=0,\quad\mathcal{H}_c'=0,\quad\mathcal{J}_c=a_0^2,\quad
    \mathcal{J}_c'=2a_0,
\end{eqnarray}
respectively, with $a_0$ being an arbitrary number.
\reply{The physical solution for $\mathcal{H}(\rho)$ and $\mathcal{J}(\rho)$ can thus be expressed as a linear combination of the two solutions with initial values specified in~Eqs.~\eqref{eq:td1} and~\eqref{eq:td2}, subject to the condition that $\mathcal{J}$ vanishes at spatial infinity.}

To numerically calculate the moment of inertia and the tidal deformability, one
needs to integrate Eqs.~\eqref{eq:moi}, \eqref{eq:H} and \eqref{eq:J}, together
with the modified TOV equations, from the stellar center to spatial infinity.
The moment of inertia is derived from the asymptotic behavior of the metric
element $g_{t\phi}$.  When $\rho$ approaches infinity, one has
\begin{eqnarray}
    g_{t\phi}\to-\frac{2J\sin^2\theta}{\rho},
\end{eqnarray}
yielding the angular momentum $J$.  The moment of inertia is subsequently
extracted as
\begin{eqnarray}\label{eq:I}
    I=\frac{J}{\Omega}=\left.
    \frac{\rho^4\omega'}{6\omega}\right|_{\rho\to\infty},
\end{eqnarray}
applying the boundary condition $\omega(\rho)\to\Omega$ as $\rho\to\infty$.
However, the infinity condition cannot be physically realized in the massive ST
theory, and directly replacing the boundary condition $\rho\to\infty$ with
$\rho\to\rho_i$ introduces significant numerical errors because $\rho_i$ is
typically comparable to the stellar radius, especially when the scalar masse is
large.  To address this issue, we assume that in the region $\rho > \rho_i$ the
solution takes the form  
\begin{eqnarray}
\omega(\rho) = \Omega - \frac{2J}{\rho^3},
\end{eqnarray}
which corresponds to the solution of Eq.~\eqref{eq:moi} when the scalar field
vanishes in the vacuum.  Then the numerical result of the moment of inertia
becomes
\begin{eqnarray}
    I=\frac{J}{\Omega}= \left. \frac{\rho^4\omega'}{6\omega+2\rho\omega'}
    \right|_{\rho\to\rho_i}.
\end{eqnarray}

In massless ST theories, NSs develop scalar charges that modify gravitational
interactions, complicating tidal deformability
calculations~\cite{Brown:2022kbw,Creci:2023cfx,Creci:2024wfu}.  Therefore, we
consider in the massive ST theories the scenario where the scalar field decays
exponentially far away ‌from‌ the NS, rendering the far zone spacetime
indistinguishable from the corresponding GR solution.  The vacuum solution of
Eq.~\eqref{eq:grtide} yields the tidal Love number $k_2$, linked to the NS
compactness $\mathscr{C}$ and its surface variable $y_s$~\cite{Hinderer:2007mb}.
Following this methodology, $\mathcal{H}(\rho)$ for $\rho>\rho_i$ is
approximated as
\begin{align}\label{eq:hinderer}
    \mathcal H &=3a_1\left(\frac{\rho}{m_i}\right)^2 \left(1-\frac{2m_i}{\rho}
    \right)\left[\frac{a_2}{a_1}-\frac{1}{2} \ln
    \left(1-\frac{2m_i}{\rho}\right)\nonumber\right.\\
    &\qquad \left.-\frac{m_i(m_i-\rho)\left(2
    m_i^2+6m_i\rho-3\rho^2\right)}{3\rho^2(2m_i-\rho)^2}\right],
\end{align}
where coefficients $a_1$ and $a_2$ are determined through asymptotic matching
between the Newtonian potential
\begin{eqnarray}
    \varPhi_\mathrm{N}=-\frac{1+g_{tt}^\mathrm{tide}}{2}
    =-\frac{1+e^\nu(1+\mathcal{H})}{2},
\end{eqnarray}
and the effective tidal external potential~\cite{Binnington:2009bb}.  The tidal
deformability is then derived as~\cite{Hinderer:2007mb,Hu:2021tyw},
\begin{align}
\lambda&= \frac{2}{3} \rho_i^5 \frac{8 C_i^5}{5}(1-2 C_i)^2 \big[2+2
C_i(y_i-1)-y_i \big]\nonumber \\
&\quad \times\Big\{4 C_i^3\big[13-11 y_i+C_i(3 y_i-2)+2
C_i^2(1+y_i)\big] \nonumber \\
&\qquad +3(1-2 C_i)^2 \big[2-y_i+2C_i(y_i-1) \big] \ln (1-2 C_i)\nonumber \\
&\qquad +2 C_i \big[6-3 y_i+3 C_i(5 y_i-8) \big] \Big\}^{-1},
\end{align}
with
\begin{eqnarray}
    C_i=\left.\frac{m}{\rho} \right|_{\rho\to\rho_i}, \quad
    y_i=\left.\frac{\rho\mathcal{H}'}{\mathcal{H}}\right|_{\rho\to\rho_i},
\end{eqnarray}
evaluated at $\rho=\rho_i$ which is serving as the infinity in our computation.

\subsection{Numerical results}

\begin{figure*}
\includegraphics[width=\linewidth]{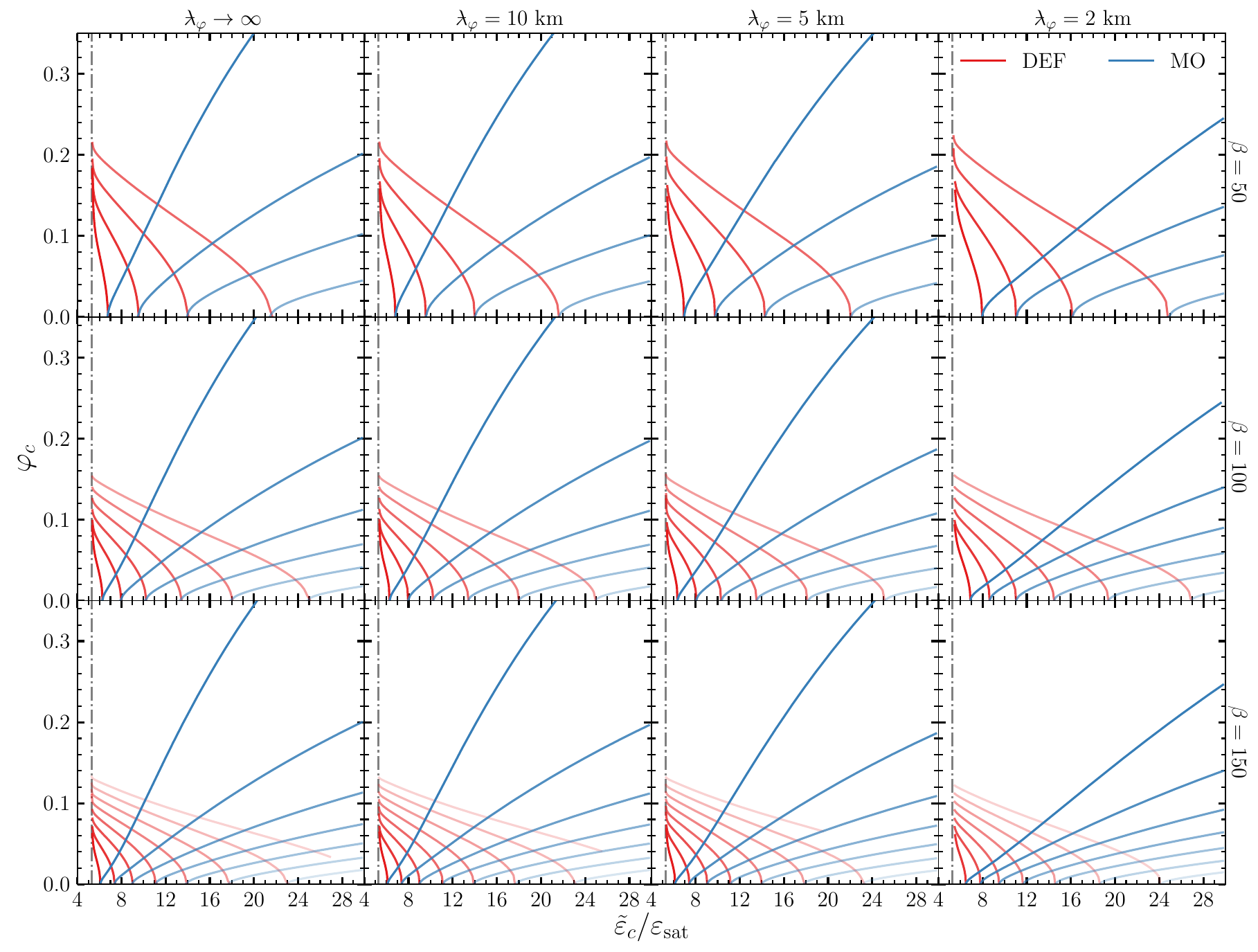}
\caption{\label{fig:scalarization} The central scalar field $\varphi_c$ versus
the central mass density $\tilde{\varepsilon}_c$ (normalized by the nuclear
saturation density ${\varepsilon}_{\rm sat}$) for scalarized solutions in DEF
(red) and MO (blue) theories with the AP4 EOS.  From top to bottom, $\beta = 50,
100, 150$; from left to right, $m_\varphi$ increases, corresponding to the
reduced Compton wavelengths of $\infty$, $10\,\mathrm{km}$, $5\,\mathrm{km}$,
and $2\,\mathrm{km}$.  The gray dot-dashed line marks the density at which the
trace of the energy-momentum tensor vanishes.  Darker curves indicate earlier
onset of scalarization as the central density increases.  Red curves in the
bottom panels truncate upon reaching the maximum pressure of the AP4 EOS.}
\end{figure*}

The central value of the scalar field, shown in Fig.~\ref{fig:scalarization} as
a function of the central mass density $\tilde{\varepsilon}_c$, serves as an
indicator of scalarization for different values of $\beta$ and reduced Compton
wavelengths, defined as $\barlambda_\varphi=\lambda_\varphi/(2\pi)$.  The red
and blue curves, representing the DEF and MO theories respectively, bifurcate
from the horizontal axis at the same points because, in the limit $\varphi \to
0$, the two theories coincide with respect to the scalarization phenomenon, that
is, $\beta_\mathrm{DEF}=\beta_\mathrm{MO}$.  After bifurcating from the same
point, the two curves diverge in different directions: the blue curve (MO
theory) moves toward larger central densities, while the red curve (DEF theory)
moves in the opposite direction, approaching the gray dot-dashed line that
represents the trace-sign-reversed condition.  This confirms  earlier results
that NSs can become scalarized in the $\beta>0$ DEF theory even when the
trace-sign-reversed region is small, provided that $\varphi_c$ is sufficiently
large.  This opposite evolution ultimately leads to different stability
properties, as reflected in the distinct outcomes following the gravitational
collapse~\cite{Mendes:2016fby}.  As the central density increases, additional
bifurcation points emerge, revealing the multi-branch structure of scalarized
solutions.  Typically, the first branch of scalarized NSs attracts the most
attention and is shown in the darkest color, while subsequent branches are
plotted with progressively lighter shades.

We now focus on the direct differences among all the subfigures.  Within each
row, all corresponding bifurcation points shift to higher central densities as
$\barlambda_\varphi$ decreases, indicating that the scalar mass suppresses the
scalarization, similar to the behavior in the $\beta<0$
regime~\cite{Doneva:2022ewd}.  Within each column, the number of branches
increases with $\beta$, reflecting the fact that a larger $\beta$ enhances
scalarization.

\begin{figure*}
\includegraphics[width=0.7\linewidth]{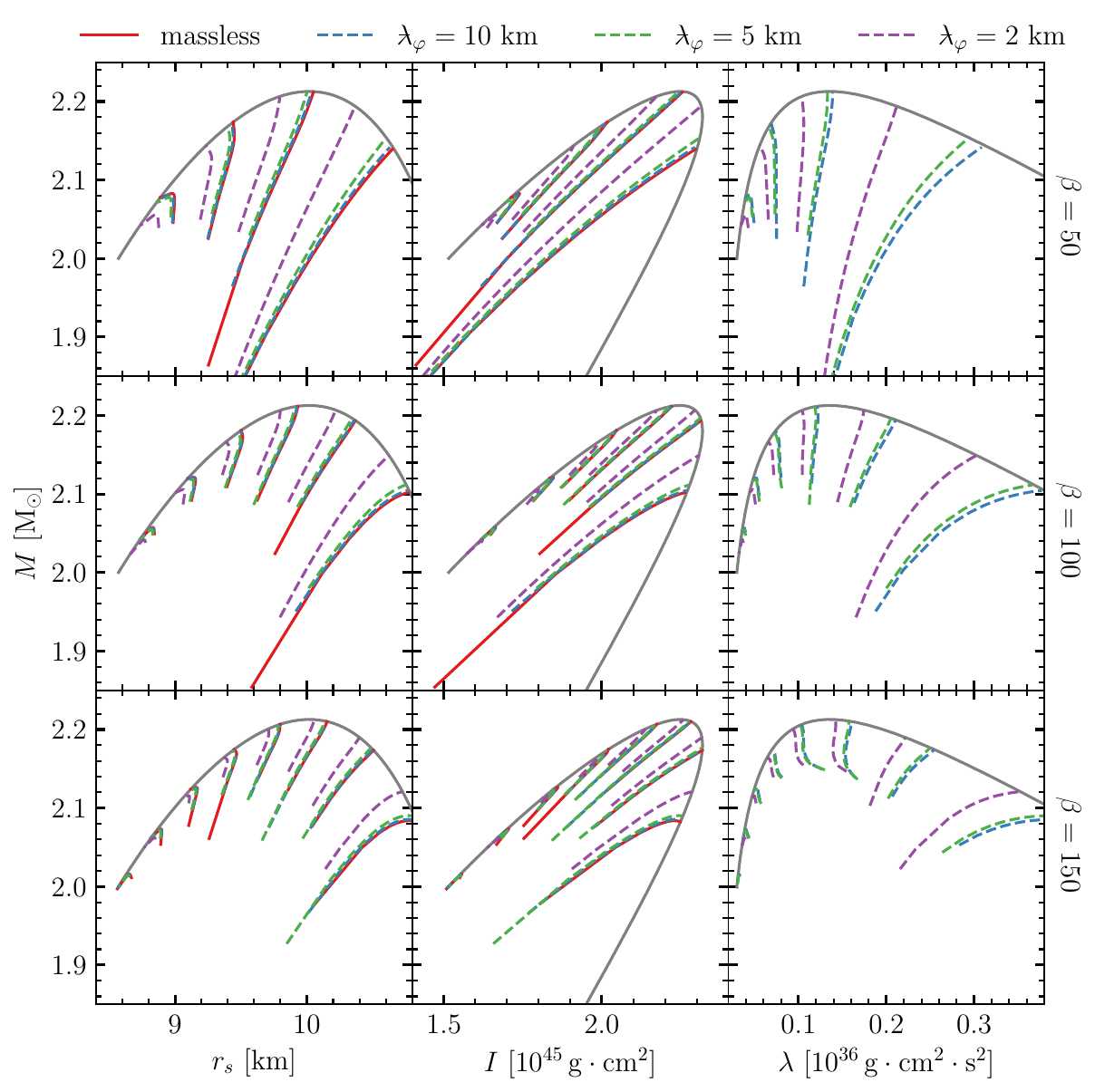}
\caption{\label{fig:DEF} Numerical properties of scalarized NSs in massless and
massive DEF theories with the AP4 EOS. From top to bottom, $\beta = 50, 100,
150$.  Each subfigure contains several branches of scalarized NSs.  Gray curves
correspond to GR solutions, while red curves represent NSs in the massless DEF
theory.  Dashed curves in blue, green, and purple correspond to massive DEF
theories with reduced Compton wavelengths of $10\,\mathrm{km}$,
$5\,\mathrm{km}$, and $2\,\mathrm{km}$, respectively.}
\end{figure*}
\begin{figure*}
\includegraphics[width=0.7\linewidth]{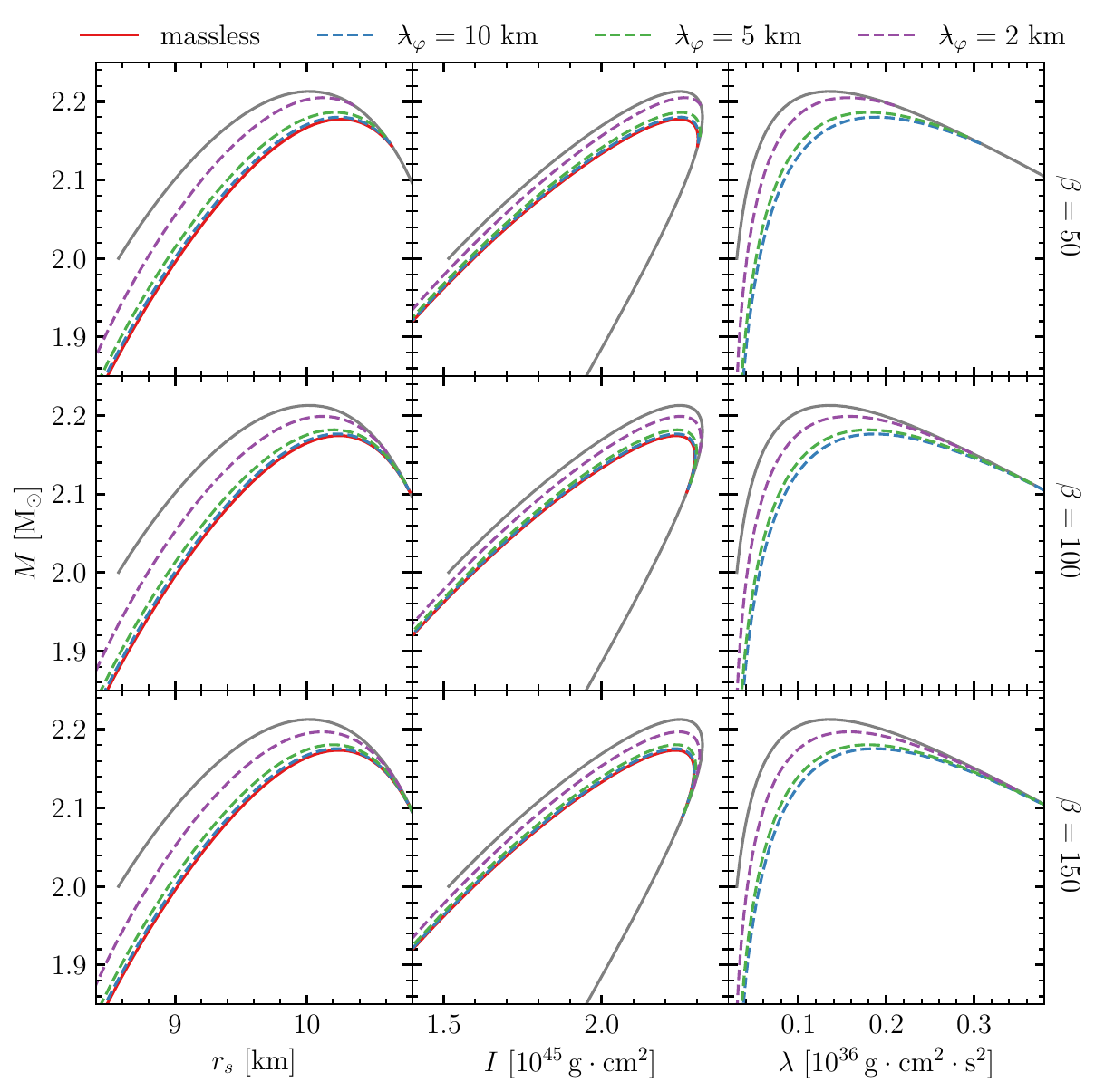}
\caption{\label{fig:MO} Numerical properties of scalarized NSs in massless and
massive MO theories with AP4 EOS.  The overall structure is similar to
Fig.~\ref{fig:DEF}.  Only the first branch of scalarized NSs is shown.}
\end{figure*}
 
The radius, moment of inertia, and tidal deformability are plotted against the
ADM mass of the NSs in the left, middle, and right columns of Fig.~\ref{fig:DEF}
and Fig.~\ref{fig:MO} for the DEF and MO theories, respectively.  The red curves
denoting the massless ST theories with $\beta=50$, $100$, and $150$ are shown in
the upper, middle, and lower rows, respectively, while the blue, green, and
purple dashed curves correspond to massive ST theories with
$\barlambda_\varphi=10\,\mathrm{km}$, $5\,\mathrm{km}$, and $2\,\mathrm{km}$,
respectively.  In Fig.~\ref{fig:DEF}, the multi-branch structure is clearly
visible, whereas in Fig.~\ref{fig:MO} only the first branch is shown for
clarity, since the second and subsequent branches, while bifurcating from the GR
curves (gray) at the same points as in the DEF theory, are essentially
indistinguishable.

Unlike in the well-studied $\beta<0$ parameter space (see
Fig.~\ref{fig:mlessscalarization} and Fig.~\ref{fig:msivescalarization}), the
curves of ST theories appear only on one side of the GR curves.  Moreover, since
scalarization occurs only at high central densities, the curves for ST theories
are confined to the high-mass region, extending even into the domain where NSs
in GR become unstable.  The mass of scalarized NSs are all smaller than the
corresponding  stars in GR, which, given an EOS, making them less likely to
explain the observed pulsars with large
masses~\cite{Antoniadis:2013pzd,Fonseca:2021wxt}.  The moment of inertia and
tidal deformability are also smaller than those of a stable star in GR with the
same ADM mass.  In the two scenarios of $\beta>0$ and $\beta<0$, the moment of
inertia of NSs lies on opposite sides of the GR curve, marking a clear
distinction between the two cases.  \citet{Hu:2021tyw} found that the
tidal deformation in ST theories can deviate from the GR prediction by more than
$10\%$, and even more for sufficiently large $\beta$ and $\barlambda_\varphi$.
In contrast, for stable NSs in the MO theories, this deviation remains small
owing to their more constrained coupling functions, as seen in Fig.~\ref{fig:MO}.

The green dashed lines ($\barlambda_\varphi = 5$\,km), and especially the blue
dashed lines ($\barlambda_\varphi = 10$\,km), lie almost in close contact with
the solid red curve ($\barlambda_\varphi = \infty$), indicating that for
relatively small masses, the properties of scalarized NSs are nearly identical
to those calculated in the massless ST theories.  The separation between the
curves representing theories with different $m_\varphi$ becomes even smaller as
$\beta$ increases, because a larger difference in $\barlambda_\varphi$ between
the curves is required to make them distinguishable in ST theories with larger
$\beta$.  Generally speaking, a massive theory with a reduced Compton wavelength
smaller than $5\,\mathrm{km}$ can be visually distinguished from the massless
theory by examining the properties of scalarized NSs.  As $\barlambda_\varphi$
decreases further, the scalar field decays exponentially more rapidly, reducing
its influence on gravity and causing the theory to gradually converge toward GR,
as seen in Fig.~\ref{fig:MO}, where the purple curves ($\barlambda_\varphi =
2$\,km) approach the GR curves.

\section{SUMMARY}\label{sec:5}

It has been more than three decades since the mechanism of spontaneous
scalarization was first discovered in the so-called DEF theory.
Theoretical~\cite{Harada:1997mr, Harada:1998ge} and
numerical~\cite{Damour:1996ke} studies of the DEF theory with negative $\beta$
establish $\beta\lesssim-4.35$ as the critical threshold for scalarization in
theory, with some sensitivity to the EOS~\cite{Novak:1997hw,
AltahaMotahar:2017ijw}.  However, recent pulsar-timing observations have
excluded the original massless DEF theory with an effective scalar charge
$\gtrsim 10^{-2}$, no matter what the EOS is~\cite{Antoniadis:2013pzd,
Kramer:2021jcw, Shao:2017gwu, Zhao:2022vig}.  Therefore, as a complement, we
investigate scalarized NSs with a positive coupling parameter $\beta$ in both
DEF and MO theories, considering both massless and massive cases, as $\beta$
would be bounded from above by $\beta\sim\mathcal{O}(10^3)$~\cite{Palenzuela:2015ima}.

For a given central density and pressure as the boundary conditions, there can
exist a series of scalarized solutions with increasing central values of the
scalar field.  This multi-branch structure arises from the positive combination
$4\pi\beta\tilde T-m_\varphi^2 > 0$, which naturally occurs in extremely dense
relativistic  stars with a core where the trace of the energy-momentum tensor is
positive. Furthermore, we point out that this multi-branch structure also
appear in scenarios with a very negative $\beta$ (see Appendix~\ref{sec:app}).

We solve the modified TOV equations to obtain the stellar masses and radii, as
well as the detailed structure of the NS spacetime.  Given a static and
spherically symmetric background solution, we further introduce slow rotation
and tidal perturbations to compute the moment of inertia and tidal
deformability.  This is the first time such calculations have been carried out
in the massive ST case with a multi-branch structure.

In the positive-$\beta$ scenario, a NS can become scalarized once its central
density exceeds a threshold, and the scalarization branches never converge back
to the corresponding GR solutions.  As a result,
all the curves of physical quantities
for scalarized NSs to lie entirely on one side of the GR curves, and
scalarized NSs exhibit smaller ADM masses.  Furthermore, compared with stable
NSs in GR  having the same ADM mass, scalarized NSs have smaller radii, lower
moments of inertia, and reduced tidal deformabilities.

The mass term has less influence on the numerical results compared with that in
the negative-$\beta$ scenario, since a large absolute value of $\beta$ is
considered here.  For a scalar field with a small mass, whose Compton wavelength
is comparable to or larger than the stellar radius, the properties of NSs are
not significantly different from those in the massless  ST theory.  Therefore,
the constraints on NS properties derived for the massless theory can also be
applied to these massive ST theories.  For a scalar field with a large mass,
whose Compton wavelength is significantly smaller than the stellar radius,
scalarization is dramatically suppressed, and the scalarized NSs are close to
the  NS solutions in GR.

\reply{\citet{Mendes:2018qwo} investigated the stability of scalarized NSs by computing the radial oscillation frequencies in both the massless DEF and MO theories with $\beta=100$.
It was found that in the massless DEF theory all NSs are unstable and collapse to black holes, while in the MO model only the first few branches of scalarized solutions (corresponding to relatively small central densities) remain stable up to the TOV mass of their respective branches and undergo scalarization~\cite{Mendes:2016fby}.
In their work, the stability criterion for scalarized solution branches can be expressed as $dM/d\tilde\varepsilon_c<0$, which is established as a necessary condition in GR~\cite{1965gtgc.book.....H} and holds approximately in modified gravity theories~\cite{Harada:1997mr, Brito:2015yfh}.
Applying this criterion to our results (see Fig.~\ref{fig:DEF}) indicates that, in the massive DEF theories with $\beta \gg 1$, all scalarized solutions are also expected to be unstable.
Nevertheless, a rigorous stability analysis in massive theories necessitates the inclusion of time-dependent perturbations to compute the frequencies of various NS modes~\cite{Kokkotas:1999bd}, particularly the radial oscillations~\cite{Chandrasekhar:1964zz}.}

As a conclusion, the existence of multiple branches of scalarized NS solutions
appears to be a generic feature of ST theories with a positive coupling
parameter $\beta$, a phenomenon that has thus far received limited attention.
\reply{The associated multi-branch structure suggests that the dynamics and evolution of an individual NS in the DEF model, as well as potential transitions between distinct branches in the MO model (primarily the first few stable ones), could be considerably more intricate than previously recognized. Stability analysis~\cite{Mendes:2018qwo} should be carefully taken into consideration.}
Furthermore, since such solutions are realized only in
highly compact stars, their investigation offers a promising avenue for
co-investigating the EOS of nuclear matter in the high-density regime.

\begin{acknowledgments}
We thank Zexin Hu and Yong Gao for helpful discussions, \reply{and the anonymous referee for invaluable feedback and constructive comments}. This work was supported
by the National SKA Program of China (2020SKA0120300), the Beijing Natural
Science Foundation (1242018), 
the National Natural Science Foundation of China (12573042),
the Max Planck Partner Group Program funded by the
Max Planck Society, and the High-Performance Computing Platform of Peking
University.
\end{acknowledgments}

\appendix
\section{MULTIPLE SCALARIZED BRANCHES IN NEGATIVE-$\beta$ SCENARIO}\label{sec:app}

\begin{figure}
\includegraphics[width=\linewidth]{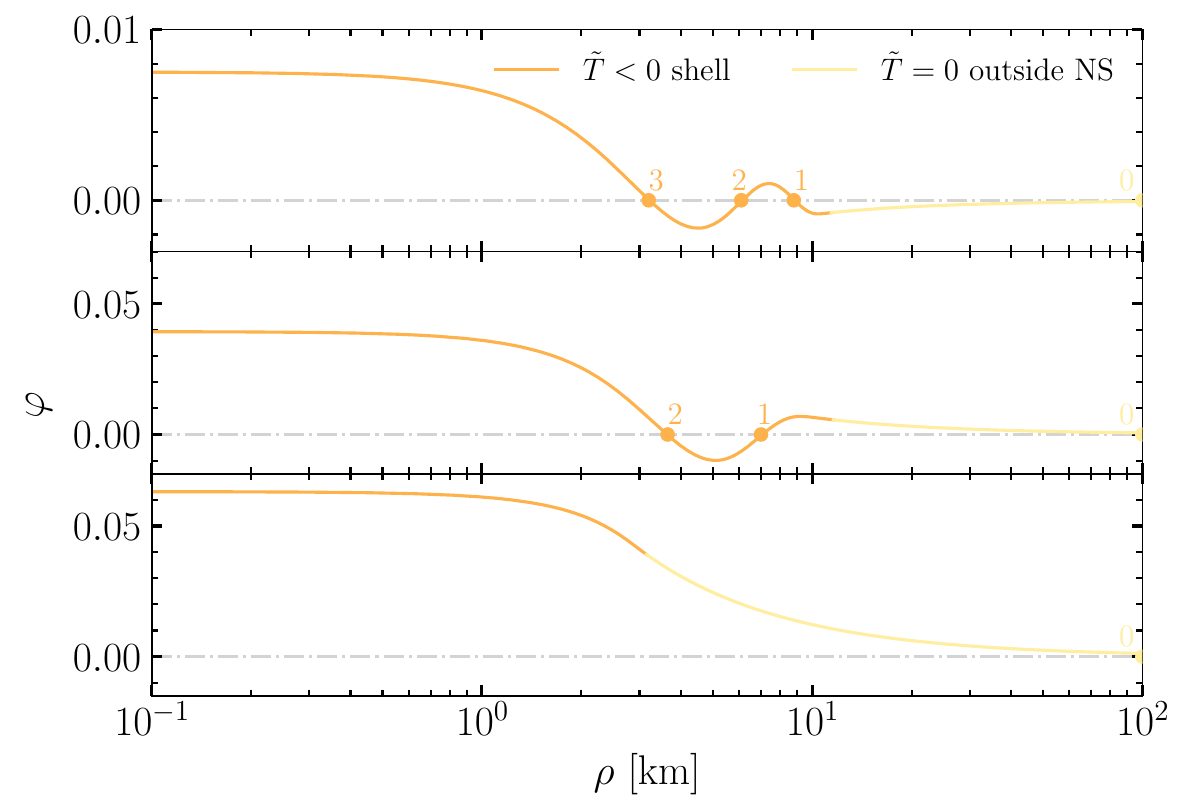}
\caption{\label{fig:app} The scalar field profiles of three scalarized NS
solutions in massless DEF theory with $\beta=-200$, and
$\tilde\varepsilon_c=10^{15}\,\mathrm{g\cdot cm^{-3}}$.  The AP4 EOS is adopted,
with the yellow dots and numbers having the same meaning as in
Fig.~\ref{fig:msDEF}, except that the color has been changed from red to orange
because the zero points lie in the $\tilde{T}<0$ region.  It should be noted
that the vertical axis range in the top panel differs from that in the bottom
two panels.}
\end{figure}

For negative $\beta$, the oscillation of the scalar field occurs in the
$\tilde{T} < 0$ region, as shown in Fig.~\ref{fig:app}.  As the central density
increases, the number of nodes in the scalar field profile decreases, similar to
the behavior observed in the MO theory (Fig.~\ref{fig:msMO}).  The missing
scalarized solution with only one node inside the NS is due to the discontinuity
of $\varphi_0$ as a function of $\varphi_c$ (like the red cliff in
Fig.~\ref{fig:msDEF}).

\begin{table*}
    \caption{\label{tab:app} Physical parameters of three scalarized NSs in the
    massless DEF theory with $\beta=-200$, including central density, central
    scalar field, stellar radius, ADM mass, baryonic mass, compactness,
    charge-to-mass ratio, and moment of inertia.}
    \renewcommand\arraystretch{1.8}
    \begin{ruledtabular}
    \begin{tabular}{cccccccc}
    $\tilde\varepsilon_c$ $(\mathrm{g\cdot cm^{-2}})$ & $\varphi_c$ & $r_s$ (km)
    & $M$ $(\mathrm{M_\odot})$ & $M_\mathrm{b}$ $(\mathrm{M_\odot})$ & $\mathscr{C}$ &
    $\eta$ & $I$ $(\mathrm{g\cdot cm^2})$\\
    \colrule
    $10^{15}$ & $0.0075$ & $11.4168$ & $1.4065$ &
    $1.5796$ & $0.1819$ & $0.00321$ & $1.3332\times10^{45}$ \\
    $10^{15}$ & $0.0394$ & $11.5515$ & $0.5677$ &
    $0.5933$ & $0.0726$ & $-0.0718$ & $0.3730\times10^{45}$ \\
    $10^{15}$ & $0.0631$ & $2.7033$ & $0.0125$ &
    $0.0131$ & $0.0068$ & $-6.6208$ & $0.0007\times10^{45}$ \\
    \end{tabular}
    \end{ruledtabular}
\end{table*}

Table~\ref{tab:app} summarizes several properties of the three scalarized NSs.
Increasing the central scalar field leads to a decrease in both mass and
compactness, whereas the scalar charge becomes more prominent for larger
$\varphi_c$.  In the case where the scalar field profile has no node inside the
star, the resulting parameters fall outside the typical range of NSs.

To our knowledge, this represents the first identification of multi-branch
scalarized NS solutions in the DEF theory for $\beta<0$.  Previous studies
overlooked this phenomenon due to their focus on small negative values $|\beta|
\sim\mathcal{O}(10)$, where the characteristic half-wavelength substantially
exceeds the stellar radius, consequently screening oscillatory behavior of the
scalar field.  However, we do not consider this scenario in detail, as the
parameter space for this $\beta$ is likely already excluded and the adopted EOS
may be unsuitable for the low nuclear density regime.


\bibliography{STT}

\begin{thebibliography}{117}%
\makeatletter
\providecommand \@ifxundefined [1]{%
 \@ifx{#1\undefined}
}%
\providecommand \@ifnum [1]{%
 \ifnum #1\expandafter \@firstoftwo
 \else \expandafter \@secondoftwo
 \fi
}%
\providecommand \@ifx [1]{%
 \ifx #1\expandafter \@firstoftwo
 \else \expandafter \@secondoftwo
 \fi
}%
\providecommand \natexlab [1]{#1}%
\providecommand \enquote  [1]{``#1''}%
\providecommand \bibnamefont  [1]{#1}%
\providecommand \bibfnamefont [1]{#1}%
\providecommand \citenamefont [1]{#1}%
\providecommand \href@noop [0]{\@secondoftwo}%
\providecommand \href [0]{\begingroup \@sanitize@url \@href}%
\providecommand \@href[1]{\@@startlink{#1}\@@href}%
\providecommand \@@href[1]{\endgroup#1\@@endlink}%
\providecommand \@sanitize@url [0]{\catcode `\\12\catcode `\$12\catcode
  `\&12\catcode `\#12\catcode `\^12\catcode `\_12\catcode `\%12\relax}%
\providecommand \@@startlink[1]{}%
\providecommand \@@endlink[0]{}%
\providecommand \url  [0]{\begingroup\@sanitize@url \@url }%
\providecommand \@url [1]{\endgroup\@href {#1}{\urlprefix }}%
\providecommand \urlprefix  [0]{URL }%
\providecommand \Eprint [0]{\href }%
\providecommand \doibase [0]{http://dx.doi.org/}%
\providecommand \selectlanguage [0]{\@gobble}%
\providecommand \bibinfo  [0]{\@secondoftwo}%
\providecommand \bibfield  [0]{\@secondoftwo}%
\providecommand \translation [1]{[#1]}%
\providecommand \BibitemOpen [0]{}%
\providecommand \bibitemStop [0]{}%
\providecommand \bibitemNoStop [0]{.\EOS\space}%
\providecommand \EOS [0]{\spacefactor3000\relax}%
\providecommand \BibitemShut  [1]{\csname bibitem#1\endcsname}%
\let\auto@bib@innerbib\@empty
\bibitem [{\citenamefont {'t~Hooft}\ and\ \citenamefont
  {Veltman}(1974)}]{tHooft:1974toh}%
  \BibitemOpen
  \bibfield  {author} {\bibinfo {author} {\bibfnamefont {G.}~\bibnamefont
  {'t~Hooft}}\ and\ \bibinfo {author} {\bibfnamefont {M.~J.~G.}\ \bibnamefont
  {Veltman}},\ }\href@noop {} {\bibfield  {journal} {\bibinfo  {journal} {Ann.
  Inst. H. Poincare A Phys. Theor.}\ }\textbf {\bibinfo {volume} {20}},\
  \bibinfo {pages} {69} (\bibinfo {year} {1974})}\BibitemShut {NoStop}%
\bibitem [{\citenamefont {Hawking}\ and\ \citenamefont
  {Penrose}(1970)}]{Hawking:1970zqf}%
  \BibitemOpen
  \bibfield  {author} {\bibinfo {author} {\bibfnamefont {S.~W.}\ \bibnamefont
  {Hawking}}\ and\ \bibinfo {author} {\bibfnamefont {R.}~\bibnamefont
  {Penrose}},\ }\href {\doibase 10.1098/rspa.1970.0021} {\bibfield  {journal}
  {\bibinfo  {journal} {Proc. Roy. Soc. Lond. A}\ }\textbf {\bibinfo {volume}
  {314}},\ \bibinfo {pages} {529} (\bibinfo {year} {1970})}\BibitemShut
  {NoStop}%
\bibitem [{\citenamefont {Hawking}(1976)}]{Hawking:1976ra}%
  \BibitemOpen
  \bibfield  {author} {\bibinfo {author} {\bibfnamefont {S.~W.}\ \bibnamefont
  {Hawking}},\ }\href {\doibase 10.1103/PhysRevD.14.2460} {\bibfield  {journal}
  {\bibinfo  {journal} {Phys. Rev. D}\ }\textbf {\bibinfo {volume} {14}},\
  \bibinfo {pages} {2460} (\bibinfo {year} {1976})}\BibitemShut {NoStop}%
\bibitem [{\citenamefont {Zwicky}(1933)}]{Zwicky:1933gu}%
  \BibitemOpen
  \bibfield  {author} {\bibinfo {author} {\bibfnamefont {F.}~\bibnamefont
  {Zwicky}},\ }\href {\doibase 10.1007/s10714-008-0707-4} {\bibfield  {journal}
  {\bibinfo  {journal} {Helv. Phys. Acta}\ }\textbf {\bibinfo {volume} {6}},\
  \bibinfo {pages} {110} (\bibinfo {year} {1933})}\BibitemShut {NoStop}%
\bibitem [{\citenamefont {Zwicky}(1937)}]{Zwicky:1937zza}%
  \BibitemOpen
  \bibfield  {author} {\bibinfo {author} {\bibfnamefont {F.}~\bibnamefont
  {Zwicky}},\ }\href {\doibase 10.1086/143864} {\bibfield  {journal} {\bibinfo
  {journal} {Astrophys. J.}\ }\textbf {\bibinfo {volume} {86}},\ \bibinfo
  {pages} {217} (\bibinfo {year} {1937})}\BibitemShut {NoStop}%
\bibitem [{\citenamefont {Riess}\ \emph {et~al.}(1998)\citenamefont {Riess}
  \emph {et~al.}}]{SupernovaSearchTeam:1998fmf}%
  \BibitemOpen
  \bibfield  {author} {\bibinfo {author} {\bibfnamefont {A.~G.}\ \bibnamefont
  {Riess}} \emph {et~al.} (\bibinfo {collaboration} {Supernova Search Team}),\
  }\href {\doibase 10.1086/300499} {\bibfield  {journal} {\bibinfo  {journal}
  {Astron. J.}\ }\textbf {\bibinfo {volume} {116}},\ \bibinfo {pages} {1009}
  (\bibinfo {year} {1998})},\ \Eprint {http://arxiv.org/abs/astro-ph/9805201}
  {arXiv:astro-ph/9805201} \BibitemShut {NoStop}%
\bibitem [{\citenamefont {Perlmutter}\ \emph {et~al.}(1999)\citenamefont
  {Perlmutter} \emph {et~al.}}]{SupernovaCosmologyProject:1998vns}%
  \BibitemOpen
  \bibfield  {author} {\bibinfo {author} {\bibfnamefont {S.}~\bibnamefont
  {Perlmutter}} \emph {et~al.} (\bibinfo {collaboration} {Supernova Cosmology
  Project}),\ }\href {\doibase 10.1086/307221} {\bibfield  {journal} {\bibinfo
  {journal} {Astrophys. J.}\ }\textbf {\bibinfo {volume} {517}},\ \bibinfo
  {pages} {565} (\bibinfo {year} {1999})},\ \Eprint
  {http://arxiv.org/abs/astro-ph/9812133} {arXiv:astro-ph/9812133} \BibitemShut
  {NoStop}%
\bibitem [{\citenamefont {Kolb}\ and\ \citenamefont
  {Turner}(2019)}]{Kolb:1990vq}%
  \BibitemOpen
  \bibfield  {author} {\bibinfo {author} {\bibfnamefont {E.~W.}\ \bibnamefont
  {Kolb}}\ and\ \bibinfo {author} {\bibfnamefont {M.~S.}\ \bibnamefont
  {Turner}},\ }\href {\doibase 10.1201/9780429492860} {\emph {\bibinfo {title}
  {{The Early Universe}}}},\ Vol.~\bibinfo {volume} {69}\ (\bibinfo
  {publisher} {Taylor and Francis},\ \bibinfo {year} {2019})\BibitemShut
  {NoStop}%
\bibitem [{\citenamefont {Bergstrom}(2012)}]{Bergstrom:2012fi}%
  \BibitemOpen
  \bibfield  {author} {\bibinfo {author} {\bibfnamefont {L.}~\bibnamefont
  {Bergstrom}},\ }\href {\doibase 10.1002/andp.201200116} {\bibfield  {journal}
  {\bibinfo  {journal} {Annalen Phys.}\ }\textbf {\bibinfo {volume} {524}},\
  \bibinfo {pages} {479} (\bibinfo {year} {2012})},\ \Eprint
  {http://arxiv.org/abs/1205.4882} {arXiv:1205.4882 [astro-ph.HE]} \BibitemShut
  {NoStop}%
\bibitem [{\citenamefont {Bertone}\ and\ \citenamefont
  {Hooper}(2018)}]{Bertone:2016nfn}%
  \BibitemOpen
  \bibfield  {author} {\bibinfo {author} {\bibfnamefont {G.}~\bibnamefont
  {Bertone}}\ and\ \bibinfo {author} {\bibfnamefont {D.}~\bibnamefont
  {Hooper}},\ }\href {\doibase 10.1103/RevModPhys.90.045002} {\bibfield
  {journal} {\bibinfo  {journal} {Rev. Mod. Phys.}\ }\textbf {\bibinfo {volume}
  {90}},\ \bibinfo {pages} {045002} (\bibinfo {year} {2018})},\ \Eprint
  {http://arxiv.org/abs/1605.04909} {arXiv:1605.04909 [astro-ph.CO]}
  \BibitemShut {NoStop}%
\bibitem [{\citenamefont {Weinberg}(1989)}]{Weinberg:1988cp}%
  \BibitemOpen
  \bibfield  {author} {\bibinfo {author} {\bibfnamefont {S.}~\bibnamefont
  {Weinberg}},\ }\href {\doibase 10.1103/RevModPhys.61.1} {\bibfield  {journal}
  {\bibinfo  {journal} {Rev. Mod. Phys.}\ }\textbf {\bibinfo {volume} {61}},\
  \bibinfo {pages} {1} (\bibinfo {year} {1989})}\BibitemShut {NoStop}%
\bibitem [{\citenamefont {Peebles}\ and\ \citenamefont
  {Ratra}(2003)}]{Peebles:2002gy}%
  \BibitemOpen
  \bibfield  {author} {\bibinfo {author} {\bibfnamefont {P.~J.~E.}\
  \bibnamefont {Peebles}}\ and\ \bibinfo {author} {\bibfnamefont
  {B.}~\bibnamefont {Ratra}},\ }\href {\doibase 10.1103/RevModPhys.75.559}
  {\bibfield  {journal} {\bibinfo  {journal} {Rev. Mod. Phys.}\ }\textbf
  {\bibinfo {volume} {75}},\ \bibinfo {pages} {559} (\bibinfo {year} {2003})},\
  \Eprint {http://arxiv.org/abs/astro-ph/0207347} {arXiv:astro-ph/0207347}
  \BibitemShut {NoStop}%
\bibitem [{\citenamefont {Damour}\ and\ \citenamefont
  {Esposito-Far\`ese}(1992)}]{Damour:1992we}%
  \BibitemOpen
  \bibfield  {author} {\bibinfo {author} {\bibfnamefont {T.}~\bibnamefont
  {Damour}}\ and\ \bibinfo {author} {\bibfnamefont {G.}~\bibnamefont
  {Esposito-Far\`ese}},\ }\href {\doibase 10.1088/0264-9381/9/9/015} {\bibfield
   {journal} {\bibinfo  {journal} {Class. Quant. Grav.}\ }\textbf {\bibinfo
  {volume} {9}},\ \bibinfo {pages} {2093} (\bibinfo {year} {1992})}\BibitemShut
  {NoStop}%
\bibitem [{\citenamefont {Fujii}\ and\ \citenamefont
  {Maeda}(2007)}]{Fujii:2003pa}%
  \BibitemOpen
  \bibfield  {author} {\bibinfo {author} {\bibfnamefont {Y.}~\bibnamefont
  {Fujii}}\ and\ \bibinfo {author} {\bibfnamefont {K.}~\bibnamefont {Maeda}},\
  }\href {\doibase 10.1017/CBO9780511535093} {\emph {\bibinfo {title} {{The
  scalar-tensor theory of gravitation}}}},\ Cambridge Monographs on
  Mathematical Physics\ (\bibinfo  {publisher} {Cambridge University Press},\
  \bibinfo {year} {2007})\BibitemShut {NoStop}%
\bibitem [{\citenamefont {Faraoni}(2004)}]{Faraoni:2004pi}%
  \BibitemOpen
  \bibfield  {author} {\bibinfo {author} {\bibfnamefont {V.}~\bibnamefont
  {Faraoni}},\ }\href {\doibase 10.1007/978-1-4020-1989-0} {\emph {\bibinfo
  {title} {{Cosmology in scalar tensor gravity}}}}\ (\bibinfo {year}
  {2004})\BibitemShut {NoStop}%
\bibitem [{\citenamefont {Kaluza}(1921)}]{Kaluza:1921tu}%
  \BibitemOpen
  \bibfield  {author} {\bibinfo {author} {\bibfnamefont {T.}~\bibnamefont
  {Kaluza}},\ }\href {\doibase 10.1142/S0218271818700017} {\bibfield  {journal}
  {\bibinfo  {journal} {Sitzungsber. Preuss. Akad. Wiss. Berlin (Math. Phys.
  )}\ }\textbf {\bibinfo {volume} {1921}},\ \bibinfo {pages} {966} (\bibinfo
  {year} {1921})},\ \Eprint {http://arxiv.org/abs/1803.08616} {arXiv:1803.08616
  [physics.hist-ph]} \BibitemShut {NoStop}%
\bibitem [{\citenamefont {La}\ and\ \citenamefont
  {Steinhardt}(1989)}]{La:1989za}%
  \BibitemOpen
  \bibfield  {author} {\bibinfo {author} {\bibfnamefont {D.}~\bibnamefont
  {La}}\ and\ \bibinfo {author} {\bibfnamefont {P.~J.}\ \bibnamefont
  {Steinhardt}},\ }\href {\doibase 10.1103/PhysRevLett.62.376} {\bibfield
  {journal} {\bibinfo  {journal} {Phys. Rev. Lett.}\ }\textbf {\bibinfo
  {volume} {62}},\ \bibinfo {pages} {376} (\bibinfo {year} {1989})},\ \bibinfo
  {note} {[Erratum: Phys.Rev.Lett. 62, 1066 (1989)]}\BibitemShut {NoStop}%
\bibitem [{\citenamefont {Steinhardt}\ and\ \citenamefont
  {Accetta}(1990)}]{Steinhardt:1990zx}%
  \BibitemOpen
  \bibfield  {author} {\bibinfo {author} {\bibfnamefont {P.~J.}\ \bibnamefont
  {Steinhardt}}\ and\ \bibinfo {author} {\bibfnamefont {F.~S.}\ \bibnamefont
  {Accetta}},\ }\href {\doibase 10.1103/PhysRevLett.64.2740} {\bibfield
  {journal} {\bibinfo  {journal} {Phys. Rev. Lett.}\ }\textbf {\bibinfo
  {volume} {64}},\ \bibinfo {pages} {2740} (\bibinfo {year}
  {1990})}\BibitemShut {NoStop}%
\bibitem [{\citenamefont {Tsujikawa}(2010)}]{Tsujikawa:2010zza}%
  \BibitemOpen
  \bibfield  {author} {\bibinfo {author} {\bibfnamefont {S.}~\bibnamefont
  {Tsujikawa}},\ }\href {\doibase 10.1007/978-3-642-10598-2_3} {\bibfield
  {journal} {\bibinfo  {journal} {Lect. Notes Phys.}\ }\textbf {\bibinfo
  {volume} {800}},\ \bibinfo {pages} {99} (\bibinfo {year} {2010})},\ \Eprint
  {http://arxiv.org/abs/1101.0191} {arXiv:1101.0191 [gr-qc]} \BibitemShut
  {NoStop}%
\bibitem [{\citenamefont {Ji}\ and\ \citenamefont {Shao}(2024)}]{Ji:2024gdc}%
  \BibitemOpen
  \bibfield  {author} {\bibinfo {author} {\bibfnamefont {P.}~\bibnamefont
  {Ji}}\ and\ \bibinfo {author} {\bibfnamefont {L.}~\bibnamefont {Shao}},\
  }\href {\doibase 10.1088/1572-9494/ad5aeb} {\bibfield  {journal} {\bibinfo
  {journal} {Commun. Theor. Phys.}\ }\textbf {\bibinfo {volume} {76}},\
  \bibinfo {pages} {107401} (\bibinfo {year} {2024})},\ \Eprint
  {http://arxiv.org/abs/2406.04954} {arXiv:2406.04954 [gr-qc]} \BibitemShut
  {NoStop}%
\bibitem [{\citenamefont {Bergmann}(1968)}]{Bergmann:1968ve}%
  \BibitemOpen
  \bibfield  {author} {\bibinfo {author} {\bibfnamefont {P.~G.}\ \bibnamefont
  {Bergmann}},\ }\href {\doibase 10.1007/BF00668828} {\bibfield  {journal}
  {\bibinfo  {journal} {Int. J. Theor. Phys.}\ }\textbf {\bibinfo {volume}
  {1}},\ \bibinfo {pages} {25} (\bibinfo {year} {1968})}\BibitemShut {NoStop}%
\bibitem [{\citenamefont {Nordtvedt}(1970)}]{Nordtvedt:1970uv}%
  \BibitemOpen
  \bibfield  {author} {\bibinfo {author} {\bibfnamefont {K.}~\bibnamefont
  {Nordtvedt}, \bibfnamefont {Jr.}},\ }\href {\doibase 10.1086/150607}
  {\bibfield  {journal} {\bibinfo  {journal} {Astrophys. J.}\ }\textbf
  {\bibinfo {volume} {161}},\ \bibinfo {pages} {1059} (\bibinfo {year}
  {1970})}\BibitemShut {NoStop}%
\bibitem [{\citenamefont {Wagoner}(1970)}]{Wagoner:1970vr}%
  \BibitemOpen
  \bibfield  {author} {\bibinfo {author} {\bibfnamefont {R.~V.}\ \bibnamefont
  {Wagoner}},\ }\href {\doibase 10.1103/PhysRevD.1.3209} {\bibfield  {journal}
  {\bibinfo  {journal} {Phys. Rev. D}\ }\textbf {\bibinfo {volume} {1}},\
  \bibinfo {pages} {3209} (\bibinfo {year} {1970})}\BibitemShut {NoStop}%
\bibitem [{\citenamefont {Sotiriou}\ and\ \citenamefont
  {Faraoni}(2010)}]{Sotiriou:2008rp}%
  \BibitemOpen
  \bibfield  {author} {\bibinfo {author} {\bibfnamefont {T.~P.}\ \bibnamefont
  {Sotiriou}}\ and\ \bibinfo {author} {\bibfnamefont {V.}~\bibnamefont
  {Faraoni}},\ }\href {\doibase 10.1103/RevModPhys.82.451} {\bibfield
  {journal} {\bibinfo  {journal} {Rev. Mod. Phys.}\ }\textbf {\bibinfo {volume}
  {82}},\ \bibinfo {pages} {451} (\bibinfo {year} {2010})},\ \Eprint
  {http://arxiv.org/abs/0805.1726} {arXiv:0805.1726 [gr-qc]} \BibitemShut
  {NoStop}%
\bibitem [{\citenamefont {Astashenok}\ \emph {et~al.}(2021)\citenamefont
  {Astashenok}, \citenamefont {Capozziello}, \citenamefont {Odintsov},\ and\
  \citenamefont {Oikonomou}}]{Astashenok:2021peo}%
  \BibitemOpen
  \bibfield  {author} {\bibinfo {author} {\bibfnamefont {A.~V.}\ \bibnamefont
  {Astashenok}}, \bibinfo {author} {\bibfnamefont {S.}~\bibnamefont
  {Capozziello}}, \bibinfo {author} {\bibfnamefont {S.~D.}\ \bibnamefont
  {Odintsov}}, \ and\ \bibinfo {author} {\bibfnamefont {V.~K.}\ \bibnamefont
  {Oikonomou}},\ }\href {\doibase 10.1016/j.physletb.2021.136222} {\bibfield
  {journal} {\bibinfo  {journal} {Phys. Lett. B}\ }\textbf {\bibinfo {volume}
  {816}},\ \bibinfo {pages} {136222} (\bibinfo {year} {2021})},\ \Eprint
  {http://arxiv.org/abs/2103.04144} {arXiv:2103.04144 [gr-qc]} \BibitemShut
  {NoStop}%
\bibitem [{\citenamefont {Capozziello}\ \emph {et~al.}(2016)\citenamefont
  {Capozziello}, \citenamefont {De~Laurentis}, \citenamefont {Farinelli},\ and\
  \citenamefont {Odintsov}}]{Capozziello:2015yza}%
  \BibitemOpen
  \bibfield  {author} {\bibinfo {author} {\bibfnamefont {S.}~\bibnamefont
  {Capozziello}}, \bibinfo {author} {\bibfnamefont {M.}~\bibnamefont
  {De~Laurentis}}, \bibinfo {author} {\bibfnamefont {R.}~\bibnamefont
  {Farinelli}}, \ and\ \bibinfo {author} {\bibfnamefont {S.~D.}\ \bibnamefont
  {Odintsov}},\ }\href {\doibase 10.1103/PhysRevD.93.023501} {\bibfield
  {journal} {\bibinfo  {journal} {Phys. Rev. D}\ }\textbf {\bibinfo {volume}
  {93}},\ \bibinfo {pages} {023501} (\bibinfo {year} {2016})},\ \Eprint
  {http://arxiv.org/abs/1509.04163} {arXiv:1509.04163 [gr-qc]} \BibitemShut
  {NoStop}%
\bibitem [{\citenamefont {Astashenok}\ \emph {et~al.}(2015)\citenamefont
  {Astashenok}, \citenamefont {Capozziello},\ and\ \citenamefont
  {Odintsov}}]{Astashenok:2014nua}%
  \BibitemOpen
  \bibfield  {author} {\bibinfo {author} {\bibfnamefont {A.~V.}\ \bibnamefont
  {Astashenok}}, \bibinfo {author} {\bibfnamefont {S.}~\bibnamefont
  {Capozziello}}, \ and\ \bibinfo {author} {\bibfnamefont {S.~D.}\ \bibnamefont
  {Odintsov}},\ }\href {\doibase 10.1088/1475-7516/2015/01/001} {\bibfield
  {journal} {\bibinfo  {journal} {JCAP}\ }\textbf {\bibinfo {volume} {01}},\
  \bibinfo {pages} {001} (\bibinfo {year} {2015})},\ \Eprint
  {http://arxiv.org/abs/1408.3856} {arXiv:1408.3856 [gr-qc]} \BibitemShut
  {NoStop}%
\bibitem [{\citenamefont {Damour}\ and\ \citenamefont
  {Nordtvedt}(1993{\natexlab{a}})}]{Damour:1992kf}%
  \BibitemOpen
  \bibfield  {author} {\bibinfo {author} {\bibfnamefont {T.}~\bibnamefont
  {Damour}}\ and\ \bibinfo {author} {\bibfnamefont {K.}~\bibnamefont
  {Nordtvedt}},\ }\href {\doibase 10.1103/PhysRevLett.70.2217} {\bibfield
  {journal} {\bibinfo  {journal} {Phys. Rev. Lett.}\ }\textbf {\bibinfo
  {volume} {70}},\ \bibinfo {pages} {2217} (\bibinfo {year}
  {1993}{\natexlab{a}})}\BibitemShut {NoStop}%
\bibitem [{\citenamefont {Jordan}(1955)}]{jordan1955schwerkraft}%
  \BibitemOpen
  \bibfield  {author} {\bibinfo {author} {\bibfnamefont {P.}~\bibnamefont
  {Jordan}},\ }\href@noop {} {\bibfield  {journal} {\bibinfo  {journal} {Die
  Wissenschaft}\ } (\bibinfo {year} {1955})}\BibitemShut {NoStop}%
\bibitem [{\citenamefont {Brans}\ and\ \citenamefont
  {Dicke}(1961)}]{Brans:1961sx}%
  \BibitemOpen
  \bibfield  {author} {\bibinfo {author} {\bibfnamefont {C.}~\bibnamefont
  {Brans}}\ and\ \bibinfo {author} {\bibfnamefont {R.~H.}\ \bibnamefont
  {Dicke}},\ }\href {\doibase 10.1103/PhysRev.124.925} {\bibfield  {journal}
  {\bibinfo  {journal} {Phys. Rev.}\ }\textbf {\bibinfo {volume} {124}},\
  \bibinfo {pages} {925} (\bibinfo {year} {1961})}\BibitemShut {NoStop}%
\bibitem [{\citenamefont {Bertotti}\ \emph {et~al.}(2003)\citenamefont
  {Bertotti}, \citenamefont {Iess},\ and\ \citenamefont
  {Tortora}}]{Bertotti:2003rm}%
  \BibitemOpen
  \bibfield  {author} {\bibinfo {author} {\bibfnamefont {B.}~\bibnamefont
  {Bertotti}}, \bibinfo {author} {\bibfnamefont {L.}~\bibnamefont {Iess}}, \
  and\ \bibinfo {author} {\bibfnamefont {P.}~\bibnamefont {Tortora}},\ }\href
  {\doibase 10.1038/nature01997} {\bibfield  {journal} {\bibinfo  {journal}
  {Nature}\ }\textbf {\bibinfo {volume} {425}},\ \bibinfo {pages} {374}
  (\bibinfo {year} {2003})}\BibitemShut {NoStop}%
\bibitem [{\citenamefont {Damour}\ and\ \citenamefont
  {Esposito-Far\`ese}(1996{\natexlab{a}})}]{Damour:1995kt}%
  \BibitemOpen
  \bibfield  {author} {\bibinfo {author} {\bibfnamefont {T.}~\bibnamefont
  {Damour}}\ and\ \bibinfo {author} {\bibfnamefont {G.}~\bibnamefont
  {Esposito-Far\`ese}},\ }\href {\doibase 10.1103/PhysRevD.53.5541} {\bibfield
  {journal} {\bibinfo  {journal} {Phys. Rev. D}\ }\textbf {\bibinfo {volume}
  {53}},\ \bibinfo {pages} {5541} (\bibinfo {year} {1996}{\natexlab{a}})},\
  \Eprint {http://arxiv.org/abs/gr-qc/9506063} {arXiv:gr-qc/9506063}
  \BibitemShut {NoStop}%
\bibitem [{\citenamefont {Will}(2014)}]{Will:2014kxa}%
  \BibitemOpen
  \bibfield  {author} {\bibinfo {author} {\bibfnamefont {C.~M.}\ \bibnamefont
  {Will}},\ }\href {\doibase 10.12942/lrr-2014-4} {\bibfield  {journal}
  {\bibinfo  {journal} {Living Rev. Rel.}\ }\textbf {\bibinfo {volume} {17}},\
  \bibinfo {pages} {4} (\bibinfo {year} {2014})},\ \Eprint
  {http://arxiv.org/abs/1403.7377} {arXiv:1403.7377 [gr-qc]} \BibitemShut
  {NoStop}%
\bibitem [{\citenamefont {Quiros}(2019)}]{Quiros:2019ktw}%
  \BibitemOpen
  \bibfield  {author} {\bibinfo {author} {\bibfnamefont {I.}~\bibnamefont
  {Quiros}},\ }\href {\doibase 10.1142/S021827181930012X} {\bibfield  {journal}
  {\bibinfo  {journal} {Int. J. Mod. Phys. D}\ }\textbf {\bibinfo {volume}
  {28}},\ \bibinfo {pages} {1930012} (\bibinfo {year} {2019})},\ \Eprint
  {http://arxiv.org/abs/1901.08690} {arXiv:1901.08690 [gr-qc]} \BibitemShut
  {NoStop}%
\bibitem [{\citenamefont {Vainshtein}(1972)}]{Vainshtein:1972sx}%
  \BibitemOpen
  \bibfield  {author} {\bibinfo {author} {\bibfnamefont {A.~I.}\ \bibnamefont
  {Vainshtein}},\ }\href {\doibase 10.1016/0370-2693(72)90147-5} {\bibfield
  {journal} {\bibinfo  {journal} {Phys. Lett. B}\ }\textbf {\bibinfo {volume}
  {39}},\ \bibinfo {pages} {393} (\bibinfo {year} {1972})}\BibitemShut
  {NoStop}%
\bibitem [{\citenamefont {Khoury}(2013)}]{Khoury:2013yya}%
  \BibitemOpen
  \bibfield  {author} {\bibinfo {author} {\bibfnamefont {J.}~\bibnamefont
  {Khoury}},\ }\href {\doibase 10.1088/0264-9381/30/21/214004} {\bibfield
  {journal} {\bibinfo  {journal} {Class. Quant. Grav.}\ }\textbf {\bibinfo
  {volume} {30}},\ \bibinfo {pages} {214004} (\bibinfo {year} {2013})},\
  \Eprint {http://arxiv.org/abs/1306.4326} {arXiv:1306.4326 [astro-ph.CO]}
  \BibitemShut {NoStop}%
\bibitem [{\citenamefont {Damour}\ and\ \citenamefont
  {Esposito-Far\`ese}(1993)}]{Damour:1993hw}%
  \BibitemOpen
  \bibfield  {author} {\bibinfo {author} {\bibfnamefont {T.}~\bibnamefont
  {Damour}}\ and\ \bibinfo {author} {\bibfnamefont {G.}~\bibnamefont
  {Esposito-Far\`ese}},\ }\href {\doibase 10.1103/PhysRevLett.70.2220}
  {\bibfield  {journal} {\bibinfo  {journal} {Phys. Rev. Lett.}\ }\textbf
  {\bibinfo {volume} {70}},\ \bibinfo {pages} {2220} (\bibinfo {year}
  {1993})}\BibitemShut {NoStop}%
\bibitem [{\citenamefont {Damour}\ and\ \citenamefont
  {Esposito-Far\`ese}(1996{\natexlab{b}})}]{Damour:1996ke}%
  \BibitemOpen
  \bibfield  {author} {\bibinfo {author} {\bibfnamefont {T.}~\bibnamefont
  {Damour}}\ and\ \bibinfo {author} {\bibfnamefont {G.}~\bibnamefont
  {Esposito-Far\`ese}},\ }\href {\doibase 10.1103/PhysRevD.54.1474} {\bibfield
  {journal} {\bibinfo  {journal} {Phys. Rev. D}\ }\textbf {\bibinfo {volume}
  {54}},\ \bibinfo {pages} {1474} (\bibinfo {year} {1996}{\natexlab{b}})},\
  \Eprint {http://arxiv.org/abs/gr-qc/9602056} {arXiv:gr-qc/9602056}
  \BibitemShut {NoStop}%
\bibitem [{\citenamefont {Doneva}\ \emph {et~al.}(2024)\citenamefont {Doneva},
  \citenamefont {Ramazano{\u{g}}lu}, \citenamefont {Silva}, \citenamefont
  {Sotiriou},\ and\ \citenamefont {Yazadjiev}}]{Doneva:2022ewd}%
  \BibitemOpen
  \bibfield  {author} {\bibinfo {author} {\bibfnamefont {D.~D.}\ \bibnamefont
  {Doneva}}, \bibinfo {author} {\bibfnamefont {F.~M.}\ \bibnamefont
  {Ramazano{\u{g}}lu}}, \bibinfo {author} {\bibfnamefont {H.~O.}\ \bibnamefont
  {Silva}}, \bibinfo {author} {\bibfnamefont {T.~P.}\ \bibnamefont {Sotiriou}},
  \ and\ \bibinfo {author} {\bibfnamefont {S.~S.}\ \bibnamefont {Yazadjiev}},\
  }\href {\doibase 10.1103/RevModPhys.96.015004} {\bibfield  {journal}
  {\bibinfo  {journal} {Rev. Mod. Phys.}\ }\textbf {\bibinfo {volume} {96}},\
  \bibinfo {pages} {015004} (\bibinfo {year} {2024})},\ \Eprint
  {http://arxiv.org/abs/2211.01766} {arXiv:2211.01766 [gr-qc]} \BibitemShut
  {NoStop}%
\bibitem [{\citenamefont {Schwarzschild}(1916)}]{Schwarzschild:1916uq}%
  \BibitemOpen
  \bibfield  {author} {\bibinfo {author} {\bibfnamefont {K.}~\bibnamefont
  {Schwarzschild}},\ }\href@noop {} {\bibfield  {journal} {\bibinfo  {journal}
  {Sitzungsber. Preuss. Akad. Wiss. Berlin (Math. Phys. )}\ }\textbf {\bibinfo
  {volume} {1916}},\ \bibinfo {pages} {189} (\bibinfo {year} {1916})},\ \Eprint
  {http://arxiv.org/abs/physics/9905030} {arXiv:physics/9905030} \BibitemShut
  {NoStop}%
\bibitem [{\citenamefont {Sennett}\ \emph {et~al.}(2017)\citenamefont
  {Sennett}, \citenamefont {Shao},\ and\ \citenamefont
  {Steinhoff}}]{Sennett:2017lcx}%
  \BibitemOpen
  \bibfield  {author} {\bibinfo {author} {\bibfnamefont {N.}~\bibnamefont
  {Sennett}}, \bibinfo {author} {\bibfnamefont {L.}~\bibnamefont {Shao}}, \
  and\ \bibinfo {author} {\bibfnamefont {J.}~\bibnamefont {Steinhoff}},\ }\href
  {\doibase 10.1103/PhysRevD.96.084019} {\bibfield  {journal} {\bibinfo
  {journal} {Phys. Rev. D}\ }\textbf {\bibinfo {volume} {96}},\ \bibinfo
  {pages} {084019} (\bibinfo {year} {2017})},\ \Eprint
  {http://arxiv.org/abs/1708.08285} {arXiv:1708.08285 [gr-qc]} \BibitemShut
  {NoStop}%
\bibitem [{\citenamefont {Harada}(1997)}]{Harada:1997mr}%
  \BibitemOpen
  \bibfield  {author} {\bibinfo {author} {\bibfnamefont {T.}~\bibnamefont
  {Harada}},\ }\href {\doibase 10.1143/PTP.98.359} {\bibfield  {journal}
  {\bibinfo  {journal} {Prog. Theor. Phys.}\ }\textbf {\bibinfo {volume}
  {98}},\ \bibinfo {pages} {359} (\bibinfo {year} {1997})},\ \Eprint
  {http://arxiv.org/abs/gr-qc/9706014} {arXiv:gr-qc/9706014} \BibitemShut
  {NoStop}%
\bibitem [{\citenamefont {Mendes}(2015)}]{Mendes:2014ufa}%
  \BibitemOpen
  \bibfield  {author} {\bibinfo {author} {\bibfnamefont {R.~F.~P.}\
  \bibnamefont {Mendes}},\ }\href {\doibase 10.1103/PhysRevD.91.064024}
  {\bibfield  {journal} {\bibinfo  {journal} {Phys. Rev. D}\ }\textbf {\bibinfo
  {volume} {91}},\ \bibinfo {pages} {064024} (\bibinfo {year} {2015})},\
  \Eprint {http://arxiv.org/abs/1412.6789} {arXiv:1412.6789 [gr-qc]}
  \BibitemShut {NoStop}%
\bibitem [{\citenamefont {Mendes}\ and\ \citenamefont
  {Ortiz}(2016)}]{Mendes:2016fby}%
  \BibitemOpen
  \bibfield  {author} {\bibinfo {author} {\bibfnamefont {R.~F.~P.}\
  \bibnamefont {Mendes}}\ and\ \bibinfo {author} {\bibfnamefont
  {N.}~\bibnamefont {Ortiz}},\ }\href {\doibase 10.1103/PhysRevD.93.124035}
  {\bibfield  {journal} {\bibinfo  {journal} {Phys. Rev. D}\ }\textbf {\bibinfo
  {volume} {93}},\ \bibinfo {pages} {124035} (\bibinfo {year} {2016})},\
  \Eprint {http://arxiv.org/abs/1604.04175} {arXiv:1604.04175 [gr-qc]}
  \BibitemShut {NoStop}%
\bibitem [{\citenamefont {Salopek}\ \emph {et~al.}(1989)\citenamefont
  {Salopek}, \citenamefont {Bond},\ and\ \citenamefont
  {Bardeen}}]{Salopek:1988qh}%
  \BibitemOpen
  \bibfield  {author} {\bibinfo {author} {\bibfnamefont {D.~S.}\ \bibnamefont
  {Salopek}}, \bibinfo {author} {\bibfnamefont {J.~R.}\ \bibnamefont {Bond}}, \
  and\ \bibinfo {author} {\bibfnamefont {J.~M.}\ \bibnamefont {Bardeen}},\
  }\href {\doibase 10.1103/PhysRevD.40.1753} {\bibfield  {journal} {\bibinfo
  {journal} {Phys. Rev. D}\ }\textbf {\bibinfo {volume} {40}},\ \bibinfo
  {pages} {1753} (\bibinfo {year} {1989})}\BibitemShut {NoStop}%
\bibitem [{\citenamefont {Shao}\ \emph {et~al.}(2017)\citenamefont {Shao},
  \citenamefont {Sennett}, \citenamefont {Buonanno}, \citenamefont {Kramer},\
  and\ \citenamefont {Wex}}]{Shao:2017gwu}%
  \BibitemOpen
  \bibfield  {author} {\bibinfo {author} {\bibfnamefont {L.}~\bibnamefont
  {Shao}}, \bibinfo {author} {\bibfnamefont {N.}~\bibnamefont {Sennett}},
  \bibinfo {author} {\bibfnamefont {A.}~\bibnamefont {Buonanno}}, \bibinfo
  {author} {\bibfnamefont {M.}~\bibnamefont {Kramer}}, \ and\ \bibinfo {author}
  {\bibfnamefont {N.}~\bibnamefont {Wex}},\ }\href {\doibase
  10.1103/PhysRevX.7.041025} {\bibfield  {journal} {\bibinfo  {journal} {Phys.
  Rev. X}\ }\textbf {\bibinfo {volume} {7}},\ \bibinfo {pages} {041025}
  (\bibinfo {year} {2017})},\ \Eprint {http://arxiv.org/abs/1704.07561}
  {arXiv:1704.07561 [gr-qc]} \BibitemShut {NoStop}%
\bibitem [{\citenamefont {Anderson}\ \emph {et~al.}(2019)\citenamefont
  {Anderson}, \citenamefont {Freire},\ and\ \citenamefont
  {Yunes}}]{Anderson:2019eay}%
  \BibitemOpen
  \bibfield  {author} {\bibinfo {author} {\bibfnamefont {D.}~\bibnamefont
  {Anderson}}, \bibinfo {author} {\bibfnamefont {P.}~\bibnamefont {Freire}}, \
  and\ \bibinfo {author} {\bibfnamefont {N.}~\bibnamefont {Yunes}},\ }\href
  {\doibase 10.1088/1361-6382/ab3a1c} {\bibfield  {journal} {\bibinfo
  {journal} {Class. Quant. Grav.}\ }\textbf {\bibinfo {volume} {36}},\ \bibinfo
  {pages} {225009} (\bibinfo {year} {2019})},\ \Eprint
  {http://arxiv.org/abs/1901.00938} {arXiv:1901.00938 [gr-qc]} \BibitemShut
  {NoStop}%
\bibitem [{\citenamefont {Kramer}\ \emph {et~al.}(2021)\citenamefont {Kramer}
  \emph {et~al.}}]{Kramer:2021jcw}%
  \BibitemOpen
  \bibfield  {author} {\bibinfo {author} {\bibfnamefont {M.}~\bibnamefont
  {Kramer}} \emph {et~al.},\ }\href {\doibase 10.1103/PhysRevX.11.041050}
  {\bibfield  {journal} {\bibinfo  {journal} {Phys. Rev. X}\ }\textbf {\bibinfo
  {volume} {11}},\ \bibinfo {pages} {041050} (\bibinfo {year} {2021})},\
  \Eprint {http://arxiv.org/abs/2112.06795} {arXiv:2112.06795 [astro-ph.HE]}
  \BibitemShut {NoStop}%
\bibitem [{\citenamefont {Zhao}\ \emph {et~al.}(2022)\citenamefont {Zhao},
  \citenamefont {Freire}, \citenamefont {Kramer}, \citenamefont {Shao},\ and\
  \citenamefont {Wex}}]{Zhao:2022vig}%
  \BibitemOpen
  \bibfield  {author} {\bibinfo {author} {\bibfnamefont {J.}~\bibnamefont
  {Zhao}}, \bibinfo {author} {\bibfnamefont {P.~C.~C.}\ \bibnamefont {Freire}},
  \bibinfo {author} {\bibfnamefont {M.}~\bibnamefont {Kramer}}, \bibinfo
  {author} {\bibfnamefont {L.}~\bibnamefont {Shao}}, \ and\ \bibinfo {author}
  {\bibfnamefont {N.}~\bibnamefont {Wex}},\ }\href {\doibase
  10.1088/1361-6382/ac69a3} {\bibfield  {journal} {\bibinfo  {journal} {Class.
  Quant. Grav.}\ }\textbf {\bibinfo {volume} {39}},\ \bibinfo {pages} {11LT01}
  (\bibinfo {year} {2022})},\ \Eprint {http://arxiv.org/abs/2201.03771}
  {arXiv:2201.03771 [astro-ph.HE]} \BibitemShut {NoStop}%
\bibitem [{\citenamefont {Shao}(2023)}]{Shao:2022izp}%
  \BibitemOpen
  \bibfield  {author} {\bibinfo {author} {\bibfnamefont {L.}~\bibnamefont
  {Shao}},\ }\href {\doibase 10.1007/978-3-031-31520-6_12} {\bibfield
  {journal} {\bibinfo  {journal} {Lect. Notes Phys.}\ }\textbf {\bibinfo
  {volume} {1017}},\ \bibinfo {pages} {385} (\bibinfo {year} {2023})},\ \Eprint
  {http://arxiv.org/abs/2206.15187} {arXiv:2206.15187 [gr-qc]} \BibitemShut
  {NoStop}%
\bibitem [{\citenamefont {Barausse}\ \emph {et~al.}(2013)\citenamefont
  {Barausse}, \citenamefont {Palenzuela}, \citenamefont {Ponce},\ and\
  \citenamefont {Lehner}}]{Barausse:2012da}%
  \BibitemOpen
  \bibfield  {author} {\bibinfo {author} {\bibfnamefont {E.}~\bibnamefont
  {Barausse}}, \bibinfo {author} {\bibfnamefont {C.}~\bibnamefont
  {Palenzuela}}, \bibinfo {author} {\bibfnamefont {M.}~\bibnamefont {Ponce}}, \
  and\ \bibinfo {author} {\bibfnamefont {L.}~\bibnamefont {Lehner}},\ }\href
  {\doibase 10.1103/PhysRevD.87.081506} {\bibfield  {journal} {\bibinfo
  {journal} {Phys. Rev. D}\ }\textbf {\bibinfo {volume} {87}},\ \bibinfo
  {pages} {081506} (\bibinfo {year} {2013})},\ \Eprint
  {http://arxiv.org/abs/1212.5053} {arXiv:1212.5053 [gr-qc]} \BibitemShut
  {NoStop}%
\bibitem [{\citenamefont {Freire}\ \emph {et~al.}(2012)\citenamefont {Freire},
  \citenamefont {Wex}, \citenamefont {Esposito-Far\`ese}, \citenamefont
  {Verbiest}, \citenamefont {Bailes}, \citenamefont {Jacoby}, \citenamefont
  {Kramer}, \citenamefont {Stairs}, \citenamefont {Antoniadis},\ and\
  \citenamefont {Janssen}}]{Freire:2012mg}%
  \BibitemOpen
  \bibfield  {author} {\bibinfo {author} {\bibfnamefont {P.~C.~C.}\
  \bibnamefont {Freire}}, \bibinfo {author} {\bibfnamefont {N.}~\bibnamefont
  {Wex}}, \bibinfo {author} {\bibfnamefont {G.}~\bibnamefont
  {Esposito-Far\`ese}}, \bibinfo {author} {\bibfnamefont {J.~P.~W.}\
  \bibnamefont {Verbiest}}, \bibinfo {author} {\bibfnamefont {M.}~\bibnamefont
  {Bailes}}, \bibinfo {author} {\bibfnamefont {B.~A.}\ \bibnamefont {Jacoby}},
  \bibinfo {author} {\bibfnamefont {M.}~\bibnamefont {Kramer}}, \bibinfo
  {author} {\bibfnamefont {I.~H.}\ \bibnamefont {Stairs}}, \bibinfo {author}
  {\bibfnamefont {J.}~\bibnamefont {Antoniadis}}, \ and\ \bibinfo {author}
  {\bibfnamefont {G.~H.}\ \bibnamefont {Janssen}},\ }\href {\doibase
  10.1111/j.1365-2966.2012.21253.x} {\bibfield  {journal} {\bibinfo  {journal}
  {Mon. Not. Roy. Astron. Soc.}\ }\textbf {\bibinfo {volume} {423}},\ \bibinfo
  {pages} {3328} (\bibinfo {year} {2012})},\ \Eprint
  {http://arxiv.org/abs/1205.1450} {arXiv:1205.1450 [astro-ph.GA]} \BibitemShut
  {NoStop}%
\bibitem [{\citenamefont {Abbott}\ \emph
  {et~al.}(2019{\natexlab{a}})\citenamefont {Abbott} \emph
  {et~al.}}]{LIGOScientific:2018dkp}%
  \BibitemOpen
  \bibfield  {author} {\bibinfo {author} {\bibfnamefont {B.~P.}\ \bibnamefont
  {Abbott}} \emph {et~al.} (\bibinfo {collaboration} {LIGO Scientific,
  Virgo}),\ }\href {\doibase 10.1103/PhysRevLett.123.011102} {\bibfield
  {journal} {\bibinfo  {journal} {Phys. Rev. Lett.}\ }\textbf {\bibinfo
  {volume} {123}},\ \bibinfo {pages} {011102} (\bibinfo {year}
  {2019}{\natexlab{a}})},\ \Eprint {http://arxiv.org/abs/1811.00364}
  {arXiv:1811.00364 [gr-qc]} \BibitemShut {NoStop}%
\bibitem [{\citenamefont {Zhang}\ \emph {et~al.}(2017)\citenamefont {Zhang},
  \citenamefont {Yu}, \citenamefont {Liu}, \citenamefont {Zhao},\ and\
  \citenamefont {Wang}}]{Zhang:2017sym}%
  \BibitemOpen
  \bibfield  {author} {\bibinfo {author} {\bibfnamefont {X.}~\bibnamefont
  {Zhang}}, \bibinfo {author} {\bibfnamefont {J.}~\bibnamefont {Yu}}, \bibinfo
  {author} {\bibfnamefont {T.}~\bibnamefont {Liu}}, \bibinfo {author}
  {\bibfnamefont {W.}~\bibnamefont {Zhao}}, \ and\ \bibinfo {author}
  {\bibfnamefont {A.}~\bibnamefont {Wang}},\ }\href {\doibase
  10.1103/PhysRevD.95.124008} {\bibfield  {journal} {\bibinfo  {journal} {Phys.
  Rev. D}\ }\textbf {\bibinfo {volume} {95}},\ \bibinfo {pages} {124008}
  (\bibinfo {year} {2017})},\ \Eprint {http://arxiv.org/abs/1703.09853}
  {arXiv:1703.09853 [gr-qc]} \BibitemShut {NoStop}%
\bibitem [{\citenamefont {Zhao}\ \emph {et~al.}(2021)\citenamefont {Zhao},
  \citenamefont {Shao}, \citenamefont {Gao}, \citenamefont {Liu}, \citenamefont
  {Cao},\ and\ \citenamefont {Ma}}]{Zhao:2021bjw}%
  \BibitemOpen
  \bibfield  {author} {\bibinfo {author} {\bibfnamefont {J.}~\bibnamefont
  {Zhao}}, \bibinfo {author} {\bibfnamefont {L.}~\bibnamefont {Shao}}, \bibinfo
  {author} {\bibfnamefont {Y.}~\bibnamefont {Gao}}, \bibinfo {author}
  {\bibfnamefont {C.}~\bibnamefont {Liu}}, \bibinfo {author} {\bibfnamefont
  {Z.}~\bibnamefont {Cao}}, \ and\ \bibinfo {author} {\bibfnamefont {B.-Q.}\
  \bibnamefont {Ma}},\ }\href {\doibase 10.1103/PhysRevD.104.084008} {\bibfield
   {journal} {\bibinfo  {journal} {Phys. Rev. D}\ }\textbf {\bibinfo {volume}
  {104}},\ \bibinfo {pages} {084008} (\bibinfo {year} {2021})},\ \Eprint
  {http://arxiv.org/abs/2106.04883} {arXiv:2106.04883 [gr-qc]} \BibitemShut
  {NoStop}%
\bibitem [{\citenamefont {Zhao}\ \emph {et~al.}(2019)\citenamefont {Zhao},
  \citenamefont {Shao}, \citenamefont {Cao},\ and\ \citenamefont
  {Ma}}]{Zhao:2019suc}%
  \BibitemOpen
  \bibfield  {author} {\bibinfo {author} {\bibfnamefont {J.}~\bibnamefont
  {Zhao}}, \bibinfo {author} {\bibfnamefont {L.}~\bibnamefont {Shao}}, \bibinfo
  {author} {\bibfnamefont {Z.}~\bibnamefont {Cao}}, \ and\ \bibinfo {author}
  {\bibfnamefont {B.-Q.}\ \bibnamefont {Ma}},\ }\href {\doibase
  10.1103/PhysRevD.100.064034} {\bibfield  {journal} {\bibinfo  {journal}
  {Phys. Rev. D}\ }\textbf {\bibinfo {volume} {100}},\ \bibinfo {pages}
  {064034} (\bibinfo {year} {2019})},\ \Eprint
  {http://arxiv.org/abs/1907.00780} {arXiv:1907.00780 [gr-qc]} \BibitemShut
  {NoStop}%
\bibitem [{\citenamefont {Guo}\ \emph {et~al.}(2021)\citenamefont {Guo},
  \citenamefont {Zhao},\ and\ \citenamefont {Shao}}]{Guo:2021leu}%
  \BibitemOpen
  \bibfield  {author} {\bibinfo {author} {\bibfnamefont {M.}~\bibnamefont
  {Guo}}, \bibinfo {author} {\bibfnamefont {J.}~\bibnamefont {Zhao}}, \ and\
  \bibinfo {author} {\bibfnamefont {L.}~\bibnamefont {Shao}},\ }\href {\doibase
  10.1103/PhysRevD.104.104065} {\bibfield  {journal} {\bibinfo  {journal}
  {Phys. Rev. D}\ }\textbf {\bibinfo {volume} {104}},\ \bibinfo {pages}
  {104065} (\bibinfo {year} {2021})},\ \Eprint
  {http://arxiv.org/abs/2106.01622} {arXiv:2106.01622 [gr-qc]} \BibitemShut
  {NoStop}%
\bibitem [{\citenamefont {Anderson}\ \emph {et~al.}(2016)\citenamefont
  {Anderson}, \citenamefont {Yunes},\ and\ \citenamefont
  {Barausse}}]{Anderson:2016aoi}%
  \BibitemOpen
  \bibfield  {author} {\bibinfo {author} {\bibfnamefont {D.}~\bibnamefont
  {Anderson}}, \bibinfo {author} {\bibfnamefont {N.}~\bibnamefont {Yunes}}, \
  and\ \bibinfo {author} {\bibfnamefont {E.}~\bibnamefont {Barausse}},\ }\href
  {\doibase 10.1103/PhysRevD.94.104064} {\bibfield  {journal} {\bibinfo
  {journal} {Phys. Rev. D}\ }\textbf {\bibinfo {volume} {94}},\ \bibinfo
  {pages} {104064} (\bibinfo {year} {2016})},\ \Eprint
  {http://arxiv.org/abs/1607.08888} {arXiv:1607.08888 [gr-qc]} \BibitemShut
  {NoStop}%
\bibitem [{\citenamefont {Anderson}\ and\ \citenamefont
  {Yunes}(2017)}]{Anderson:2017phb}%
  \BibitemOpen
  \bibfield  {author} {\bibinfo {author} {\bibfnamefont {D.}~\bibnamefont
  {Anderson}}\ and\ \bibinfo {author} {\bibfnamefont {N.}~\bibnamefont
  {Yunes}},\ }\href {\doibase 10.1103/PhysRevD.96.064037} {\bibfield  {journal}
  {\bibinfo  {journal} {Phys. Rev. D}\ }\textbf {\bibinfo {volume} {96}},\
  \bibinfo {pages} {064037} (\bibinfo {year} {2017})},\ \Eprint
  {http://arxiv.org/abs/1705.06351} {arXiv:1705.06351 [gr-qc]} \BibitemShut
  {NoStop}%
\bibitem [{\citenamefont {Damour}\ and\ \citenamefont
  {Nordtvedt}(1993{\natexlab{b}})}]{Damour:1993id}%
  \BibitemOpen
  \bibfield  {author} {\bibinfo {author} {\bibfnamefont {T.}~\bibnamefont
  {Damour}}\ and\ \bibinfo {author} {\bibfnamefont {K.}~\bibnamefont
  {Nordtvedt}},\ }\href {\doibase 10.1103/PhysRevD.48.3436} {\bibfield
  {journal} {\bibinfo  {journal} {Phys. Rev. D}\ }\textbf {\bibinfo {volume}
  {48}},\ \bibinfo {pages} {3436} (\bibinfo {year}
  {1993}{\natexlab{b}})}\BibitemShut {NoStop}%
\bibitem [{\citenamefont {Antoniou}\ \emph {et~al.}(2021)\citenamefont
  {Antoniou}, \citenamefont {Bordin},\ and\ \citenamefont
  {Sotiriou}}]{Antoniou:2020nax}%
  \BibitemOpen
  \bibfield  {author} {\bibinfo {author} {\bibfnamefont {G.}~\bibnamefont
  {Antoniou}}, \bibinfo {author} {\bibfnamefont {L.}~\bibnamefont {Bordin}}, \
  and\ \bibinfo {author} {\bibfnamefont {T.~P.}\ \bibnamefont {Sotiriou}},\
  }\href {\doibase 10.1103/PhysRevD.103.024012} {\bibfield  {journal} {\bibinfo
   {journal} {Phys. Rev. D}\ }\textbf {\bibinfo {volume} {103}},\ \bibinfo
  {pages} {024012} (\bibinfo {year} {2021})},\ \Eprint
  {http://arxiv.org/abs/2004.14985} {arXiv:2004.14985 [gr-qc]} \BibitemShut
  {NoStop}%
\bibitem [{\citenamefont {Erices}\ \emph {et~al.}(2022)\citenamefont {Erices},
  \citenamefont {Riquelme},\ and\ \citenamefont {Zalaquett}}]{Erices:2022bws}%
  \BibitemOpen
  \bibfield  {author} {\bibinfo {author} {\bibfnamefont {C.}~\bibnamefont
  {Erices}}, \bibinfo {author} {\bibfnamefont {S.}~\bibnamefont {Riquelme}}, \
  and\ \bibinfo {author} {\bibfnamefont {N.}~\bibnamefont {Zalaquett}},\ }\href
  {\doibase 10.1103/PhysRevD.106.044046} {\bibfield  {journal} {\bibinfo
  {journal} {Phys. Rev. D}\ }\textbf {\bibinfo {volume} {106}},\ \bibinfo
  {pages} {044046} (\bibinfo {year} {2022})},\ \Eprint
  {http://arxiv.org/abs/2203.06030} {arXiv:2203.06030 [gr-qc]} \BibitemShut
  {NoStop}%
\bibitem [{\citenamefont {Ramazano{\u{g}}lu}\ and\ \citenamefont
  {Pretorius}(2016)}]{Ramazanoglu:2016kul}%
  \BibitemOpen
  \bibfield  {author} {\bibinfo {author} {\bibfnamefont {F.~M.}\ \bibnamefont
  {Ramazano{\u{g}}lu}}\ and\ \bibinfo {author} {\bibfnamefont {F.}~\bibnamefont
  {Pretorius}},\ }\href {\doibase 10.1103/PhysRevD.93.064005} {\bibfield
  {journal} {\bibinfo  {journal} {Phys. Rev. D}\ }\textbf {\bibinfo {volume}
  {93}},\ \bibinfo {pages} {064005} (\bibinfo {year} {2016})},\ \Eprint
  {http://arxiv.org/abs/1601.07475} {arXiv:1601.07475 [gr-qc]} \BibitemShut
  {NoStop}%
\bibitem [{\citenamefont {Yazadjiev}\ \emph {et~al.}(2016)\citenamefont
  {Yazadjiev}, \citenamefont {Doneva},\ and\ \citenamefont
  {Popchev}}]{Yazadjiev:2016pcb}%
  \BibitemOpen
  \bibfield  {author} {\bibinfo {author} {\bibfnamefont {S.~S.}\ \bibnamefont
  {Yazadjiev}}, \bibinfo {author} {\bibfnamefont {D.~D.}\ \bibnamefont
  {Doneva}}, \ and\ \bibinfo {author} {\bibfnamefont {D.}~\bibnamefont
  {Popchev}},\ }\href {\doibase 10.1103/PhysRevD.93.084038} {\bibfield
  {journal} {\bibinfo  {journal} {Phys. Rev. D}\ }\textbf {\bibinfo {volume}
  {93}},\ \bibinfo {pages} {084038} (\bibinfo {year} {2016})},\ \Eprint
  {http://arxiv.org/abs/1602.04766} {arXiv:1602.04766 [gr-qc]} \BibitemShut
  {NoStop}%
\bibitem [{\citenamefont {Xu}\ \emph {et~al.}(2020)\citenamefont {Xu},
  \citenamefont {Gao},\ and\ \citenamefont {Shao}}]{Xu:2020vbs}%
  \BibitemOpen
  \bibfield  {author} {\bibinfo {author} {\bibfnamefont {R.}~\bibnamefont
  {Xu}}, \bibinfo {author} {\bibfnamefont {Y.}~\bibnamefont {Gao}}, \ and\
  \bibinfo {author} {\bibfnamefont {L.}~\bibnamefont {Shao}},\ }\href {\doibase
  10.1103/PhysRevD.102.064057} {\bibfield  {journal} {\bibinfo  {journal}
  {Phys. Rev. D}\ }\textbf {\bibinfo {volume} {102}},\ \bibinfo {pages}
  {064057} (\bibinfo {year} {2020})},\ \Eprint
  {http://arxiv.org/abs/2007.10080} {arXiv:2007.10080 [gr-qc]} \BibitemShut
  {NoStop}%
\bibitem [{\citenamefont {Hu}\ \emph {et~al.}(2021)\citenamefont {Hu},
  \citenamefont {Gao}, \citenamefont {Xu},\ and\ \citenamefont
  {Shao}}]{Hu:2021tyw}%
  \BibitemOpen
  \bibfield  {author} {\bibinfo {author} {\bibfnamefont {Z.}~\bibnamefont
  {Hu}}, \bibinfo {author} {\bibfnamefont {Y.}~\bibnamefont {Gao}}, \bibinfo
  {author} {\bibfnamefont {R.}~\bibnamefont {Xu}}, \ and\ \bibinfo {author}
  {\bibfnamefont {L.}~\bibnamefont {Shao}},\ }\href {\doibase
  10.1103/PhysRevD.104.104014} {\bibfield  {journal} {\bibinfo  {journal}
  {Phys. Rev. D}\ }\textbf {\bibinfo {volume} {104}},\ \bibinfo {pages}
  {104014} (\bibinfo {year} {2021})},\ \bibinfo {note} {[Erratum: Phys.Rev.D
  111, 109903 (2025)]},\ \Eprint {http://arxiv.org/abs/2109.13453}
  {arXiv:2109.13453 [gr-qc]} \BibitemShut {NoStop}%
\bibitem [{\citenamefont {Faraoni}(2009)}]{Faraoni:2009km}%
  \BibitemOpen
  \bibfield  {author} {\bibinfo {author} {\bibfnamefont {V.}~\bibnamefont
  {Faraoni}},\ }\href {\doibase 10.1088/0264-9381/26/14/145014} {\bibfield
  {journal} {\bibinfo  {journal} {Class. Quant. Grav.}\ }\textbf {\bibinfo
  {volume} {26}},\ \bibinfo {pages} {145014} (\bibinfo {year} {2009})},\
  \Eprint {http://arxiv.org/abs/0906.1901} {arXiv:0906.1901 [gr-qc]}
  \BibitemShut {NoStop}%
\bibitem [{\citenamefont {Alsing}\ \emph {et~al.}(2012)\citenamefont {Alsing},
  \citenamefont {Berti}, \citenamefont {Will},\ and\ \citenamefont
  {Zaglauer}}]{Alsing:2011er}%
  \BibitemOpen
  \bibfield  {author} {\bibinfo {author} {\bibfnamefont {J.}~\bibnamefont
  {Alsing}}, \bibinfo {author} {\bibfnamefont {E.}~\bibnamefont {Berti}},
  \bibinfo {author} {\bibfnamefont {C.~M.}\ \bibnamefont {Will}}, \ and\
  \bibinfo {author} {\bibfnamefont {H.}~\bibnamefont {Zaglauer}},\ }\href
  {\doibase 10.1103/PhysRevD.85.064041} {\bibfield  {journal} {\bibinfo
  {journal} {Phys. Rev. D}\ }\textbf {\bibinfo {volume} {85}},\ \bibinfo
  {pages} {064041} (\bibinfo {year} {2012})},\ \Eprint
  {http://arxiv.org/abs/1112.4903} {arXiv:1112.4903 [gr-qc]} \BibitemShut
  {NoStop}%
\bibitem [{\citenamefont {Liu}\ \emph {et~al.}(2020)\citenamefont {Liu},
  \citenamefont {Zhao},\ and\ \citenamefont {Wang}}]{Liu:2020moh}%
  \BibitemOpen
  \bibfield  {author} {\bibinfo {author} {\bibfnamefont {T.}~\bibnamefont
  {Liu}}, \bibinfo {author} {\bibfnamefont {W.}~\bibnamefont {Zhao}}, \ and\
  \bibinfo {author} {\bibfnamefont {Y.}~\bibnamefont {Wang}},\ }\href {\doibase
  10.1103/PhysRevD.102.124035} {\bibfield  {journal} {\bibinfo  {journal}
  {Phys. Rev. D}\ }\textbf {\bibinfo {volume} {102}},\ \bibinfo {pages}
  {124035} (\bibinfo {year} {2020})},\ \Eprint
  {http://arxiv.org/abs/2007.10068} {arXiv:2007.10068 [gr-qc]} \BibitemShut
  {NoStop}%
\bibitem [{\citenamefont {Chen}\ \emph {et~al.}(2015)\citenamefont {Chen},
  \citenamefont {Suyama},\ and\ \citenamefont {Yokoyama}}]{Chen:2015zmx}%
  \BibitemOpen
  \bibfield  {author} {\bibinfo {author} {\bibfnamefont {P.}~\bibnamefont
  {Chen}}, \bibinfo {author} {\bibfnamefont {T.}~\bibnamefont {Suyama}}, \ and\
  \bibinfo {author} {\bibfnamefont {J.}~\bibnamefont {Yokoyama}},\ }\href
  {\doibase 10.1103/PhysRevD.92.124016} {\bibfield  {journal} {\bibinfo
  {journal} {Phys. Rev. D}\ }\textbf {\bibinfo {volume} {92}},\ \bibinfo
  {pages} {124016} (\bibinfo {year} {2015})},\ \Eprint
  {http://arxiv.org/abs/1508.01384} {arXiv:1508.01384 [gr-qc]} \BibitemShut
  {NoStop}%
\bibitem [{\citenamefont {Morisaki}\ and\ \citenamefont
  {Suyama}(2017)}]{Morisaki:2017nit}%
  \BibitemOpen
  \bibfield  {author} {\bibinfo {author} {\bibfnamefont {S.}~\bibnamefont
  {Morisaki}}\ and\ \bibinfo {author} {\bibfnamefont {T.}~\bibnamefont
  {Suyama}},\ }\href {\doibase 10.1103/PhysRevD.96.084026} {\bibfield
  {journal} {\bibinfo  {journal} {Phys. Rev. D}\ }\textbf {\bibinfo {volume}
  {96}},\ \bibinfo {pages} {084026} (\bibinfo {year} {2017})},\ \Eprint
  {http://arxiv.org/abs/1707.02809} {arXiv:1707.02809 [gr-qc]} \BibitemShut
  {NoStop}%
\bibitem [{\citenamefont {Doneva}\ and\ \citenamefont
  {Yazadjiev}(2016)}]{Doneva:2016xmf}%
  \BibitemOpen
  \bibfield  {author} {\bibinfo {author} {\bibfnamefont {D.~D.}\ \bibnamefont
  {Doneva}}\ and\ \bibinfo {author} {\bibfnamefont {S.~S.}\ \bibnamefont
  {Yazadjiev}},\ }\href {\doibase 10.1088/1475-7516/2016/11/019} {\bibfield
  {journal} {\bibinfo  {journal} {JCAP}\ }\textbf {\bibinfo {volume} {11}},\
  \bibinfo {pages} {019} (\bibinfo {year} {2016})},\ \Eprint
  {http://arxiv.org/abs/1607.03299} {arXiv:1607.03299 [gr-qc]} \BibitemShut
  {NoStop}%
\bibitem [{\citenamefont {Maggiore}(2018)}]{Maggiore:2018sht}%
  \BibitemOpen
  \bibfield  {author} {\bibinfo {author} {\bibfnamefont {M.}~\bibnamefont
  {Maggiore}},\ }\href@noop {} {\emph {\bibinfo {title} {{Gravitational Waves.
  Vol. 2: Astrophysics and Cosmology}}}}\ (\bibinfo  {publisher} {Oxford
  University Press},\ \bibinfo {year} {2018})\BibitemShut {NoStop}%
\bibitem [{\citenamefont {Ferrari}\ \emph {et~al.}(2020)\citenamefont
  {Ferrari}, \citenamefont {Gualtieri},\ and\ \citenamefont
  {Pani}}]{Ferrari:2020nzo}%
  \BibitemOpen
  \bibfield  {author} {\bibinfo {author} {\bibfnamefont {V.}~\bibnamefont
  {Ferrari}}, \bibinfo {author} {\bibfnamefont {L.}~\bibnamefont {Gualtieri}},
  \ and\ \bibinfo {author} {\bibfnamefont {P.}~\bibnamefont {Pani}},\ }\href
  {\doibase 10.1201/9780429491405} {\emph {\bibinfo {title} {{General
  Relativity and its Applications}}}}\ (\bibinfo  {publisher} {CRC Press},\
  \bibinfo {year} {2020})\BibitemShut {NoStop}%
\bibitem [{\citenamefont {Lattimer}\ and\ \citenamefont
  {Prakash}(2001)}]{Lattimer:2000nx}%
  \BibitemOpen
  \bibfield  {author} {\bibinfo {author} {\bibfnamefont {J.~M.}\ \bibnamefont
  {Lattimer}}\ and\ \bibinfo {author} {\bibfnamefont {M.}~\bibnamefont
  {Prakash}},\ }\href {\doibase 10.1086/319702} {\bibfield  {journal} {\bibinfo
   {journal} {Astrophys. J.}\ }\textbf {\bibinfo {volume} {550}},\ \bibinfo
  {pages} {426} (\bibinfo {year} {2001})},\ \Eprint
  {http://arxiv.org/abs/astro-ph/0002232} {arXiv:astro-ph/0002232} \BibitemShut
  {NoStop}%
\bibitem [{\citenamefont {Podkowka}\ \emph {et~al.}(2018)\citenamefont
  {Podkowka}, \citenamefont {Mendes},\ and\ \citenamefont
  {Poisson}}]{Podkowka:2018gib}%
  \BibitemOpen
  \bibfield  {author} {\bibinfo {author} {\bibfnamefont {D.~M.}\ \bibnamefont
  {Podkowka}}, \bibinfo {author} {\bibfnamefont {R.~F.~P.}\ \bibnamefont
  {Mendes}}, \ and\ \bibinfo {author} {\bibfnamefont {E.}~\bibnamefont
  {Poisson}},\ }\href {\doibase 10.1103/PhysRevD.98.064057} {\bibfield
  {journal} {\bibinfo  {journal} {Phys. Rev. D}\ }\textbf {\bibinfo {volume}
  {98}},\ \bibinfo {pages} {064057} (\bibinfo {year} {2018})},\ \Eprint
  {http://arxiv.org/abs/1807.01565} {arXiv:1807.01565 [gr-qc]} \BibitemShut
  {NoStop}%
\bibitem [{\citenamefont {Haensel}\ \emph {et~al.}(2007)\citenamefont
  {Haensel}, \citenamefont {Potekhin},\ and\ \citenamefont
  {Yakovlev}}]{Haensel:2007yy}%
  \BibitemOpen
  \bibfield  {author} {\bibinfo {author} {\bibfnamefont {P.}~\bibnamefont
  {Haensel}}, \bibinfo {author} {\bibfnamefont {A.~Y.}\ \bibnamefont
  {Potekhin}}, \ and\ \bibinfo {author} {\bibfnamefont {D.~G.}\ \bibnamefont
  {Yakovlev}},\ }\href {\doibase 10.1007/978-0-387-47301-7} {\emph {\bibinfo
  {title} {{Neutron stars 1: Equation of state and structure}}}},\ Vol.\
  \bibinfo {volume} {326}\ (\bibinfo  {publisher} {Springer},\ \bibinfo
  {address} {New York, USA},\ \bibinfo {year} {2007})\BibitemShut {NoStop}%
\bibitem [{\citenamefont {Read}\ \emph {et~al.}(2009)\citenamefont {Read},
  \citenamefont {Lackey}, \citenamefont {Owen},\ and\ \citenamefont
  {Friedman}}]{Read:2008iy}%
  \BibitemOpen
  \bibfield  {author} {\bibinfo {author} {\bibfnamefont {J.~S.}\ \bibnamefont
  {Read}}, \bibinfo {author} {\bibfnamefont {B.~D.}\ \bibnamefont {Lackey}},
  \bibinfo {author} {\bibfnamefont {B.~J.}\ \bibnamefont {Owen}}, \ and\
  \bibinfo {author} {\bibfnamefont {J.~L.}\ \bibnamefont {Friedman}},\ }\href
  {\doibase 10.1103/PhysRevD.79.124032} {\bibfield  {journal} {\bibinfo
  {journal} {Phys. Rev. D}\ }\textbf {\bibinfo {volume} {79}},\ \bibinfo
  {pages} {124032} (\bibinfo {year} {2009})},\ \Eprint
  {http://arxiv.org/abs/0812.2163} {arXiv:0812.2163 [astro-ph]} \BibitemShut
  {NoStop}%
\bibitem [{\citenamefont {Douchin}\ and\ \citenamefont
  {Haensel}(2001)}]{Douchin:2001sv}%
  \BibitemOpen
  \bibfield  {author} {\bibinfo {author} {\bibfnamefont {F.}~\bibnamefont
  {Douchin}}\ and\ \bibinfo {author} {\bibfnamefont {P.}~\bibnamefont
  {Haensel}},\ }\href {\doibase 10.1051/0004-6361:20011402} {\bibfield
  {journal} {\bibinfo  {journal} {Astron. Astrophys.}\ }\textbf {\bibinfo
  {volume} {380}},\ \bibinfo {pages} {151} (\bibinfo {year} {2001})},\ \Eprint
  {http://arxiv.org/abs/astro-ph/0111092} {arXiv:astro-ph/0111092} \BibitemShut
  {NoStop}%
\bibitem [{\citenamefont {{Damour}}(2009)}]{Damour:2007uf}%
  \BibitemOpen
  \bibfield  {author} {\bibinfo {author} {\bibfnamefont {T.}~\bibnamefont
  {{Damour}}},\ }in\ \href@noop {} {\emph {\bibinfo {booktitle} {{Physics of
  Relativistic Objects in Compact Binaries: From Birth to Coalescence}}}},\
  Vol.\ \bibinfo {volume} {359},\ \bibinfo {editor} {edited by\ \bibinfo
  {editor} {\bibfnamefont {M.}~\bibnamefont {Colpi}}, \bibinfo {editor}
  {\bibfnamefont {P.}~\bibnamefont {Casella}}, \bibinfo {editor} {\bibfnamefont
  {V.}~\bibnamefont {Gorini}}, \bibinfo {editor} {\bibfnamefont
  {U.}~\bibnamefont {Moschella}}, \ and\ \bibinfo {editor} {\bibfnamefont
  {A.}~\bibnamefont {Possenti}}}\ (\bibinfo  {publisher} {Springer,
  Dordrecht},\ \bibinfo {year} {2009})\ p.~\bibinfo {pages} {1},\ \Eprint
  {http://arxiv.org/abs/0704.0749} {arXiv:0704.0749 [gr-qc]} \BibitemShut
  {NoStop}%
\bibitem [{\citenamefont {Arnowitt}\ \emph {et~al.}(1959)\citenamefont
  {Arnowitt}, \citenamefont {Deser},\ and\ \citenamefont
  {Misner}}]{Arnowitt:1959ah}%
  \BibitemOpen
  \bibfield  {author} {\bibinfo {author} {\bibfnamefont {R.~L.}\ \bibnamefont
  {Arnowitt}}, \bibinfo {author} {\bibfnamefont {S.}~\bibnamefont {Deser}}, \
  and\ \bibinfo {author} {\bibfnamefont {C.~W.}\ \bibnamefont {Misner}},\
  }\href {\doibase 10.1103/PhysRev.116.1322} {\bibfield  {journal} {\bibinfo
  {journal} {Phys. Rev.}\ }\textbf {\bibinfo {volume} {116}},\ \bibinfo {pages}
  {1322} (\bibinfo {year} {1959})}\BibitemShut {NoStop}%
\bibitem [{\citenamefont {Ji}\ \emph {et~al.}(2024)\citenamefont {Ji},
  \citenamefont {Li}, \citenamefont {Yang}, \citenamefont {Xu}, \citenamefont
  {Hu},\ and\ \citenamefont {Shao}}]{Ji:2024aeg}%
  \BibitemOpen
  \bibfield  {author} {\bibinfo {author} {\bibfnamefont {P.}~\bibnamefont
  {Ji}}, \bibinfo {author} {\bibfnamefont {Z.}~\bibnamefont {Li}}, \bibinfo
  {author} {\bibfnamefont {L.}~\bibnamefont {Yang}}, \bibinfo {author}
  {\bibfnamefont {R.}~\bibnamefont {Xu}}, \bibinfo {author} {\bibfnamefont
  {Z.}~\bibnamefont {Hu}}, \ and\ \bibinfo {author} {\bibfnamefont
  {L.}~\bibnamefont {Shao}},\ }\href {\doibase 10.1103/PhysRevD.110.104057}
  {\bibfield  {journal} {\bibinfo  {journal} {Phys. Rev. D}\ }\textbf {\bibinfo
  {volume} {110}},\ \bibinfo {pages} {104057} (\bibinfo {year} {2024})},\
  \Eprint {http://arxiv.org/abs/2409.04805} {arXiv:2409.04805 [gr-qc]}
  \BibitemShut {NoStop}%
\bibitem [{\citenamefont {Raithel}\ \emph {et~al.}(2016)\citenamefont
  {Raithel}, \citenamefont {Ozel},\ and\ \citenamefont
  {Psaltis}}]{Raithel:2016vtt}%
  \BibitemOpen
  \bibfield  {author} {\bibinfo {author} {\bibfnamefont {C.~A.}\ \bibnamefont
  {Raithel}}, \bibinfo {author} {\bibfnamefont {F.}~\bibnamefont {Ozel}}, \
  and\ \bibinfo {author} {\bibfnamefont {D.}~\bibnamefont {Psaltis}},\ }\href
  {\doibase 10.1103/PhysRevC.93.032801} {\bibfield  {journal} {\bibinfo
  {journal} {Phys. Rev. C}\ }\textbf {\bibinfo {volume} {93}},\ \bibinfo
  {pages} {032801} (\bibinfo {year} {2016})},\ \bibinfo {note} {[Addendum:
  Phys.Rev.C 93, 049905 (2016)]},\ \Eprint {http://arxiv.org/abs/1603.06594}
  {arXiv:1603.06594 [astro-ph.HE]} \BibitemShut {NoStop}%
\bibitem [{\citenamefont {Breu}\ and\ \citenamefont
  {Rezzolla}(2016)}]{Breu:2016ufb}%
  \BibitemOpen
  \bibfield  {author} {\bibinfo {author} {\bibfnamefont {C.}~\bibnamefont
  {Breu}}\ and\ \bibinfo {author} {\bibfnamefont {L.}~\bibnamefont
  {Rezzolla}},\ }\href {\doibase 10.1093/mnras/stw575} {\bibfield  {journal}
  {\bibinfo  {journal} {Mon. Not. Roy. Astron. Soc.}\ }\textbf {\bibinfo
  {volume} {459}},\ \bibinfo {pages} {646} (\bibinfo {year} {2016})},\ \Eprint
  {http://arxiv.org/abs/1601.06083} {arXiv:1601.06083 [gr-qc]} \BibitemShut
  {NoStop}%
\bibitem [{\citenamefont {Gao}\ \emph {et~al.}(2021)\citenamefont {Gao},
  \citenamefont {Lai}, \citenamefont {Shao},\ and\ \citenamefont
  {Xu}}]{Gao:2021uus}%
  \BibitemOpen
  \bibfield  {author} {\bibinfo {author} {\bibfnamefont {Y.}~\bibnamefont
  {Gao}}, \bibinfo {author} {\bibfnamefont {X.-Y.}\ \bibnamefont {Lai}},
  \bibinfo {author} {\bibfnamefont {L.}~\bibnamefont {Shao}}, \ and\ \bibinfo
  {author} {\bibfnamefont {R.-X.}\ \bibnamefont {Xu}},\ }\href {\doibase
  10.1093/mnras/stab3181} {\bibfield  {journal} {\bibinfo  {journal} {Mon. Not.
  Roy. Astron. Soc.}\ }\textbf {\bibinfo {volume} {509}},\ \bibinfo {pages}
  {2758} (\bibinfo {year} {2021})},\ \Eprint {http://arxiv.org/abs/2109.13234}
  {arXiv:2109.13234 [gr-qc]} \BibitemShut {NoStop}%
\bibitem [{\citenamefont {Yagi}\ and\ \citenamefont
  {Yunes}(2013{\natexlab{a}})}]{Yagi:2013awa}%
  \BibitemOpen
  \bibfield  {author} {\bibinfo {author} {\bibfnamefont {K.}~\bibnamefont
  {Yagi}}\ and\ \bibinfo {author} {\bibfnamefont {N.}~\bibnamefont {Yunes}},\
  }\href {\doibase 10.1103/PhysRevD.88.023009} {\bibfield  {journal} {\bibinfo
  {journal} {Phys. Rev. D}\ }\textbf {\bibinfo {volume} {88}},\ \bibinfo
  {pages} {023009} (\bibinfo {year} {2013}{\natexlab{a}})},\ \Eprint
  {http://arxiv.org/abs/1303.1528} {arXiv:1303.1528 [gr-qc]} \BibitemShut
  {NoStop}%
\bibitem [{\citenamefont {Shao}\ and\ \citenamefont
  {Yagi}(2022)}]{Shao:2022koz}%
  \BibitemOpen
  \bibfield  {author} {\bibinfo {author} {\bibfnamefont {L.}~\bibnamefont
  {Shao}}\ and\ \bibinfo {author} {\bibfnamefont {K.}~\bibnamefont {Yagi}},\
  }\href {\doibase 10.1016/j.scib.2022.09.018} {\bibfield  {journal} {\bibinfo
  {journal} {Sci. Bull.}\ }\textbf {\bibinfo {volume} {67}},\ \bibinfo {pages}
  {1946} (\bibinfo {year} {2022})},\ \Eprint {http://arxiv.org/abs/2209.03351}
  {arXiv:2209.03351 [gr-qc]} \BibitemShut {NoStop}%
\bibitem [{\citenamefont {Landry}\ and\ \citenamefont
  {Kumar}(2018)}]{Landry:2018jyg}%
  \BibitemOpen
  \bibfield  {author} {\bibinfo {author} {\bibfnamefont {P.}~\bibnamefont
  {Landry}}\ and\ \bibinfo {author} {\bibfnamefont {B.}~\bibnamefont {Kumar}},\
  }\href {\doibase 10.3847/2041-8213/aaee76} {\bibfield  {journal} {\bibinfo
  {journal} {Astrophys. J. Lett.}\ }\textbf {\bibinfo {volume} {868}},\
  \bibinfo {pages} {L22} (\bibinfo {year} {2018})},\ \Eprint
  {http://arxiv.org/abs/1807.04727} {arXiv:1807.04727 [gr-qc]} \BibitemShut
  {NoStop}%
\bibitem [{\citenamefont {Flanagan}\ and\ \citenamefont
  {Hinderer}(2008)}]{Flanagan:2007ix}%
  \BibitemOpen
  \bibfield  {author} {\bibinfo {author} {\bibfnamefont {E.~E.}\ \bibnamefont
  {Flanagan}}\ and\ \bibinfo {author} {\bibfnamefont {T.}~\bibnamefont
  {Hinderer}},\ }\href {\doibase 10.1103/PhysRevD.77.021502} {\bibfield
  {journal} {\bibinfo  {journal} {Phys. Rev. D}\ }\textbf {\bibinfo {volume}
  {77}},\ \bibinfo {pages} {021502} (\bibinfo {year} {2008})},\ \Eprint
  {http://arxiv.org/abs/0709.1915} {arXiv:0709.1915 [astro-ph]} \BibitemShut
  {NoStop}%
\bibitem [{\citenamefont {Damour}\ and\ \citenamefont
  {Nagar}(2009)}]{Damour:2009vw}%
  \BibitemOpen
  \bibfield  {author} {\bibinfo {author} {\bibfnamefont {T.}~\bibnamefont
  {Damour}}\ and\ \bibinfo {author} {\bibfnamefont {A.}~\bibnamefont {Nagar}},\
  }\href {\doibase 10.1103/PhysRevD.80.084035} {\bibfield  {journal} {\bibinfo
  {journal} {Phys. Rev. D}\ }\textbf {\bibinfo {volume} {80}},\ \bibinfo
  {pages} {084035} (\bibinfo {year} {2009})},\ \Eprint
  {http://arxiv.org/abs/0906.0096} {arXiv:0906.0096 [gr-qc]} \BibitemShut
  {NoStop}%
\bibitem [{\citenamefont {Hinderer}\ \emph {et~al.}(2010)\citenamefont
  {Hinderer}, \citenamefont {Lackey}, \citenamefont {Lang},\ and\ \citenamefont
  {Read}}]{Hinderer:2009ca}%
  \BibitemOpen
  \bibfield  {author} {\bibinfo {author} {\bibfnamefont {T.}~\bibnamefont
  {Hinderer}}, \bibinfo {author} {\bibfnamefont {B.~D.}\ \bibnamefont
  {Lackey}}, \bibinfo {author} {\bibfnamefont {R.~N.}\ \bibnamefont {Lang}}, \
  and\ \bibinfo {author} {\bibfnamefont {J.~S.}\ \bibnamefont {Read}},\ }\href
  {\doibase 10.1103/PhysRevD.81.123016} {\bibfield  {journal} {\bibinfo
  {journal} {Phys. Rev. D}\ }\textbf {\bibinfo {volume} {81}},\ \bibinfo
  {pages} {123016} (\bibinfo {year} {2010})},\ \Eprint
  {http://arxiv.org/abs/0911.3535} {arXiv:0911.3535 [astro-ph.HE]} \BibitemShut
  {NoStop}%
\bibitem [{\citenamefont {Silva}\ \emph {et~al.}(2021)\citenamefont {Silva},
  \citenamefont {Holgado}, \citenamefont {C{\'a}rdenas-Avenda{\~n}o},\ and\
  \citenamefont {Yunes}}]{Silva:2020acr}%
  \BibitemOpen
  \bibfield  {author} {\bibinfo {author} {\bibfnamefont {H.~O.}\ \bibnamefont
  {Silva}}, \bibinfo {author} {\bibfnamefont {A.~M.}\ \bibnamefont {Holgado}},
  \bibinfo {author} {\bibfnamefont {A.}~\bibnamefont
  {C{\'a}rdenas-Avenda{\~n}o}}, \ and\ \bibinfo {author} {\bibfnamefont
  {N.}~\bibnamefont {Yunes}},\ }\href {\doibase 10.1103/PhysRevLett.126.181101}
  {\bibfield  {journal} {\bibinfo  {journal} {Phys. Rev. Lett.}\ }\textbf
  {\bibinfo {volume} {126}},\ \bibinfo {pages} {181101} (\bibinfo {year}
  {2021})},\ \Eprint {http://arxiv.org/abs/2004.01253} {arXiv:2004.01253
  [gr-qc]} \BibitemShut {NoStop}%
\bibitem [{\citenamefont {Pani}\ and\ \citenamefont
  {Berti}(2014)}]{Pani:2014jra}%
  \BibitemOpen
  \bibfield  {author} {\bibinfo {author} {\bibfnamefont {P.}~\bibnamefont
  {Pani}}\ and\ \bibinfo {author} {\bibfnamefont {E.}~\bibnamefont {Berti}},\
  }\href {\doibase 10.1103/PhysRevD.90.024025} {\bibfield  {journal} {\bibinfo
  {journal} {Phys. Rev. D}\ }\textbf {\bibinfo {volume} {90}},\ \bibinfo
  {pages} {024025} (\bibinfo {year} {2014})},\ \Eprint
  {http://arxiv.org/abs/1405.4547} {arXiv:1405.4547 [gr-qc]} \BibitemShut
  {NoStop}%
\bibitem [{\citenamefont {Brown}(2023)}]{Brown:2022kbw}%
  \BibitemOpen
  \bibfield  {author} {\bibinfo {author} {\bibfnamefont {S.~M.}\ \bibnamefont
  {Brown}},\ }\href {\doibase 10.3847/1538-4357/acfbe5} {\bibfield  {journal}
  {\bibinfo  {journal} {Astrophys. J.}\ }\textbf {\bibinfo {volume} {958}},\
  \bibinfo {pages} {125} (\bibinfo {year} {2023})},\ \Eprint
  {http://arxiv.org/abs/2210.14025} {arXiv:2210.14025 [gr-qc]} \BibitemShut
  {NoStop}%
\bibitem [{\citenamefont {Abbott}\ \emph {et~al.}(2017)\citenamefont {Abbott}
  \emph {et~al.}}]{LIGOScientific:2017vwq}%
  \BibitemOpen
  \bibfield  {author} {\bibinfo {author} {\bibfnamefont {B.~P.}\ \bibnamefont
  {Abbott}} \emph {et~al.} (\bibinfo {collaboration} {LIGO Scientific,
  Virgo}),\ }\href {\doibase 10.1103/PhysRevLett.119.161101} {\bibfield
  {journal} {\bibinfo  {journal} {Phys. Rev. Lett.}\ }\textbf {\bibinfo
  {volume} {119}},\ \bibinfo {pages} {161101} (\bibinfo {year} {2017})},\
  \Eprint {http://arxiv.org/abs/1710.05832} {arXiv:1710.05832 [gr-qc]}
  \BibitemShut {NoStop}%
\bibitem [{\citenamefont {Abbott}\ \emph
  {et~al.}(2019{\natexlab{b}})\citenamefont {Abbott} \emph
  {et~al.}}]{LIGOScientific:2018hze}%
  \BibitemOpen
  \bibfield  {author} {\bibinfo {author} {\bibfnamefont {B.~P.}\ \bibnamefont
  {Abbott}} \emph {et~al.} (\bibinfo {collaboration} {LIGO Scientific,
  Virgo}),\ }\href {\doibase 10.1103/PhysRevX.9.011001} {\bibfield  {journal}
  {\bibinfo  {journal} {Phys. Rev. X}\ }\textbf {\bibinfo {volume} {9}},\
  \bibinfo {pages} {011001} (\bibinfo {year} {2019}{\natexlab{b}})},\ \Eprint
  {http://arxiv.org/abs/1805.11579} {arXiv:1805.11579 [gr-qc]} \BibitemShut
  {NoStop}%
\bibitem [{\citenamefont {Yagi}\ and\ \citenamefont
  {Yunes}(2013{\natexlab{b}})}]{Yagi:2013bca}%
  \BibitemOpen
  \bibfield  {author} {\bibinfo {author} {\bibfnamefont {K.}~\bibnamefont
  {Yagi}}\ and\ \bibinfo {author} {\bibfnamefont {N.}~\bibnamefont {Yunes}},\
  }\href {\doibase 10.1126/science.1236462} {\bibfield  {journal} {\bibinfo
  {journal} {Science}\ }\textbf {\bibinfo {volume} {341}},\ \bibinfo {pages}
  {365} (\bibinfo {year} {2013}{\natexlab{b}})},\ \Eprint
  {http://arxiv.org/abs/1302.4499} {arXiv:1302.4499 [gr-qc]} \BibitemShut
  {NoStop}%
\bibitem [{\citenamefont {Doneva}\ \emph {et~al.}(2013)\citenamefont {Doneva},
  \citenamefont {Yazadjiev}, \citenamefont {Stergioulas},\ and\ \citenamefont
  {Kokkotas}}]{Doneva:2013rha}%
  \BibitemOpen
  \bibfield  {author} {\bibinfo {author} {\bibfnamefont {D.~D.}\ \bibnamefont
  {Doneva}}, \bibinfo {author} {\bibfnamefont {S.~S.}\ \bibnamefont
  {Yazadjiev}}, \bibinfo {author} {\bibfnamefont {N.}~\bibnamefont
  {Stergioulas}}, \ and\ \bibinfo {author} {\bibfnamefont {K.~D.}\ \bibnamefont
  {Kokkotas}},\ }\href {\doibase 10.1088/2041-8205/781/1/L6} {\bibfield
  {journal} {\bibinfo  {journal} {Astrophys. J. Lett.}\ }\textbf {\bibinfo
  {volume} {781}},\ \bibinfo {pages} {L6} (\bibinfo {year} {2013})},\ \Eprint
  {http://arxiv.org/abs/1310.7436} {arXiv:1310.7436 [gr-qc]} \BibitemShut
  {NoStop}%
\bibitem [{\citenamefont {Doneva}\ \emph {et~al.}(2014)\citenamefont {Doneva},
  \citenamefont {Yazadjiev}, \citenamefont {Staykov},\ and\ \citenamefont
  {Kokkotas}}]{Doneva:2014faa}%
  \BibitemOpen
  \bibfield  {author} {\bibinfo {author} {\bibfnamefont {D.~D.}\ \bibnamefont
  {Doneva}}, \bibinfo {author} {\bibfnamefont {S.~S.}\ \bibnamefont
  {Yazadjiev}}, \bibinfo {author} {\bibfnamefont {K.~V.}\ \bibnamefont
  {Staykov}}, \ and\ \bibinfo {author} {\bibfnamefont {K.~D.}\ \bibnamefont
  {Kokkotas}},\ }\href {\doibase 10.1103/PhysRevD.90.104021} {\bibfield
  {journal} {\bibinfo  {journal} {Phys. Rev. D}\ }\textbf {\bibinfo {volume}
  {90}},\ \bibinfo {pages} {104021} (\bibinfo {year} {2014})},\ \Eprint
  {http://arxiv.org/abs/1408.1641} {arXiv:1408.1641 [gr-qc]} \BibitemShut
  {NoStop}%
\bibitem [{\citenamefont {Mendes}\ and\ \citenamefont
  {Ottoni}(2019)}]{Mendes:2019zpw}%
  \BibitemOpen
  \bibfield  {author} {\bibinfo {author} {\bibfnamefont {R.~F.~P.}\
  \bibnamefont {Mendes}}\ and\ \bibinfo {author} {\bibfnamefont
  {T.}~\bibnamefont {Ottoni}},\ }\href {\doibase 10.1103/PhysRevD.99.124003}
  {\bibfield  {journal} {\bibinfo  {journal} {Phys. Rev. D}\ }\textbf {\bibinfo
  {volume} {99}},\ \bibinfo {pages} {124003} (\bibinfo {year} {2019})},\
  \Eprint {http://arxiv.org/abs/1903.11638} {arXiv:1903.11638 [gr-qc]}
  \BibitemShut {NoStop}%
\bibitem [{\citenamefont {Hartle}(1967)}]{Hartle:1967he}%
  \BibitemOpen
  \bibfield  {author} {\bibinfo {author} {\bibfnamefont {J.~B.}\ \bibnamefont
  {Hartle}},\ }\href {\doibase 10.1086/149400} {\bibfield  {journal} {\bibinfo
  {journal} {Astrophys. J.}\ }\textbf {\bibinfo {volume} {150}},\ \bibinfo
  {pages} {1005} (\bibinfo {year} {1967})}\BibitemShut {NoStop}%
\bibitem [{\citenamefont {Regge}\ and\ \citenamefont
  {Wheeler}(1957)}]{Regge:1957td}%
  \BibitemOpen
  \bibfield  {author} {\bibinfo {author} {\bibfnamefont {T.}~\bibnamefont
  {Regge}}\ and\ \bibinfo {author} {\bibfnamefont {J.~A.}\ \bibnamefont
  {Wheeler}},\ }\href {\doibase 10.1103/PhysRev.108.1063} {\bibfield  {journal}
  {\bibinfo  {journal} {Phys. Rev.}\ }\textbf {\bibinfo {volume} {108}},\
  \bibinfo {pages} {1063} (\bibinfo {year} {1957})}\BibitemShut {NoStop}%
\bibitem [{\citenamefont {Hinderer}(2008)}]{Hinderer:2007mb}%
  \BibitemOpen
  \bibfield  {author} {\bibinfo {author} {\bibfnamefont {T.}~\bibnamefont
  {Hinderer}},\ }\href {\doibase 10.1086/533487} {\bibfield  {journal}
  {\bibinfo  {journal} {Astrophys. J.}\ }\textbf {\bibinfo {volume} {677}},\
  \bibinfo {pages} {1216} (\bibinfo {year} {2008})},\ \bibinfo {note}
  {[Erratum: Astrophys.J. 697, 964 (2009)]},\ \Eprint
  {http://arxiv.org/abs/0711.2420} {arXiv:0711.2420 [astro-ph]} \BibitemShut
  {NoStop}%
\bibitem [{\citenamefont {Creci}\ \emph {et~al.}(2023)\citenamefont {Creci},
  \citenamefont {Hinderer},\ and\ \citenamefont {Steinhoff}}]{Creci:2023cfx}%
  \BibitemOpen
  \bibfield  {author} {\bibinfo {author} {\bibfnamefont {G.}~\bibnamefont
  {Creci}}, \bibinfo {author} {\bibfnamefont {T.}~\bibnamefont {Hinderer}}, \
  and\ \bibinfo {author} {\bibfnamefont {J.}~\bibnamefont {Steinhoff}},\ }\href
  {\doibase 10.1103/PhysRevD.108.124073} {\bibfield  {journal} {\bibinfo
  {journal} {Phys. Rev. D}\ }\textbf {\bibinfo {volume} {108}},\ \bibinfo
  {pages} {124073} (\bibinfo {year} {2023})},\ \bibinfo {note} {[Erratum:
  Phys.Rev.D 111, 089901 (2025)]},\ \Eprint {http://arxiv.org/abs/2308.11323}
  {arXiv:2308.11323 [gr-qc]} \BibitemShut {NoStop}%
\bibitem [{\citenamefont {Creci}\ \emph {et~al.}(2025)\citenamefont {Creci},
  \citenamefont {van Gemeren}, \citenamefont {Hinderer},\ and\ \citenamefont
  {Steinhoff}}]{Creci:2024wfu}%
  \BibitemOpen
  \bibfield  {author} {\bibinfo {author} {\bibfnamefont {G.}~\bibnamefont
  {Creci}}, \bibinfo {author} {\bibfnamefont {I.}~\bibnamefont {van Gemeren}},
  \bibinfo {author} {\bibfnamefont {T.}~\bibnamefont {Hinderer}}, \ and\
  \bibinfo {author} {\bibfnamefont {J.}~\bibnamefont {Steinhoff}},\ }\href
  {\doibase 10.21468/SciPostPhysCore.8.2.042} {\bibfield  {journal} {\bibinfo
  {journal} {SciPost Phys. Core}\ }\textbf {\bibinfo {volume} {8}},\ \bibinfo
  {pages} {042} (\bibinfo {year} {2025})},\ \Eprint
  {http://arxiv.org/abs/2412.06620} {arXiv:2412.06620 [gr-qc]} \BibitemShut
  {NoStop}%
\bibitem [{\citenamefont {Binnington}\ and\ \citenamefont
  {Poisson}(2009)}]{Binnington:2009bb}%
  \BibitemOpen
  \bibfield  {author} {\bibinfo {author} {\bibfnamefont {T.}~\bibnamefont
  {Binnington}}\ and\ \bibinfo {author} {\bibfnamefont {E.}~\bibnamefont
  {Poisson}},\ }\href {\doibase 10.1103/PhysRevD.80.084018} {\bibfield
  {journal} {\bibinfo  {journal} {Phys. Rev. D}\ }\textbf {\bibinfo {volume}
  {80}},\ \bibinfo {pages} {084018} (\bibinfo {year} {2009})},\ \Eprint
  {http://arxiv.org/abs/0906.1366} {arXiv:0906.1366 [gr-qc]} \BibitemShut
  {NoStop}%
\bibitem [{\citenamefont {Antoniadis}\ \emph {et~al.}(2013)\citenamefont
  {Antoniadis} \emph {et~al.}}]{Antoniadis:2013pzd}%
  \BibitemOpen
  \bibfield  {author} {\bibinfo {author} {\bibfnamefont {J.}~\bibnamefont
  {Antoniadis}} \emph {et~al.},\ }\href {\doibase 10.1126/science.1233232}
  {\bibfield  {journal} {\bibinfo  {journal} {Science}\ }\textbf {\bibinfo
  {volume} {340}},\ \bibinfo {pages} {6131} (\bibinfo {year} {2013})},\ \Eprint
  {http://arxiv.org/abs/1304.6875} {arXiv:1304.6875 [astro-ph.HE]} \BibitemShut
  {NoStop}%
\bibitem [{\citenamefont {Fonseca}\ \emph {et~al.}(2021)\citenamefont {Fonseca}
  \emph {et~al.}}]{Fonseca:2021wxt}%
  \BibitemOpen
  \bibfield  {author} {\bibinfo {author} {\bibfnamefont {E.}~\bibnamefont
  {Fonseca}} \emph {et~al.},\ }\href {\doibase 10.3847/2041-8213/ac03b8}
  {\bibfield  {journal} {\bibinfo  {journal} {Astrophys. J. Lett.}\ }\textbf
  {\bibinfo {volume} {915}},\ \bibinfo {pages} {L12} (\bibinfo {year}
  {2021})},\ \Eprint {http://arxiv.org/abs/2104.00880} {arXiv:2104.00880
  [astro-ph.HE]} \BibitemShut {NoStop}%
\bibitem [{\citenamefont {Harada}(1998)}]{Harada:1998ge}%
  \BibitemOpen
  \bibfield  {author} {\bibinfo {author} {\bibfnamefont {T.}~\bibnamefont
  {Harada}},\ }\href {\doibase 10.1103/PhysRevD.57.4802} {\bibfield  {journal}
  {\bibinfo  {journal} {Phys. Rev. D}\ }\textbf {\bibinfo {volume} {57}},\
  \bibinfo {pages} {4802} (\bibinfo {year} {1998})},\ \Eprint
  {http://arxiv.org/abs/gr-qc/9801049} {arXiv:gr-qc/9801049} \BibitemShut
  {NoStop}%
\bibitem [{\citenamefont {Novak}(1998)}]{Novak:1997hw}%
  \BibitemOpen
  \bibfield  {author} {\bibinfo {author} {\bibfnamefont {J.}~\bibnamefont
  {Novak}},\ }\href {\doibase 10.1103/PhysRevD.57.4789} {\bibfield  {journal}
  {\bibinfo  {journal} {Phys. Rev. D}\ }\textbf {\bibinfo {volume} {57}},\
  \bibinfo {pages} {4789} (\bibinfo {year} {1998})},\ \Eprint
  {http://arxiv.org/abs/gr-qc/9707041} {arXiv:gr-qc/9707041} \BibitemShut
  {NoStop}%
\bibitem [{\citenamefont {Altaha~Motahar}\ \emph {et~al.}(2017)\citenamefont
  {Altaha~Motahar}, \citenamefont {Bl{\'a}zquez-Salcedo}, \citenamefont
  {Kleihaus},\ and\ \citenamefont {Kunz}}]{AltahaMotahar:2017ijw}%
  \BibitemOpen
  \bibfield  {author} {\bibinfo {author} {\bibfnamefont {Z.}~\bibnamefont
  {Altaha~Motahar}}, \bibinfo {author} {\bibfnamefont {J.~L.}\ \bibnamefont
  {Bl{\'a}zquez-Salcedo}}, \bibinfo {author} {\bibfnamefont {B.}~\bibnamefont
  {Kleihaus}}, \ and\ \bibinfo {author} {\bibfnamefont {J.}~\bibnamefont
  {Kunz}},\ }\href {\doibase 10.1103/PhysRevD.96.064046} {\bibfield  {journal}
  {\bibinfo  {journal} {Phys. Rev. D}\ }\textbf {\bibinfo {volume} {96}},\
  \bibinfo {pages} {064046} (\bibinfo {year} {2017})},\ \Eprint
  {http://arxiv.org/abs/1707.05280} {arXiv:1707.05280 [gr-qc]} \BibitemShut
  {NoStop}%
\bibitem [{\citenamefont {Palenzuela}\ and\ \citenamefont
  {Liebling}(2016)}]{Palenzuela:2015ima}%
  \BibitemOpen
  \bibfield  {author} {\bibinfo {author} {\bibfnamefont {C.}~\bibnamefont
  {Palenzuela}}\ and\ \bibinfo {author} {\bibfnamefont {S.~L.}\ \bibnamefont
  {Liebling}},\ }\href {\doibase 10.1103/PhysRevD.93.044009} {\bibfield
  {journal} {\bibinfo  {journal} {Phys. Rev. D}\ }\textbf {\bibinfo {volume}
  {93}},\ \bibinfo {pages} {044009} (\bibinfo {year} {2016})},\ \Eprint
  {http://arxiv.org/abs/1510.03471} {arXiv:1510.03471 [gr-qc]} \BibitemShut
  {NoStop}%
\bibitem [{\citenamefont {Mendes}\ and\ \citenamefont
  {Ortiz}(2018)}]{Mendes:2018qwo}%
  \BibitemOpen
  \bibfield  {author} {\bibinfo {author} {\bibfnamefont {R.~F.~P.}\
  \bibnamefont {Mendes}}\ and\ \bibinfo {author} {\bibfnamefont
  {N.}~\bibnamefont {Ortiz}},\ }\href {\doibase 10.1103/PhysRevLett.120.201104}
  {\bibfield  {journal} {\bibinfo  {journal} {Phys. Rev. Lett.}\ }\textbf
  {\bibinfo {volume} {120}},\ \bibinfo {pages} {201104} (\bibinfo {year}
  {2018})},\ \Eprint {http://arxiv.org/abs/1802.07847} {arXiv:1802.07847
  [gr-qc]} \BibitemShut {NoStop}%
\bibitem [{\citenamefont {{Harrison}}\ \emph {et~al.}(1965)\citenamefont
  {{Harrison}}, \citenamefont {{Thorne}}, \citenamefont {{Wakano}},\ and\
  \citenamefont {{Wheeler}}}]{1965gtgc.book.....H}%
  \BibitemOpen
  \bibfield  {author} {\bibinfo {author} {\bibfnamefont {B.~K.}\ \bibnamefont
  {{Harrison}}}, \bibinfo {author} {\bibfnamefont {K.~S.}\ \bibnamefont
  {{Thorne}}}, \bibinfo {author} {\bibfnamefont {M.}~\bibnamefont {{Wakano}}},
  \ and\ \bibinfo {author} {\bibfnamefont {J.~A.}\ \bibnamefont {{Wheeler}}},\
  }\href@noop {} {\emph {\bibinfo {title} {{Gravitation Theory and
  Gravitational Collapse}}}}\ (\bibinfo {year} {1965})\BibitemShut {NoStop}%
\bibitem [{\citenamefont {Brito}\ \emph {et~al.}(2016)\citenamefont {Brito},
  \citenamefont {Cardoso}, \citenamefont {Macedo}, \citenamefont {Okawa},\ and\
  \citenamefont {Palenzuela}}]{Brito:2015yfh}%
  \BibitemOpen
  \bibfield  {author} {\bibinfo {author} {\bibfnamefont {R.}~\bibnamefont
  {Brito}}, \bibinfo {author} {\bibfnamefont {V.}~\bibnamefont {Cardoso}},
  \bibinfo {author} {\bibfnamefont {C.~F.~B.}\ \bibnamefont {Macedo}}, \bibinfo
  {author} {\bibfnamefont {H.}~\bibnamefont {Okawa}}, \ and\ \bibinfo {author}
  {\bibfnamefont {C.}~\bibnamefont {Palenzuela}},\ }\href {\doibase
  10.1103/PhysRevD.93.044045} {\bibfield  {journal} {\bibinfo  {journal} {Phys.
  Rev. D}\ }\textbf {\bibinfo {volume} {93}},\ \bibinfo {pages} {044045}
  (\bibinfo {year} {2016})},\ \Eprint {http://arxiv.org/abs/1512.00466}
  {arXiv:1512.00466 [astro-ph.SR]} \BibitemShut {NoStop}%
\bibitem [{\citenamefont {Kokkotas}\ and\ \citenamefont
  {Schmidt}(1999)}]{Kokkotas:1999bd}%
  \BibitemOpen
  \bibfield  {author} {\bibinfo {author} {\bibfnamefont {K.~D.}\ \bibnamefont
  {Kokkotas}}\ and\ \bibinfo {author} {\bibfnamefont {B.~G.}\ \bibnamefont
  {Schmidt}},\ }\href {\doibase 10.12942/lrr-1999-2} {\bibfield  {journal}
  {\bibinfo  {journal} {Living Rev. Rel.}\ }\textbf {\bibinfo {volume} {2}},\
  \bibinfo {pages} {2} (\bibinfo {year} {1999})},\ \Eprint
  {http://arxiv.org/abs/gr-qc/9909058} {arXiv:gr-qc/9909058} \BibitemShut
  {NoStop}%
\bibitem [{\citenamefont {Chandrasekhar}(1964)}]{Chandrasekhar:1964zz}%
  \BibitemOpen
  \bibfield  {author} {\bibinfo {author} {\bibfnamefont {S.}~\bibnamefont
  {Chandrasekhar}},\ }\href {\doibase 10.1086/147938} {\bibfield  {journal}
  {\bibinfo  {journal} {Astrophys. J.}\ }\textbf {\bibinfo {volume} {140}},\
  \bibinfo {pages} {417} (\bibinfo {year} {1964})},\ \bibinfo {note} {[Erratum:
  Astrophys.J. 140, 1342 (1964)]}\BibitemShut {NoStop}%
\end{thebibliography}%

\end{document}